\documentclass[useAMS,usenatbib,twocolumn, fleqn]{mn2e} \usepackage{natbib,
graphics, epsfig, color}

\usepackage{times} 
\usepackage{amsmath} 
\usepackage{amssymb}
\usepackage{textcomp} 
\usepackage{verbatim}
\usepackage{hyperref}
\newcommand{\kpc}{\,{\rm kpc}}

\newcommand{\msun}{{\,\rm M_\odot}}
\newcommand{\kms}{\,{\rm km}\,{\rm s}^{-1}}

\newcommand{\yr}{\,{\rm yr}}
\newcommand{\Gyr}{\,{\rm Gyr}}
\newcommand{\Myr}{\,{\rm Myr}}
\newcommand{\pc}{\,{\rm pc}}

\newcommand{\mpc}{\,{\rm Mpc}}

\newcommand{\apj}{ Astrophys. J.}
\newcommand{\apjl}{Astrophys. J. Let.}
\newcommand{\apjs}{Astrophys. J. Suppl.}
\newcommand{\aj}{Astron. J.}
\newcommand{\mnras}{MNRAS}
\newcommand{\nat}{Nature}
\newcommand{\nar}{NewAR}
\newcommand{\aap}{A\&A}

\newcommand{\pasp}{Pub. Astron. Soc. Pac.}
\newcommand{\araa}{ARA\&A}

\newcommand{\physrep}{Phys. Rep.}

\title[Introducing the Illustris Project] 
      {Introducing the Illustris Project: Simulating the coevolution of dark and visible matter in the Universe} 
       \author[M. Vogelsberger et al.]
      {\parbox{20cm}{ 
       Mark Vogelsberger$^{1}$, Shy Genel$^{2}$, Volker Springel$^{3,4}$, Paul Torrey$^{2}$, Debora Sijacki$^{5}$, 
       \\Dandan Xu$^{3}$, Greg Snyder$^{6}$, Dylan Nelson$^{2}$ , and Lars Hernquist$^{2}$}\vspace{0.3cm}\\ 
       $^1$ Department of Physics, Kavli Institute for Astrophysics and Space Research, Massachusetts Institute of Technology, Cambridge, MA 02139, USA\\
       $^2$ Harvard-Smithsonian Center for Astrophysics, 60 Garden Street, Cambridge, MA 02138, USA\\
       $^3$ Heidelberg Institute for Theoretical Studies, Schloss-Wolfsbrunnenweg 35, 69118 Heidelberg, Germany\\
       $^4$ Zentrum f\"{u}r Astronomie der Universit\"{a}t Heidelberg, ARI, M\"onchhofstr. 12-14, 69120 Heidelberg, Germany\\
       $^5$ Kavli Institute for Cosmology, Cambridge, and Institute of Astronomy, Madingley Road, Cambridge, CB3 0HA, UK\\
       $^6$ Space Telescope Science Institute, 3700 San Martin Drive, Baltimore, MD 21218, USA\\}

\begin{document}

\maketitle

\begin{abstract}
We introduce the Illustris Project, a series of large-scale hydrodynamical
simulations of galaxy formation. The highest resolution simulation,
Illustris-1, covers a volume of $(106.5\mpc)^3$, has a dark mass resolution of
${6.26 \times 10^{6}\msun}$, and an initial baryonic matter mass resolution of
${1.26 \times 10^{6}\msun}$.  At $z=0$ gravitational forces are softened on
scales of $710\pc$, and the smallest hydrodynamical gas cells have an extent of
$48\pc$.  We follow the dynamical evolution of $2\times 1820^3$ resolution
elements and in addition passively evolve $1820^3$  Monte Carlo tracer
particles reaching a total particle count of more than $18$ billion. The galaxy
formation model includes: primordial and metal-line cooling with self-shielding
corrections, stellar evolution, stellar feedback, gas recycling, chemical
enrichment, supermassive black hole growth, and feedback from active galactic
nuclei. Here we describe the simulation suite, and contrast basic predictions
of our model for the present day galaxy population with observations of the
local universe. At $z=0$ our simulation volume contains about $40,000$
well-resolved galaxies covering a diverse range of morphologies and colours
including early-type, late-type and irregular galaxies. The simulation
reproduces reasonably well the cosmic star formation rate density, the galaxy luminosity
function, and baryon conversion efficiency at $z=0$.  It also qualitatively
captures the impact of galaxy environment on the red fractions of galaxies.
The internal velocity structure of selected well-resolved disk galaxies obeys
the stellar and baryonic Tully-Fisher relation together with flat circular
velocity curves. In the well-resolved regime the simulation reproduces the
observed mix of early-type and late-type galaxies.  Our model predicts a halo
mass dependent impact of baryonic effects on the halo mass function and the
masses of haloes caused by feedback from supernova and active galactic nuclei.
\end{abstract}

\begin{keywords} methods: numerical -- cosmology: theory -- cosmology: galaxy formation
\end{keywords}

\section{Introduction}

Over the last decade the amount of available astronomical data on the
large-scale structure of the Universe has grown rapidly, locally, as in the the
Two-degree-Field Galaxy Redshift Survey (2dfGRS)~\citep[][]{Colless2001} and
Sloan Digital Sky Survey (SDSS)~\citep[][]{York2000}, and in surveys at higher
redshifts, such as DEEP~\citep[][]{Weiner2005}, DEEP2~\citep[][]{Davis2003} and the
Cosmic Assembly Near-IR Deep Extragalactic Legacy Survey
(CANDELS)~\citep[][]{Grogin2011,Koekemoer2011}.  Upcoming programs and
instruments like the Large Synoptic Survey Telescope
(LSST)~\citep[][]{LSST2009} promise maps of the distribution and properties of
galaxies in even higher detail.  This increasingly precise observational
picture requires equally accurate theoretical models for understanding
structure formation and the nature of dark matter (DM) and dark energy
(DE)~\citep[][]{Riess1999, Perlmutter1999}.  The prospects for such theoretical
calculations are in principle bright, because the initial conditions for cosmic
structure formation are tightly constrained through measurements of
anisotropies in the cosmic microwave background
radiation~\citep[][]{PLANCK2013}.  The $\Lambda$ cold dark matter
($\Lambda$CDM) paradigm favored by these data implies that the cosmos is filled
with three distinct components: baryons, DM and DE, and although the physical
nature of DM and DE remain unknown, their evolution can be followed far into
the non-linear regime using collisionless N-body
simulations~\citep[e.g.,][]{Springel2005, Springel2008, Boylan-Kolchin2009,
Klypin2011, Angulo2012, Wu2013}.  However, such simulations by themselves
cannot predict the distribution of galaxies made up of baryonic matter,
severely limiting their utility as a means to directly connect with the
observations.

There are two approaches for establishing this link: (i) Through a
post-processing procedure known as semi-analytic modelling, in which the output
of N-body simulations is combined with simple physical prescriptions to
estimate the distribution of galaxies~\citep[e.g.,][]{White1991, Kauffmann1993,
Somerville1999, Cole2000, Croton2006, Bower2006, Guo2011}.  (ii) Through
directly accounting for the baryonic component (gas, stars, supermassive black
holes, etc.) in cosmological simulations that include hydrodynamics and
gravity~\citep[e.g.,][]{Katz1992, Springel2005a, Crain2009,
Schaye2010, Khandai2014,Dubois2014}, in principle offering a
self-consistent and fully predictive approach.  However, the vast computational
challenges related to simulations of the baryonic component have thus far
precluded wide-spread adoption of this approach.  Large-scale predictions of
the galaxy population have thus mainly been obtained with the semi-analytic
method during the past few decades.

Hydrodynamical simulations of galaxy formation have become more prominent
over the last several years, and a significant number of results have been
published from them~\citep[e.g.,][]{Crain2009, Schaye2010,  Puchwein2013, Kannan2013}. These cosmological
simulations of the formation of representative galaxy populations are
different from studies where the computational power is focused on a
single galaxy~\citep[e.g.,][]{Guedes2011,
  Scannapieco2012, Aumer2013, Marinacci2014a, Kim2014}.  Although such
``zoom-in'' simulations offer higher mass and spatial resolution, they
do not yield a statistical sample of objects to compare to
observations, which limits their ability to falsify certain model aspects.
However, when used in conjunction with large-scale simulations, the
zoom-in procedure is clearly essential for constructing more reliable
sub-resolution models for characterising entire galaxy populations.
Ideally both approaches should be pursued simultaneously to learn more
about the complex problem of galaxy formation.

Hydrodynamical galaxy formation simulations typically include 
sub-resolution models for star formation~\citep[e.g.,][]{
Springel2003a, Schaye2008}, radiative cooling
processes (primordial and metal-line)~\citep[e.g.,][]{Katz1996,Wiersma2009a}, stellar evolution and chemical
enrichment procedures based on stellar
synthesis calculations~\citep[e.g.,][]{
Wiersma2009b,Few2012}, stellar feedback~\citep[e.g.,][]{
Navarro1993,Springel2003a,Stinson2013}, and feedback from active galactic nuclei~\citep[e.g.,][]{Springel2005, DiMatteo2005,
Sijacki2007, Booth2009, Teyssier2011, Dubois2012}.
Small-scale simulations also have also included
magnetic fields~\citep[e.g.,][]{
Pakmor2014}, radiative transfer~\citep[e.g.,][]{
Petkova2010}, cosmic ray physics~\citep[][]{Jubelgas2008} and
thermal conduction~\citep{Jubelgas2004, Dolag2004}.  Recently, more ``explicit''
forms of feedback have been included~\citep[e.g.,][]{Hopkins2014}, which
require however much higher spatial and mass resolution, something which is
currently out of reach for large-scale hydrodynamical simulations. Nevertheless,
such detailed models might lead to new effective models, which can then be
adapted for large-scale simulations. 

Here we present first results of the Illustris Simulation, the most ambitious
large-scale hydrodynamical calculation of structure formation to date, first
described in \cite{Nature2014}. Our model incorporates a broad range of
astrophysical processes that are believed to be relevant to galaxy
formation~\citep[see][for details]{Vogelsberger2013}: gas cooling with
radiative self-shielding corrections, energetic feedback from growing
supermassive black holes (SMBHs) and exploding supernovae (SNe), stellar
evolution and associated chemical enrichment and stellar mass loss, and
radiation proximity effects for active galactic nuclei (AGNs).  The simulation
starts at redshift $z=127$ ($\sim 12$ million years after the Big Bang) and
follows the evolution of $2\times 1820^3$ resolution elements in a volume of
$(106.5\mpc)^3$.  We achieve a DM mass resolution of $6.26 \times 10^{6}\msun$,
and an initial baryonic mass resolution of $1.26 \times 10^{6}\msun$.  The
smallest scale (corresponding to the smallest fiducial cell radius) over which
the hydrodynamical Euler equations are solved with the moving-mesh technique
{\tt AREPO}~\citep[described in][]{Springel2010} is $48\pc$ at $z=0$.  The mass
and spatial resolution are thus comparable to state-of-the-art large-scale
N-body simulations~\citep[e.g.,][]{Boylan-Kolchin2009,Klypin2011} which do not
include baryons. Further details of the redshift zero population of Illustris
are presented in a companion paper~\citep[][]{Nature2014}, where we focus on
deep field mock observations, the distribution of satellites in clusters, the
metal content of damped Lyman-$\alpha$ absorbers, the column density
distribution of neutral hydrogen, and the impact of baryons on the total matter
power spectrum. Here we extend this comparison to a broader range of
observables of the local Universe. The high redshift galaxy population is
discussed in another companion paper~\cite[][]{Genel2014}, where we
demonstrate that many observables at high $z$ are also reproduced by our model.
Another companion paper~(Sijacki et al., in prep) presents results on the
population of SMBHs in Illustris.

This paper is organised as follows. In Section~2, we describe our galaxy
formation model, the initial conditions and the simulation suite including
post-processing procedures.  Section~3 presents first results on the predicted
large-scale structure focusing on the halo mass function and the impact of
baryons on halo masses. The galaxy population is inspected more closely in
Section~4. We demonstrate here that our model reproduces key observables of the
local Universe and the cosmic star formation rate density over cosmic time. We
also discuss the colours of galaxies and the impact of environment on galaxy
evolution and specifically on the red fraction of galaxies. Section~5 focuses
on the internal structure of galaxies based on well-resolved sample galaxies
taken from the simulation, demonstrating that our model correctly predicts the
velocity structure of galaxies and the morphological mix of early- and
late-type galaxies.  We give a detailed summary of our results in Section~6 and
our conclusions in Section~7. Readers interested in the main results are
encouraged to read Section~6, which summaries the most significant findings of
the present work. Finally, Section~8 briefly describes upcoming data releases
and details about distribution of the simulation data to the community. All
data distribution and project updates will be handled through
\url{http://www.illustris-project.org}.

\section{Simulation code and galaxy formation model}

\subsection{Simulation method}

Following the dynamics of DM and gas accurately requires a highly reliable and
efficient numerical scheme to solve the coupled system of partial differential
equations describing gravity and hydrodynamics. We use the recently developed
moving-mesh code {\tt AREPO} code~\citep[][]{Springel2010} to solve the
inviscid Euler equations. It has been demonstrated that such a moving-mesh approach is
typically superior to traditional smoothed particle hydrodynamics (SPH), and
especially in the presence of large bulk velocities also to Eulerian adaptive
mesh refinement (AMR)~\citep[see][for details]{Vogelsberger2012, Keres2012, Sijacki2012,
Torrey2012, Genel2013, Nelson2013}.  Gravitational forces are calculated using
a Tree-PM scheme~\citep[][]{Xu1995} such that long-range forces are determined
using a particle-mesh method while short-range forces are computed via a
hierarchical octree algorithm~\citep[][]{Barnes1986}.

\subsection{Galaxy formation model}

Besides calculating gravitational forces and hydrodynamical fluxes, our
simulations account for astrophysical process known to be crucial for galaxy
formation. Our implementation is described in detail in~\cite{Vogelsberger2013,
Torrey2013}. This model includes gas cooling (primordial and metal line
cooling), a sub-resolution ISM model, stochastic star formation with a density
threshold of $0.13\,{\rm cm}^{-3}$, stellar evolution, gas recycling, chemical
enrichment, kinetic stellar feedback driven by SNe, procedures for supermassive
black hole (SMBH) seeding, SMBH accretion and SMBH merging, and related AGN
feedback (radio-mode, quasar-mode, and radiative).  The model includes some
free parameters, which were constrained based on the overall star formation
efficiency using smaller scale simulations~\citep[][]{Vogelsberger2013}. This
galaxy formation model was also explored with minor modifications in
high-resolution zoom-in simulations of individual Milky Way-like
haloes~\citep[][]{Marinacci2014a, Marinacci2014b}, and magneto-hydrodynamical
simulations~\citep[][]{Pakmor2014}. 

\subsection{Initial conditions and simulation suite}

\begin{table*}
\begin{tabular}{lllllllll}
\hline
\noalign{\vskip 0.5mm}
name                     & volume         & DM particles / hydro cells /    & $\epsilon_{\rm baryon}/\epsilon_{\rm DM}$  & $m_\mathrm{baryon}/m_\mathrm{DM}$      & $r_{\rm cell}^{\rm min}$    & $m_{\rm cell}^{\rm min}$       & description\\
\noalign{\vskip 0.5mm}
                         & [$(\mpc)^3$]   & MC tracers                 & [$\pc$]                                    & [$10^5\msun$]                          & [$\pc$]           & [$10^5\msun$]                            & \\
\noalign{\vskip 0.5mm}
\hline
\hline 
\noalign{\vskip 0.5mm}  
Illustris-1              & $106.5^3$      & $3\times1,820^3 \cong 18.1\times 10^9$     & $710/1,420$                                 & $12.6/62.6   $                           & $48$            & $0.15$                      & full physics \\
\noalign{\vskip 0.5mm}
Illustris-2              & $106.5^3$      & $3\times910^3   \cong 2.3\times 10^9$      & $1,420/2,840$                                & $100.7/501.0 $                           & $98$            & $1.3$                      & full physics \\
\noalign{\vskip 0.5mm}
Illustris-3              & $106.5^3$      & $3\times455^3  \cong 0.3 \times 10^9$      & $2,840/5,680$                                & $805.2/4008.2$                           & $273$           & $15.3$                     & full physics \\
\hline
\noalign{\vskip 0.5mm}
Illustris-Dark-1         & $106.5^3$      & $1\times1,820^3$     & $710/1,420$                                 & $-/75.2  $                              & $-$              & $-$ 
                         & DM only \\
\noalign{\vskip 0.5mm}
Illustris-Dark-2         & $106.5^3$      & $1\times910^3$      & $1,420/2,840$                                & $-/601.7 $                              & $-$              & $-$ 
                         & DM only \\
\noalign{\vskip 0.5mm}
Illustris-Dark-3         & $106.5^3$      & $1\times455^3$      & $2,840/5,680$                                & $-/4813.3$                              & $-$              & $-$ 
                         & DM only \\
\noalign{\vskip 0.5mm}
\hline
Illustris-NR-2           & $106.5^3$      & $2\times910^3   \cong 1.5\times 10^9$      & $1,420/2,840$         & $100.7/501.0 $                          & $893.8$         & $6.6$                       & no cooling/SF/feedback\\
Illustris-NR-3           & $106.5^3$      & $2\times455^3  \cong 0.2 \times 10^9$      & $2,840/5,680$         & $805.2/4008.2$                          & $2322.8$            & $39.4$                  & no cooling/SF/feedback\\
\hline
\end{tabular}
\caption{Details of the Illustris simulation suite. Illustris-(1,2,3) are
  hydrodynamical simulations including our model for galaxy formation
  physics.  Illustris-Dark-(1,2,3) are DM-only versions of the
  original Illustris simulations. They have the same initial
  conditions but do not include baryonic matter. Illustris-NR-(2,3)
  include baryons, but do not account for any feedback or cooling processes (non-radiative). 
  Illustris-1 follows the evolution of $12,057,136,000$ DM particles and
  hydrodynamical cells in total with the smallest fiducial cell size ($r_{\rm
    cell}^{\rm min}$) below $50\pc$ and the least massive cells having
  masses of a few times $10^4\msun$ ($m_{\rm cell}^{\rm min}$). In addition, we 
  follow the evolution of $1,820^3$ Monte-Carlo tracer particles (MC tracers).}
\label{table:illustris_sims}
\end{table*}

For our simulations we adopt the following cosmological parameters:
${\Omega_{\rm m}=0.2726}$, ${\Omega_{\Lambda}=0.7274}$, ${\Omega_{\rm
b}=0.0456}$, ${\sigma_8=0.809}$, ${n_{\rm s}=0.963}$, and ${H_0=100\,h\,{\rm
km}\,{\rm s}^{\rm -1}\,{\rm Mpc}^{\rm -1}}$ with ${h=0.704}$. These parameters
are consistent with the latest Wilkinson Microwave Anisotropy Probe (WMAP)-9
measurements~\citep[][]{Hinshaw2013}. The construction of the initial
conditions is described in detail in \cite{Nature2014}.

The main simulations of the Illustris project are summarised in
Table~\ref{table:illustris_sims}. Illustris-(1,2,3) represent our principal
simulations which include hydrodynamics and our galaxy formation model. The
different resolution levels (1,2,3) provide a convergence study in which the
mass resolution is changed by a factor of $64$ in total, and the spatial force
resolution by a factor of $4$. The level-1 simulation initially consists of
$6,028,568,000$ hydrodynamic cells and the same number of DM particles. At
$z=0$, the simulation contains $5,280,615,062$ gas resolution elements, $595,243,070$
stellar particles, and $32,552$ SMBH particles. The number of DM and tracer
particles (see below) is fixed as a function of time (i.e.~$1,820^3$ each).  We
employ a fixed comoving softening length for DM particles ($\epsilon_{\rm
DM}$).  The softening length for baryonic collisionless particles (stars and
SMBHs) is further limited to a maximum physical scale ($\epsilon_{\rm
baryon}$).  For gas cells we use an adaptive softening length which is
proportional to the fiducial cell size with an additional floor applied. This floor is
given by the softening length of the other collisionless baryonic particles.
Our {(de-)~refinement} scheme keeps the masses of cells close to within a
factor of two of a target mass ($m_\mathrm{baryon}$). Furthermore, we employ a
regularisation scheme, which steers the mesh towards its centroidal
configuration~\citep[][]{Springel2010, Vogelsberger2012, Vogelsberger2013}.
Illustris-1 has a large dynamic range, resolving gravitational dynamics down to
about $710\pc$ (at $z=0$), while at the same time following the large-scale
evolution in the full $(106.5\mpc)^3$ volume. At $z=0$ the smallest cells in
Illustris-1 have a typical extent (fiducial radius) of only $48\pc$, which is
the smallest scale over which the gas hydrodynamics and baryonic processes are
resolved. For the least massive cells we achieve a mass resolution of ${1.5
\times 10^{4}\msun}$.  

The Illustris-Dark-(1,2,3) simulations use the same initial conditions but do
not include the baryonic component. Illustris-NR-(2,3) include baryons, but do
not include any feedback, SF, or cooling processes (non-radiative). Through
this combined set of simulations we can, for example, study the impact of
baryons and feedback processes on the distribution of DM, as we will
demonstrate below.

\subsection{Post-processing}

We use a friends-of-friends (FOF) algorithm~\citep[][]{Davis1985} (linking
length of $0.2$ times the mean particle separation) to identify DM haloes with
a minimum particle number of $32$.  Other particle types (stellar particles,
gas cells, SMBH particles) were attached to these FOF primaries in a secondary
linking stage~\citep[][]{Dolag2009}.  The {\tt SUBFIND} algorithm was then used
to identify gravitationally bound
structures~\citep[][]{Springel2001,Dolag2009}. Each subhalo has a well defined
mass based on the particles which are bound to it.  However, haloes are
composed of many subhaloes and different mass definitions for haloes are
commonly used.  The most basic definition assigns the total FOF mass to a halo
($M_{\rm FOF}$).  Another common approach defines the halo mass to be the mass
$M_\Delta$, defined as the mass which is contained in a spherical region with
average density $\Delta$ times the critical density of the universe at that
time.  There are different choices for $\Delta$ possible, for example:
$\Delta=200$, $\Delta=200\,\Omega_m(z)$, or $\Delta_v(z)$, where the last value
comes from a spherical top-hat collapse model~\citep[][]{Bryan1998}. In the
following we will mainly use $\Delta=200$, and refer to $M_{\rm 200,crit}$ for
the corresponding mass and $R_{\rm 200,crit}$ for the enclosing radius.
Whenever we refer to a virial mass and virial radius we mean $M_{\rm 200,crit}$
and $R_{\rm 200,crit}$, respectively.  For more massive haloes (groups and
clusters) we will also quote $M_{\rm 500,crit}$ and $R_{\rm 500,crit}$, which
are more closely related to observationally derived quantities of clusters.

The structure finding procedure yields $7,713,601$ FOF groups with more than
$32$ particles and $4,366,546$ individual (sub)haloes at $z = 0$ in
Illustris-1. The largest FOF group at $z = 0$, a cluster-mass halo with a total
mass of ${4.6 \times 10^{14}\msun}$, contains over $66.5$ million DM particles,
$22.2$ million Voronoi cells, and $14.4$ million stellar particles.  Based on
the snapshots and group catalogues, we have constructed different types of
merger trees, for example using an extended and updated version of {\tt
LHaloTree}~\citep[also used for the Millennium simulations][]{Springel2005},
and {\tt ROCKSTAR}~\citep[][]{Behroozi2013r}, and the newly developed {\tt
SubLink} code~(Rodriguez-Gomez et al., in prep). Each snapshot of Illustris-1
amounts to about $1.5\,{\rm TB}$ of raw particle data, whereas a snapshot of
Illustris-Dark-1 adds up to $0.2\,{\rm TB}$. For each simulation we generate
$136$ snapshots, yielding a cumulative data volume of more than $230\,{\rm
TB}$. To facilitate efficient access to individual groups, (sub)haloes, and
galaxies, the raw particle data are stored in sorted order in output files
according to the {\tt SUBFIND} grouping. With the help of an offset table, this
allows for a quick retrieval of any bound structure in the snapshots with
minimal I/O requirements. We have furthermore extended {\tt SUBFIND} to also
group non-FOF particle/cell data such that the environment of specific haloes
can be easily read from disk. Furthermore, {\tt SUBFIND} was extended to
compute and store many galaxy properties, such as stellar luminosities in
various broadband filters (SDSS and Johnson), galaxy masses, galaxy stellar
metallicities, gas metallicities, neutral hydrogen content, etc.,  which
greatly simplifies subsequent analysis.

\begin{figure*}
\centering
\includegraphics[width=1.0\textwidth]{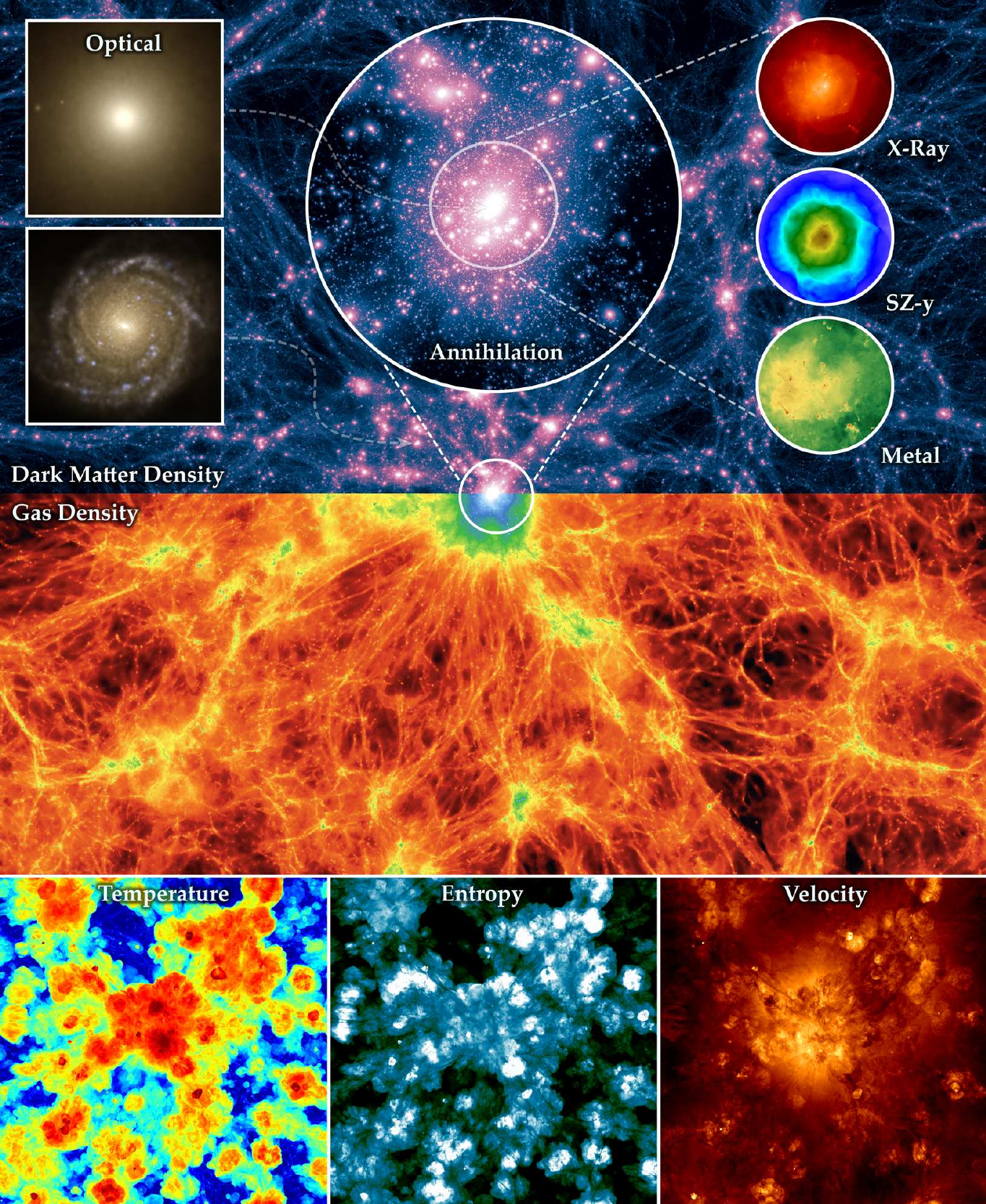}
\caption{Simulated present-day dark and baryonic matter structures. Top panel:
Dark matter mass distribution in a slice ($21.3\mpc$ thickness, $106.5\mpc$
width) centered on the most massive halo. Lower panels: gas distribution shown
in density, temperature, entropy, and velocity.  On the right in the top panel
we show (from top to bottom): X-ray emission of hot intra-cluster gas; thermal
Sunyaev-Zel'dovich signal; and the distribution of metals in the gas (all
within one virial radius). The central circle shows the expected annihilation
signal from self-annihilating DM particles within three virial radii.  On the
left in the top panel we present optical images (g,r,i SDSS broadband filter
composites) of the central galaxy of the cluster (top) and a random disk field
galaxy (bottom).}
\label{fig:illustris}
\end{figure*}

\begin{figure*}
\centering
\includegraphics[width=1.0\textwidth]{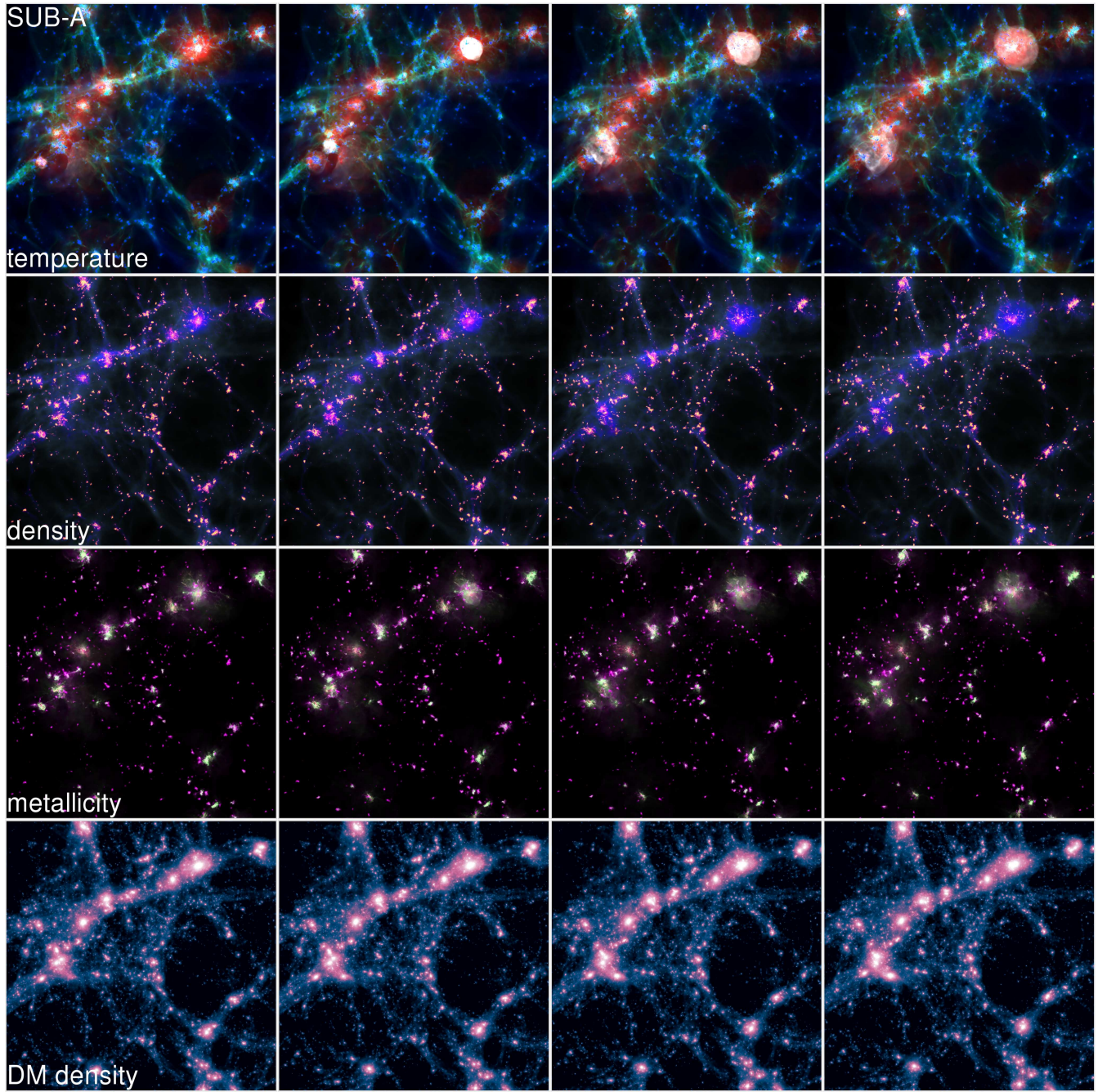}
\caption{Short-time evolution of subbox SUB-A (top to bottom: gas temperature, gas density
gas metallicity, DM density). The time evolution starts at $z=1.47$ (left) and ends at
$z=1.36$ (right). We output the subbox data $100$ times over this time span yielding a
time resolution of less then $3\,{\rm Myr}$. The more massive halo in the upper
right shows strong AGN activity leading to heating and expansion of large
amounts of gas.}
\label{fig:subbox_radio}
\end{figure*}

For definiteness, we derive the various galaxy properties from the
gravitationally bound mass that is contained within a radius $r_\star$ that
equals twice the stellar half mass radius of each {\tt SUBFIND} (sub)halo.
Unless mentioned otherwise, we will always measure galactic properties (masses,
light, SF rates, etc.) within the $r_\star$ radius.

\begin{figure*}
\centering
\includegraphics[width=1.0\textwidth]{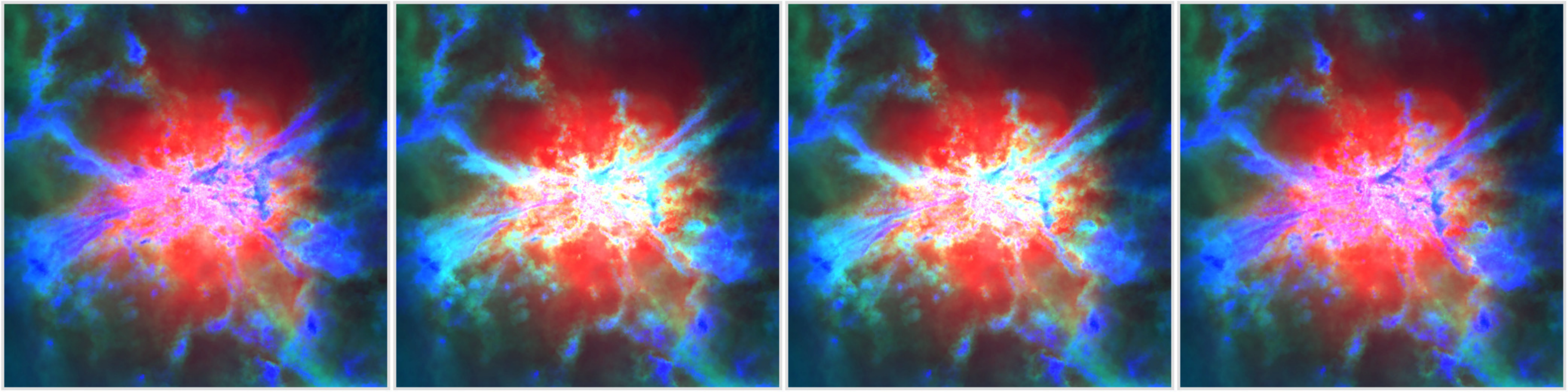}
\vskip0.5cm
\includegraphics[width=0.48\textwidth]{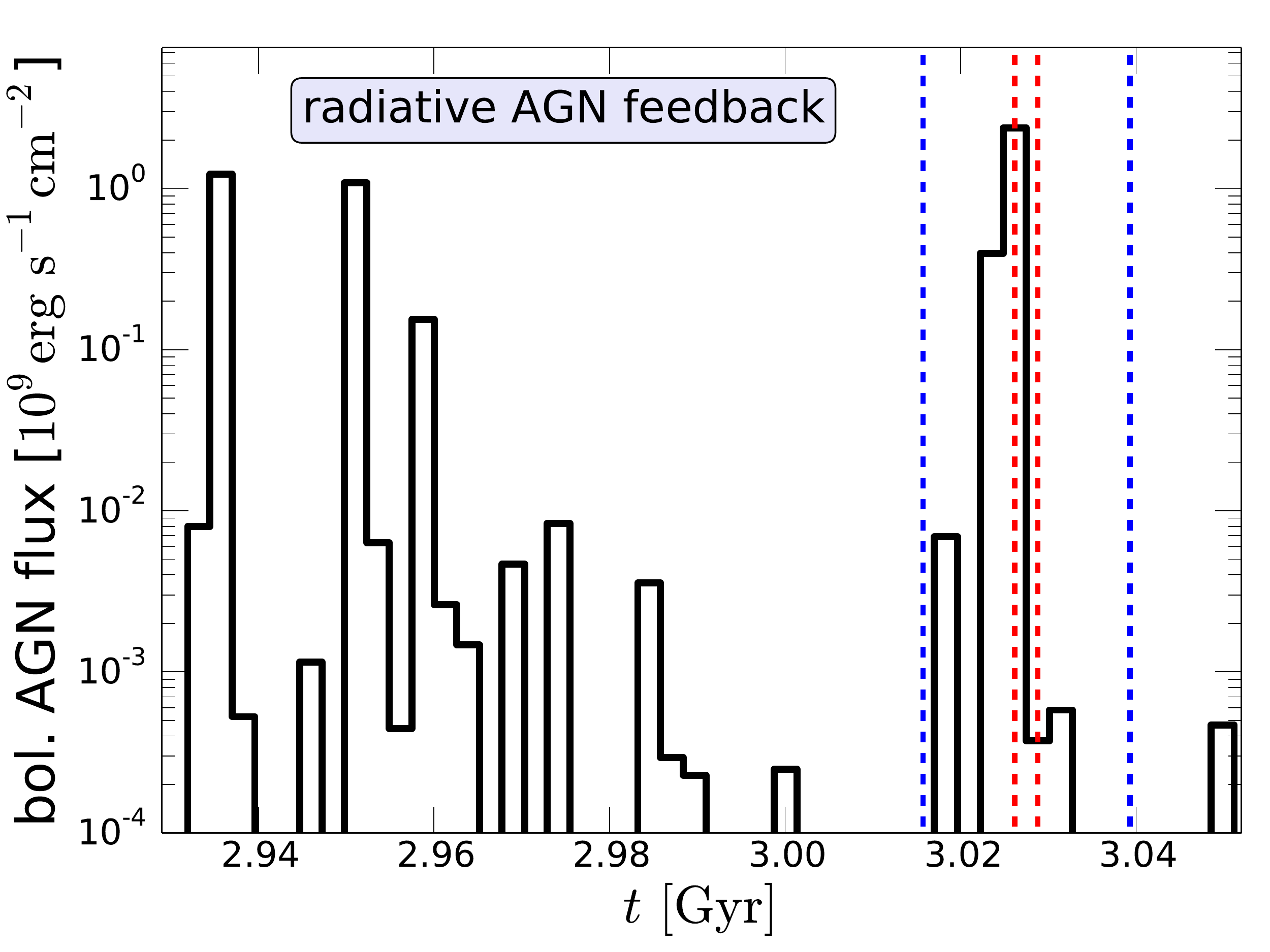}
\includegraphics[width=0.48\textwidth]{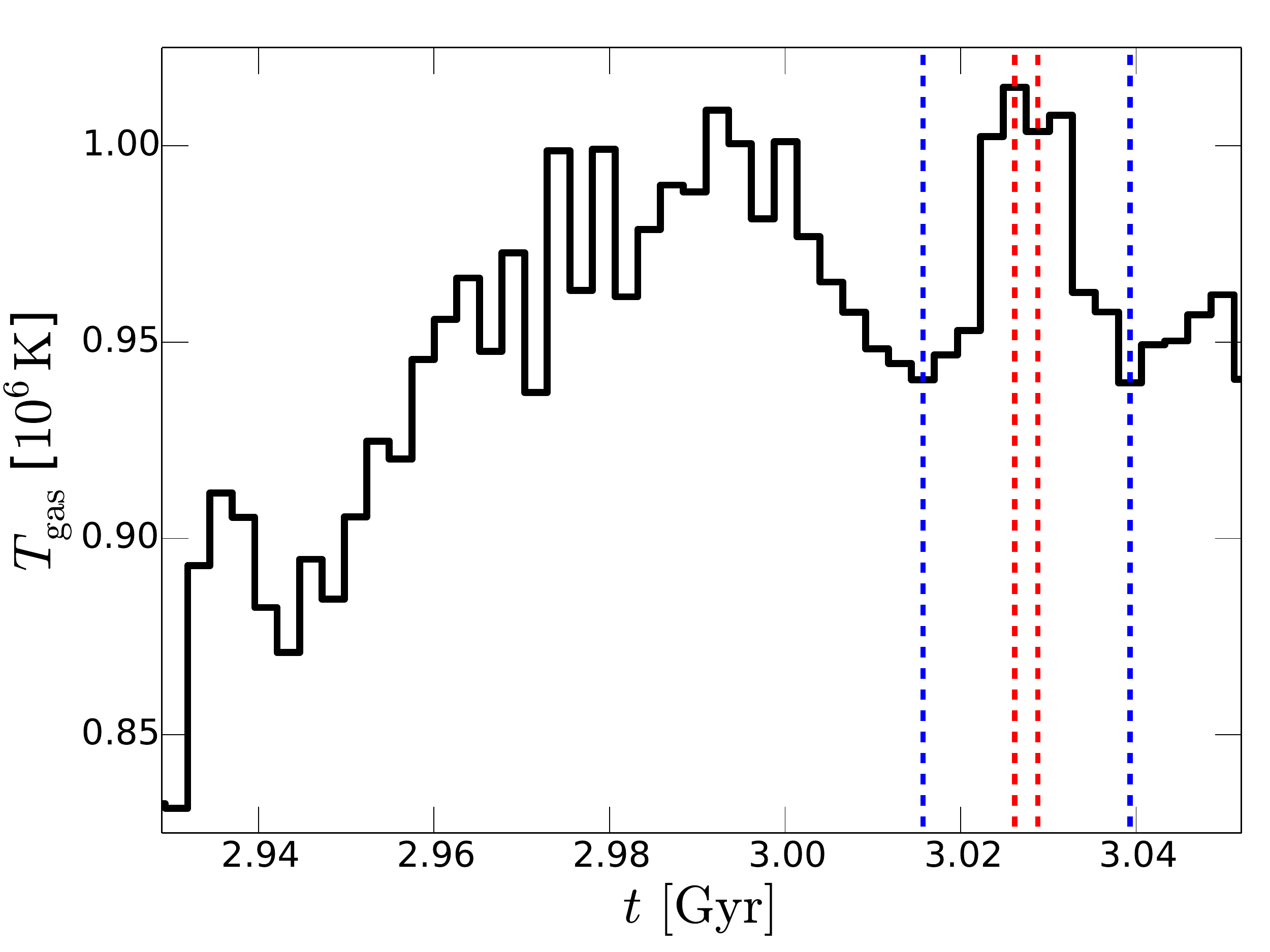}
\caption{Impact of radiative AGN feedback on halo gas. Top panels: Gas
temperature projections of a halo within SUB-A. The time separation of the
three temperature projections is (in ${\rm Myr}$): 10.46, 2.62, 10.51 capturing
a brief episode of quasar activity, where AGN radiation alters the cooling
rates and heats up the gas. Bottom left panel: Bolometric AGN flux within this
halo. Right panel: Temperature evolution of gas belonging to the halo. The four
dashed vertical lines correspond to the times of the temperature projections
shown in the top. The blue dashed lines capture the system at times of low
radiative AGN activity, whereas the red dashed lines show moments of high
activity (especially the first dashed red line). AGN radiation heats up the gas
for a short amount of time during maximum quasar activity. However, the heating
phase is rather short such that this energy injection does not influence the
gas significantly on longer time scales due to efficient cooling.}
\label{fig:subbox_radiation}
\end{figure*}

Besides writing snapshots for the total simulation volume at $136$ specific
times, we have selected four disjoint small subboxes (denoted as SUB-A, SUB-B,
SUB-C, SUB-D) for writing their data at every high-level synchronisation point
of the simulations. This yields $3,976$ snapshots for each of the four subboxes
in Illustris-1; i.e.~the time spacing is about $30$ times finer than in our
standard snapshots. At $z=2$ the time spacing between two subbox outputs is
therefore about $3\Myr$, and at $z=0$ about $8\Myr$. The different subboxes
cover nearly the same volumes at different places in the simulation volume:
SUB-A: $(10.65\mpc)^3$ , SUB-B: $(11.36\mpc)^3$, SUB-C: $(7.10\mpc)^3$, SUB-D:
$(7.10\mpc)^3$.  At $z=0$, these subboxes contain in SUB-A $33,638,720$, in
SUB-B $4,361,433$, in SUB-C $1,872,848$, and in SUB-D $1,651,277$ DM particles.
The large variation in the number of resolution elements in these subboxes is
mainly due to the very different regions they sample, although the box sizes
are also slightly different.  The present day matter density $\Omega_{\rm
m}^{\rm sub}$ in the volumes is $1.47$, $0.16$, $0.29$ and $0.25$, for SUB-A to
SUB-D, respectively.  The stellar content $\Omega_{\star}^{\rm sub}$ is also
different, amounting to $0.0244$, $0.0013$, $0.0025$, and $0.0027$ for SUB-A to
SUB-D. These numbers demonstrate that SUB-A samples an overdense region,
whereas the other regions are underdense with the most extreme case given by
SUB-B. The fine time spacing of these subboxes can be used for detailed studies
of dynamical processes in different environments. 

All hydrodynamical Illustris simulations include Monte Carlo tracer particles
to follow the gas flow accurately~\citep[][]{Nelson2013, Genel2013,
Vogelsberger2013}.  For each tracer, we record $13$ different hydrodynamical
quantities (e.g.,~maximum past temperature and maximum past density), and we
follow the evolution of $1,820^3 \sim 6$~billion Monte Carlo tracer particles
in total. These tracers faithfully follow all conversion and exchange processes
in the baryonic component of the simulation (e.g. gas turning into stars, stars
returning mass to the gas phase, ejection into the stellar wind state,
accretion into SMBHs, etc). The inclusion of tracer particles increases the
total particle number of Illustris-1 to more than $18$ billion. We note that we
use explicitly only MC tracer particles instead of traditional velocity field
tracer particles typically employed in AMR codes to avoid biases in the mass
fluxes~\citep[see][for details]{Genel2013}.

We need to transform stellar information of our simulations into photometric
properties, i.e. stellar light in different filters, to compare the simulated
galaxy population with observations.  Stellar population synthesis
models~\citep[e.g.,][]{Bruzual2003} provide a straightforward way to associate
the stars in our simulations with observable spectra or broad band
luminosities. Colours were assigned using an RGB mapping of the (g,r,i) bands
using a  $\rm asinh$ scaling~\citep[see][for details]{Lupton2004}.

\section{Large-scale structure}

The volume of Illustris is sufficiently large to measure large-scale
statistics. We demonstrate this below for a few selected quantities, and
especially highlight the importance of taking baryonic effects into account
since they affect structure formation in a way that cannot be captured by
DM-only simulations or semi-analytic models applied to them in post-processing.
Before presenting these statistics we will first give some visual overview of
the simulation and one of the subboxes, demonstrating their power in
disentangling physical processes on short time-scales.

\subsection{Visual impression}

\begin{figure*}
\centering
\includegraphics[width=1.0\textwidth]{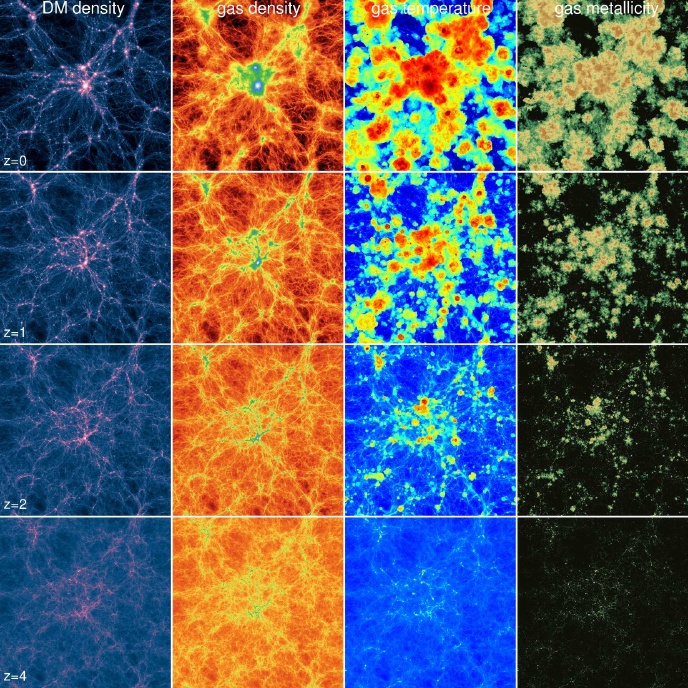}
\caption{Redshift evolution of the DM density (dark blue: $\sim
 10^{0}\msun/\kpc^3$, white: $\sim 10^{3}\msun/\kpc^3$), gas density
 (dark red: $\sim 10^{0}\msun/\kpc^3$, white: $\sim 10^{4}\msun/\kpc^3$),
 gas temperature  (dark blue: $\sim 10^{3}\,{\rm K}$, dark red: $\sim
 10^{7}\,{\rm K}$), and gas metallicity (dark green: $\sim 10^{-4}$, dark orange:
 $\sim 10^{-1}$) for $z=0$, $1$, $2$, and $4$ in a projected slice with a thickness of
 $21.3\mpc$ (comoving). The slice is centered on the most massive cluster at
 $z=0$ and covers the entire simulation volume ($106.5\mpc$ width). This is the
 same slice that was presented in Figure~1. All distributions are fairly uniform
 at early times. Gravity drives the collapse of haloes as a function of time.
 Gas heats up when it falls into these haloes, and it then cools down and
 eventually forms stars.  Stellar SN feedback leads to the enrichment of the
 IGM, as can be seen in the evolving
 metal distribution.}
\label{fig:dmevolve}
\end{figure*}

Figure~\ref{fig:illustris} gives a pictorial overview of our simulation
volume at the present day, redshift $z=0$.  Dark matter forms the backbone
of the cosmic structure, as displayed in the top portion of the figure.  This
invisible form of matter, potentially made up of neutralinos or axions, is
arranged in an intricate cosmic web~\citep[e.g.][]{Bond1996}, with tenuous filaments surrounding large
voids and connecting gravitationally bound haloes in which galaxies form
through gas condensation and star formation~\citep[][]{White1978}.  On large
scales, the gas density follows the evolution of the DM (middle panel), which
dictates the gravitational field with its five times higher mass density.
Hydrodynamical simulations provide important additional information about the
gas besides the density, e.g., its temperature, entropy and velocity fields,
as illustrated (from left to right) in the bottom panels of the figure.

The DM and gas slice shown in Figure~\ref{fig:illustris} is centered on a
galaxy cluster with a virial radius ($R_{\rm 200,crit}$) of $1.26\mpc$ and a
virial mass ($M_{\rm 200,crit}$) of ${2.32\times 10^{14}\msun}$.  This
particular cluster, shown enlarged in different figure insets with some of its
predicted observational signals, contains $5,116$ resolved subhaloes within its
virial radius ($16,937$ within its larger FOF group), consistent with the
expected population of such subhaloes in a CDM model~\citep[][]{Moore1999,
Springel2008}. The large inset on top shows the cluster's expected $\gamma$-ray
annihilation signal within three virial radii if the elusive DM  consists of a
Majorana particle that can self-annihilate~\citep[see][for a
review]{Jungman1996}. Annihilation is a two-body process; i.e. the annihilation
signal is proportional to $\propto \int\!{\rm d}l\,\rho^2$, where $l$ is the
distance along the line-of-sight.

\begin{figure*}
\centering
\includegraphics[width=1.0\textwidth]{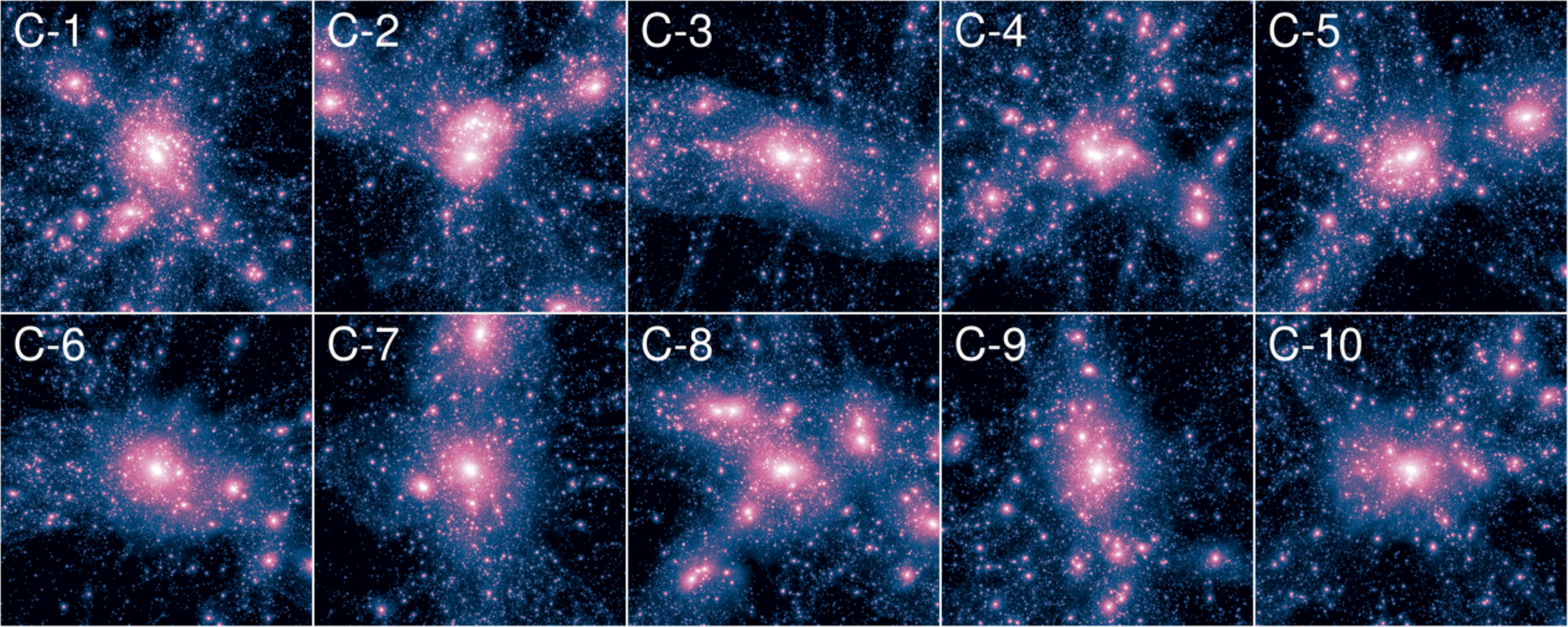}
\caption{ Annihilation maps ($\propto \int\!{\rm d}l\,\rho^2$, where $l$ is the distance along the line-of-sight) of the ten most
massive haloes (C-1 to C-10, ordered by mass) at $z=0$. The most massive
cluster C-1 has a mass of ${2.32\times 10^{14}\msun}$ and the least massive
(lower right) has a mass of ${1.04\times 10^{14}\msun}$. They have virial radii
of the order of $\sim 1\mpc$. Basic parameters of these haloes are summarised
in Table~\ref{table:clusters}.  In Illustris-1 the clusters are resolved with
up to $\sim 30$ million DM particles within the virial radius, which is
sufficient to resolve up to $\sim 7,000$ subhaloes within this region. Each
cluster projection shows a cube with side length of ten times the cluster's
virial radius.}
\label{fig:clusters}
\end{figure*}

The three insets in the top right of Figure~\ref{fig:illustris} show maps of
various observational diagnostics of the central cluster.  They include the X-ray emission from hot gas in the cluster within
one virial radius (top right inset), where the main contribution comes from
bremsstrahlung ($\propto \sqrt{T}$) emission from the hot ($T>10^6\,{\rm K}$)
intra-cluster gas. We also show the expected thermal
Sunyaev-Zel'dovich signal~\citep[][]{Sunyaev1980}, which is due to inverse
Compton scattering of background photons of the cosmic microwave background by
hot electrons in the cluster plasma.  The thermal Sunyaev-Zel'dovich signal of
clusters is an important cosmological probe that is being measured, for
example, by the Planck satellite~\citep[][]{PLANCK2013}, the South Pole
Telescope (SPT)~\citep[][]{Ruhl2004}, and the Atacama Cosmology Telescope (ACT)~\citep[][]{Kosowsky2003}. Thirdly, we show the expected distribution of metals
in cluster gas in the lower right inset.  These predictions are possible
because our simulation traces the evolution of nine different chemical elements
as they are synthesised in stars and dispersed by galactic winds. Ionisation
states of these elements can be calculated in post-processing and can be compared in detail
to quasar spectra probing the circum-galactic medium~(Suresh et al, in prep).

Finally, in the upper left panel we show an optical image (in g, r, i SDSS
broadband filters) of the central galaxy that formed in the cluster; a massive
red elliptical galaxy.  For comparison, we also show a typical disk galaxy
(lower on the left) that we preferentially find in the field, away from 
clusters. These two galaxies differ not only in their morphologies and colours,
but also in their present-day star formation rates. The red elliptical galaxy
has essentially no ongoing star formation whereas the blue disk still forms
stars. 

\begin{table*}
\centering
\begin{tabular}{llllllllllll}
\hline
\noalign{\vskip 0.5mm}
cluster                 & $M_{\rm 200,crit}$  & $M_{\rm 500,crit}$    & $R_{\rm 200,crit}$   & $R_{\rm 500,crit}$  & $N_{\rm sub}$  & $N_{\rm DM}^{\rm FOF}$  & $N_{\rm stars}^{\rm FOF}$   & $N_{\rm cells}^{\rm FOF}$  & $N_{\rm DM}^{\rm 200}$  & $N_{\rm stars}^{\rm 200}$   & $N_{\rm cells}^{\rm 200}$\\
\noalign{\vskip 0.5mm}
                        & [$10^{14}\msun$]    & [$10^{14}\msun$]      & [$\mpc$]             & [$\mpc$]            &                & [$10^6$]                & [$10^6$]                    & [$10^6$]                 & [$10^6$]                & [$10^6$]                    & [$10^6$] \\
\noalign{\vskip 0.5mm}
\hline
\hline
\noalign{\vskip 0.5mm}
C-1     &         2.32          &         1.54          &         1.26          &         0.81          &         5,116         &         66.53         &         14.40         &         22.20         &  32.94        &         8.33          &         12.68 \\
C-2     &         2.23          &         0.87          &         1.25          &         0.67          &         6,792         &         56.13         &         10.69         &         14.74         &  32.83        &         5.94          &         8.12 \\
C-3     &         2.18          &         1.24          &         1.24          &         0.75          &         5,285         &         44.44         &         9.08          &         11.22         &  32.25        &         7.21          &         6.40 \\
C-4     &         1.69          &         0.94          &         1.14          &         0.69          &         4,906         &         32.58         &         5.90          &         7.48          &  24.96        &         4.74          &         5.55 \\
C-5     &         1.36          &         0.85          &         1.06          &         0.67          &         3,716         &         29.29         &         6.13          &         7.66          &  19.90        &         4.84          &         4.64 \\
C-6     &         1.28          &         0.80          &         1.04          &         0.65          &         2,753         &         30.29         &         6.77          &         7.36          &  18.80        &         4.73          &         4.38 \\
C-7     &         1.17          &         0.77          &         1.01          &         0.64          &         2,150         &         27.82         &         6.39          &         7.65          &  17.17        &         4.19          &         4.66 \\
C-8     &         1.17          &         0.80          &         1.01          &         0.65          &         2,722         &         42.53         &         8.54          &         11.01         &  17.08        &         3.25          &         5.17 \\
C-9     &         1.12          &         0.71          &         0.99          &         0.63          &         2,841         &         27.29         &         5.47          &         5.78          &  16.67        &         3.22          &         2.98 \\
C-10    &         1.04          &         0.76          &         0.97          &         0.64          &         2,369         &         24.24         &         4.95          &         5.50          &  15.34        &         3.48          &         2.74 \\

\noalign{\vskip 0.5mm}
\hline
\end{tabular}
\caption{Basic characteristics of the ten most massive
  clusters presented in the ten panels of
  Figure~\ref{fig:clusters}. The different columns list:  mass
  ($200$ critical),  mass ($500$ critical), radius ($200$
  critical), radius ($500$ critical), number of resolved
  subhaloes within $R_{\rm 200,crit}$, number of DM particles in FOF group, number of stellar
  particles in FOF group, number of hydrodynamical cells in FOF group,
  number of DM particles within virial radius, number of stellar
  particles within virial radius, and number of hydrodynamical cells
  within virial radius.}
\label{table:clusters}
\end{table*}

The subboxes can be used to study the time evolution of different large-scale
environments in great detail. We demonstrate this in
Figure~\ref{fig:subbox_radio}, where we show for SUB-A the time evolution of
the full temperature projection starting at $z=1.47$ ending at $z=1.36$ (top
row). We output the subbox data $100$ times over this time span, yielding a
time resolution of less then $3\,{\rm Myr}$.  One halo in the upper right shows
strong radio-mode AGN activity leading to significant heating and expansion of
gas around the central SMBH; an effect that can be studied in detail due to the
fine time-spacing. The other rows in Figure~\ref{fig:subbox_radio} show
density, metallicity, and DM density projections. Some of these quantities, for
example the metallicity distribution, is also affected by the outflows
generated by SMBH feedback.

The high time resolution of the subbox outputs also makes it possible to understand
the impact of physical effects, which occur on even shorter time scales. One
such effect is the impact of radiative electro-magnetic AGN feedback on nearby halo gas, which is
included in our model using various approximations~\citep[see][for more
details]{Vogelsberger2013}. The top panels of Figure~\ref{fig:subbox_radiation}
show the projected gas temperature of a halo within SUB-A. The time separation
of the four temperature projections is (from left to right in ${\rm Myr}$):
10.46, 2.62, 10.51. This is sufficiently short to capture a brief episode of
strong quasar activity, where AGN radiation alters the cooling rates of nearby
gas and heats up this gas on a very short time scale. The bottom left panel of
Figure~\ref{fig:subbox_radiation} shows the bolometric flux within this halo,
and the right panel shows the temperature evolution of gas belonging to the
halo. The four dashed vertical lines correspond to the times of the
temperature projections above. The blue dashed lines capture the system at times of low radiative
AGN activity, whereas the red dashed lines show moments of high activity
(especially the first dashed red line). The temperature panel demonstrates that
this also increases the gas temperature within the halo. However, this effect
is still rather weak compared to thermal and mechanical AGN feedback and
furthermore it is only relevant on a very short time scale of about $10\,{\rm
Myr}$ due to efficient gas cooling. This form of feedback is therefore not able
to inject significant amounts of energy into the gas on longer time scales, and
it is particularly not able to quench SF in massive clusters~\citep[see also][]{Vogelsberger2013}. This is in disagreement with 
previous statements that this form of feedback has a significant impact on the
thermodynamic state of halo gas~\cite[e.g.,][]{Gnedin2012}.

The full simulation volume of Illustris is stored at the $136$ major output times.
This is sufficient for studying the large-scale evolution and allows also
for the computation of detailed merger trees (Illustris stored twice as many snapshots as the
Millennium-I and Millennium-II simulations). An overview of the time evolution
of the large-scale structure of the Illustris-1 simulation is presented in
Figure~\ref{fig:dmevolve} where we show snapshots of the DM density, gas
density, gas temperature, and gas metallicity for redshifts $z=0$, $1$, $2$,
and $4$. The displayed slice is centered on the same most massive halo in the
simulation volume and has a thickness of $21.3\mpc$ (comoving). The
gradual build-up of cosmic structure is evident in all panels.  The matter
distribution at $z=4$ is still comparatively smooth and uniform, and the gas
temperature is mostly cold around $10^4\,{\rm K}$.  Metals are not spread
around much yet; instead they are mainly concentrated at the centers of forming
haloes, where they were created as a product of SF and subsequent stellar
evolution processes.  The gas is significantly heated over time by virial
shocks and feedback, and a population of haloes forms through the hierarchical
growth process as predicted by the $\Lambda$CDM cosmological model. SN and AGN
feedback enrich the circum-galactic medium (CGM) and the intergalactic medium
(IGM) by dispersing metals even into regions outside of
haloes.  This can clearly be seen in the metallicity map, where the metal
distribution at $z = 0$ appears more diffuse compared to earlier times,
indicating that metals have become much less concentrated towards halo centers.
Gas heating at $z \gtrsim 2$ is largely driven by virial shocks and partially
also by SN heating.  At later times around $z=1$ and $z=0$, radio-mode AGN
feedback leads to significant gas heating especially in more massive haloes,
where SMBH feedback is relevant to regulating SF. We note that this feedback is
only relevant for low SMBH accretion rates, and it is crucial in regulating SF
in massive clusters.  Quasar-mode feedback is not efficient in quenching these
systems~\citep[see][for details]{Sijacki2007, Vogelsberger2013}.

At the present epoch, Illustris-1 contains $3,237,813,076$ DM particles in FOF
groups ($53.7\%$) and $2,790,754,924$  DM particles ($46.3\%$) which do not
belong to any FOF group. The most massive haloes in our simulation reach
cluster scale masses and are particularly well-resolved.  We demonstrate this
in Figure~\ref{fig:clusters}, where we show annihilation radiation maps
of the ten most massive haloes. C-1 is the most massive halo, which is at
the center of Figure~\ref{fig:illustris}.  Table~\ref{table:clusters} contains
some basic simulation characteristics of these clusters.  All of them have
virial masses ($M_{\rm 200,crit}$) above $10^{14}\msun$ and virial radii
($R_{\rm 200,crit}$) close to or above $\sim 1\mpc$.  The most well-resolved
cluster (C-1) contains nearly $33$ million DM particles within the virial
radius, and more than $12.6$ million gas resolution elements.  SF led to the
creation of more than $8.3$ million stellar particles by $z=0$.

\subsection{Large-scale structure statistics}

\begin{figure}
\centering
\includegraphics[width=0.49\textwidth]{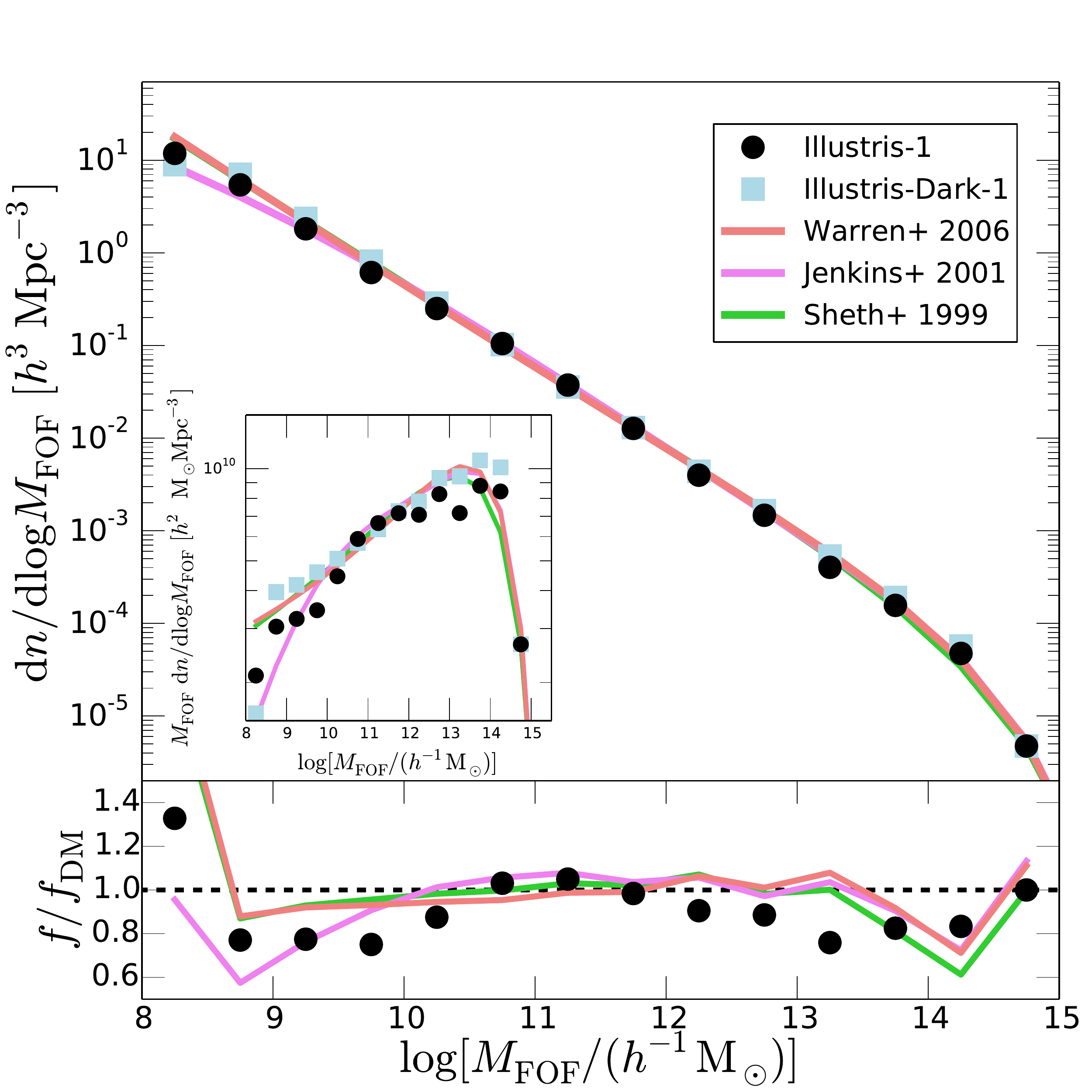}
\caption{Differential halo mass function of
  Illustris-1 and Illustris-Dark-1, where we define the halo mass as
  the total FOF mass, compared to some recent empirical fitting
  formulae~\protect\citep[][]{Sheth1999,Jenkins2001,Warren2006}. The inset shows
  the halo mass function multiplied with $M_{\rm FOF}$ to reduce the dynamic range and show differences more clearly. The lower panel
  shows the relative differences between the various curves in the
  upper panel. The halo mass function of the full physics simulation
  deviates at low and high halo masses from the Illustris-Dark-1
  prediction. Both differences are due to efficient feedback affecting the
  structure of low mass and high mass haloes.}
\label{fig:fof_mass_function}
\end{figure}

\begin{figure}
\centering
\includegraphics[width=0.49\textwidth]{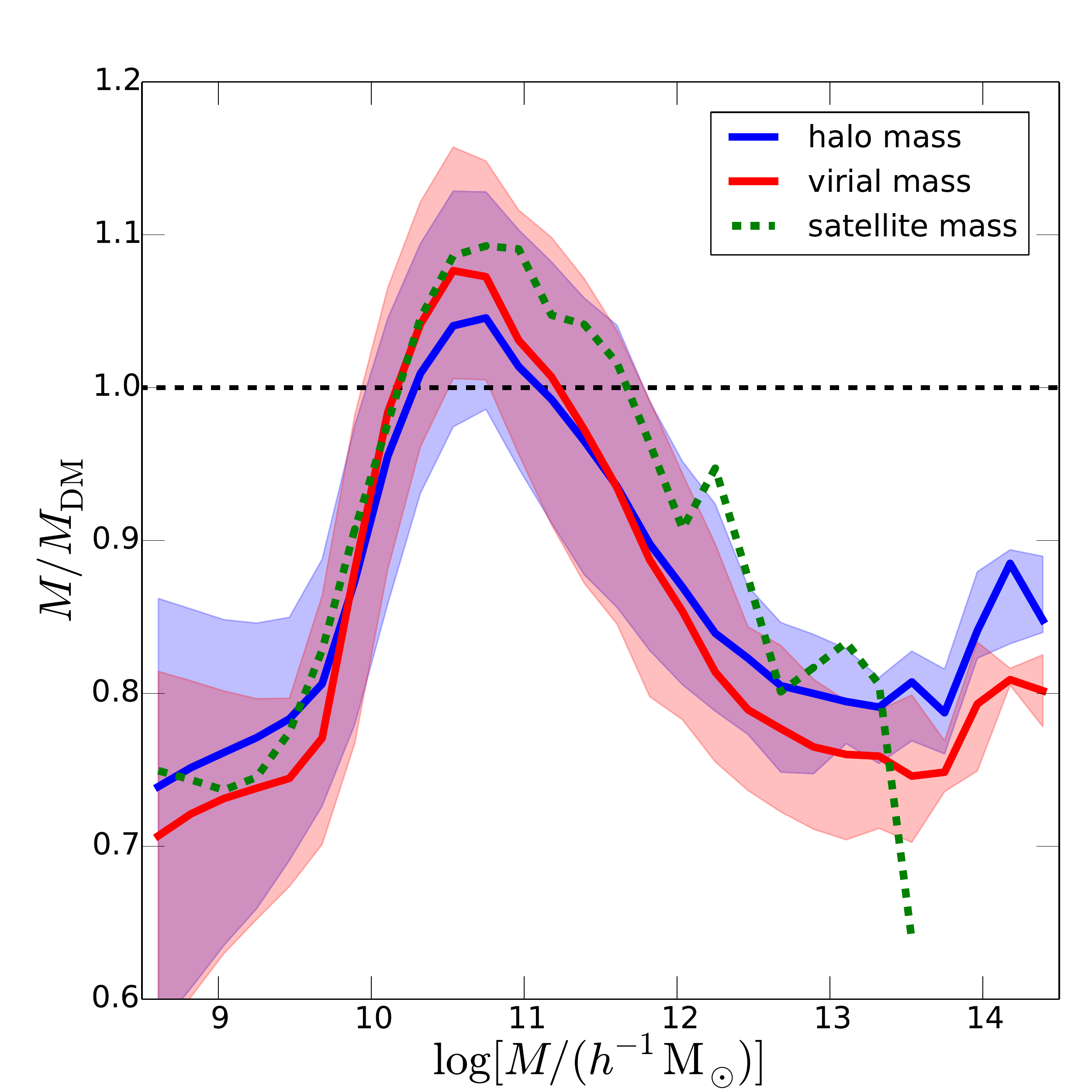}
\caption{Feedback induced changes in gravitationally bound and virial halo masses
between Illustris-1 and Illustris-Dark-1 for matched haloes between the two
simulations. Unlike most previous studies we find that feedback reduces the
halo masses not only for low mass systems, but also in massive haloes, which is
caused by our AGN feedback implementation. In both cases the mass deviates by
up to $\sim 30\%$ compared to the DM-only result of the matched
Illustris-Dark-1 haloes. The blue line shows this relation for all haloes (main
and subhaloes), whereas the red lines focuses on the virial masses of main
haloes (centrals). The green line shows the result for satellites. Blue and red
contours show the $1\sigma$ spread.}
\label{fig:halo_masses}
\end{figure}

The total matter power spectrum of Illustris-1 and Illustris-Dark-1 was presented and discussed
in~\cite{Nature2014}. There we have demonstrated that baryons can lead to significant changes in the matter distribution even on larger scales due to energetic AGN feedback, which affects, for
example, future weak lensing based precision measurements of cosmological parameters. 

Another important way to infer cosmological parameters lies in measurements of
the halo mass function, which are ultimately based on a ``counting procedure''
of some kind, for example using surveys of galaxy clusters in X-ray emission
(e.g., the upcoming eROSITA mission~\citep[][]{Predehl2006}, or thermal
Sunyaev-Zel'dovich observations (e.g., by the South Pole Telescope
(SPT)~\citep[][]{Ruhl2004}, and the Atacama Cosmology Telescope
(ACT)~\citep[][]{Kosowsky2003}.  However, taking advantage of current and
upcoming surveys to improve cosmological parameter constraints requires
theoretical mass function predictions that are accurate to high
precision~\citep[][]{Wu2010}.  Recent theoretical predictions of the halo mass
function based on dissipationless N-body simulations have reduced the
statistical uncertainties to a very small level~\citep[][]{Tinker2008}, but
they have not yet addressed the systematic uncertainty associated with the
effect of baryons on DM haloes, which depends sensitively on the feedback
models~\citep[][]{Rudd2008, Stanek2009, Cui2012}.

\begin{figure}
\centering
\includegraphics[width=0.49\textwidth]{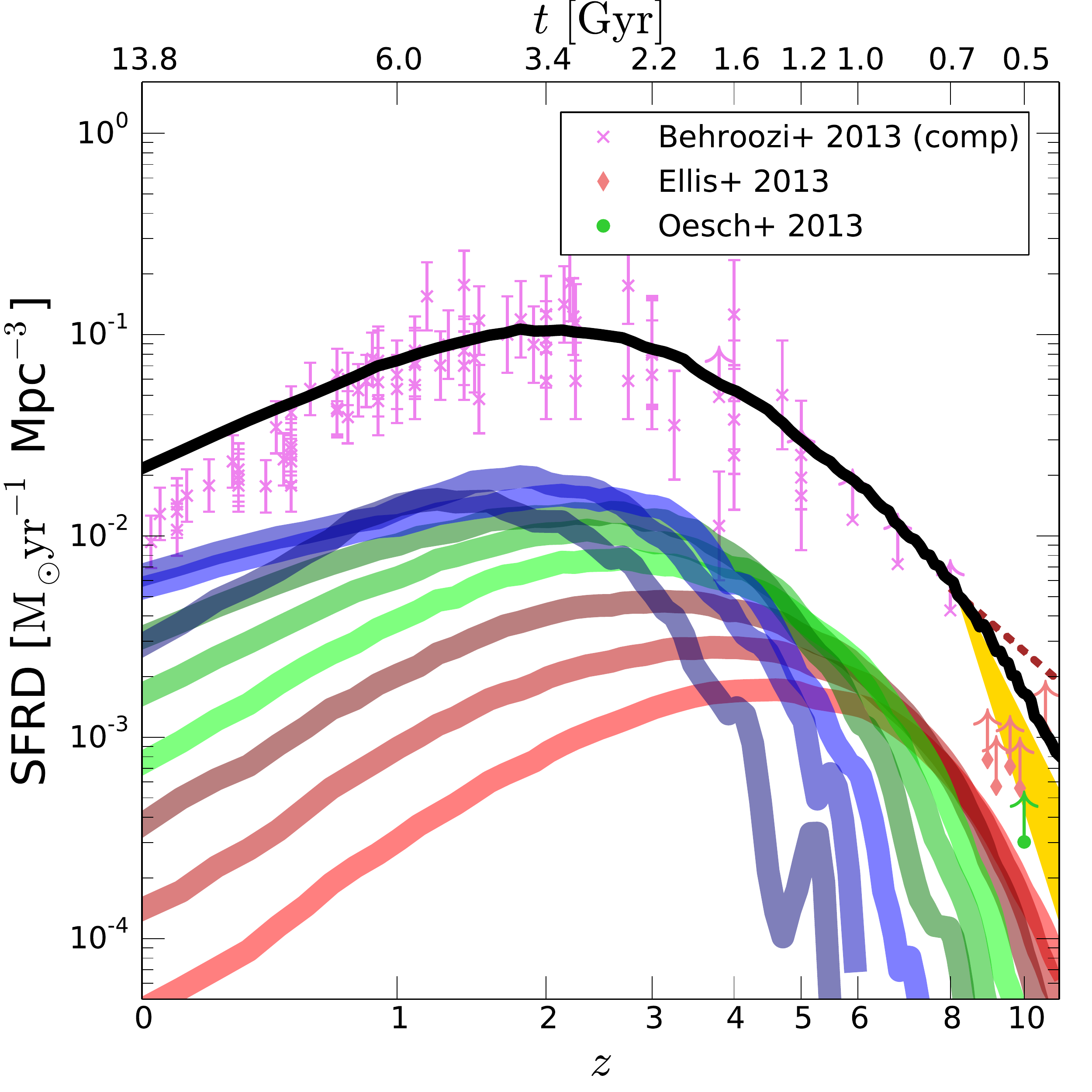}
\includegraphics[width=0.49\textwidth]{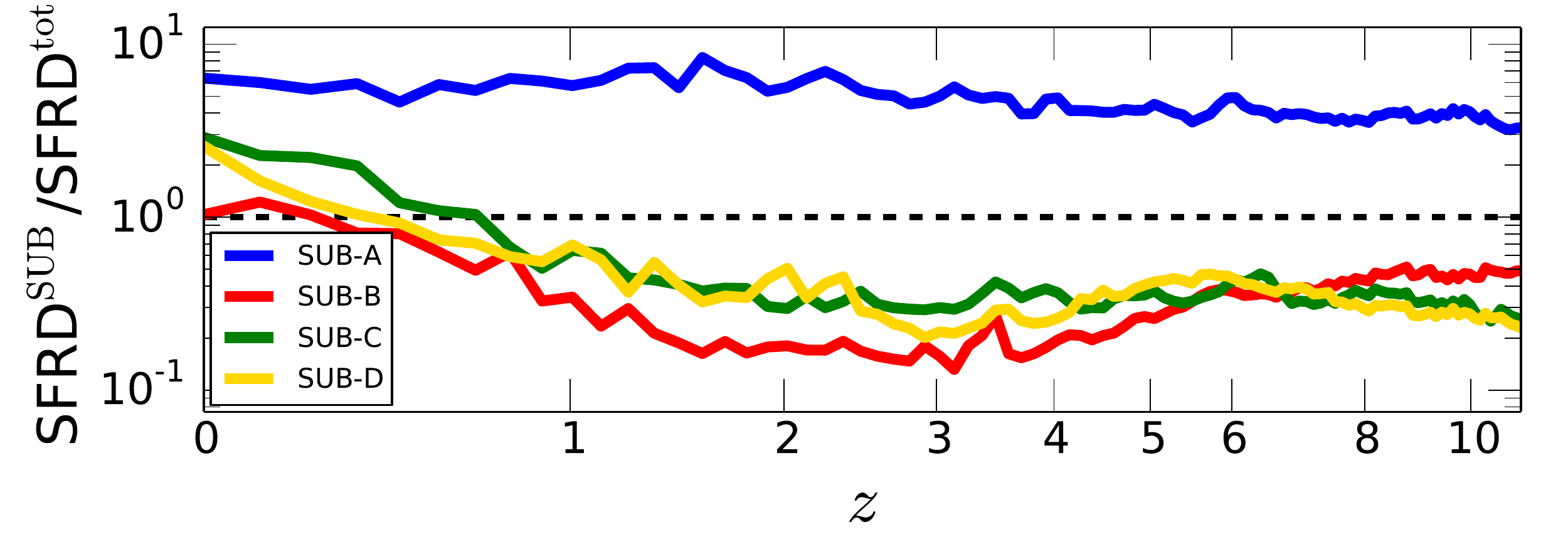}
\caption{ Top panel: Cosmic star formation rate density (SFRD) compared to
observations~\protect\citep[][]{Behroozi2013, Ellis2013, Oesch2013} (arrows
show lower observational limits). The black solid line shows the total SFRD;
coloured thick lines show the contributions from galaxies of different stellar
masses ($10^{7}\msun$ to $10^{11}\msun$) ranging from low mass (red lines:
${M_\star=10^{7,7.5,8}\msun}$) to intermediate mass (green lines:
${M_\star=10^{8.5,9,9.5}\msun}$) to the most massive systems (blue lines:
${M_\star=10^{10,10.5,11}\msun}$). The thick coloured lines have colours from
bright to dark to distinguish the different masses (low to high) within the
range.  In general, star formation at early times is dominated by low mass
systems (red lines), whereas late time star formation is mainly caused by
massive systems (blue lines), although the highest mass systems are quenched
by AGN feedback. The dashed line at high $z$ shows the extrapolated
observational high redshift trend ($\propto (1+z)^{-3.6}$) and the yellow
contour brackets the best-fit slopes for combined $z\geq8$ measurements
($\propto (1+z)^{-10.9 \pm 2.5}$) in this redshift
regime~\protect\citep[][]{Oesch2013}. For these high redshifts beyond $z\sim 8$
our model agrees best with the shallowest slope ($-10.9+2.5$)
of~\protect\cite{Oesch2013}.  Bottom panel: Ratio of SFRD in individual
subboxes (SUB-A to SUB-D) and total SFRD. SUB-A has a high overdensity, and
therefore produces more stars compared to the total volume. SUB-B, SUB-C, and
SUB-D are less dense and produce fewer stars until $z ~\sim 1$.}
\label{fig:sfrd}
\end{figure}

In Figure~\ref{fig:fof_mass_function} we present the differential halo mass
functions of Illustris-1 and Illustris-Dark-1 along with empirical fitting
formulae~\citep[][]{Sheth1999,Jenkins2001,Warren2006}.  Here we define halo
mass as FOF mass. The main panel of Figure~\ref{fig:fof_mass_function} shows
the unscaled differential halo mass function.  Although the dynamic range of
this figure is large, some differences between Illustris-1 and Illustris-1-Dark
are already visible.  Multiplying the halo mass function by mass reduces the
displayed dynamic range significantly and makes the differences clearer (see
inset). The bottom panel shows the relative differences between the
simulations. This reveals some interesting features in the halo mass function.
Baryons affect the mass function strongly at low and high halo masses, leading
to a reduction of the overall abundance at both ends. This is the regime where
baryonic feedback processes are strongest: reionisation, SN feedback, and AGN
feedback.  The effects are different if no strong feedback processes are
included. For example, neglecting AGN feedback leads to an increase of the halo
mass function at the massive end~\citep[e.g.,][]{Cui2012}.

Related effects can be identified when comparing the masses of matched
individual haloes as presented in Figure~\ref{fig:halo_masses}, where the
matching is based on the number of common DM particles and their binding
energies within candidate haloes. Here we find that haloes with masses around
$10^{11}\msun$ are typically more massive than their DM-only counterparts. This
is close to the halo mass scale where SF is most efficient (see below).
However, below and above this mass scale we find that feedback processes lead
to a significant redistribution of matter such that lower and higher mass
haloes tend to be less massive in Illustris-1 compared to the matched haloes
from Illustris-Dark-1.  This decrease in mass can reach $\sim 30\%$ for the
lowest and highest halo masses probed by our simulation.  We note that the
findings in Figure~\ref{fig:halo_masses} are different from many previous
predictions \citep[e.g.,][]{Sawala2013, Velliscig2014} since we find mass
suppression also for massive haloes, which is caused by our strong AGN
feedback. This also leads to a significant baryon fraction suppression in these
systems~\citep[][]{Genel2014}. We caution that this effect likely depends on
the detailed implementation of AGN feedback, specifically, the form of
radio-mode AGN feedback, and may not be present for implementations that do not
lead to a significant suppression of the baryon fraction~\citep[see also][]{Genel2014}. A detailed study of the impact of baryonic processes on the halo
masses will be presented in Rodriguez-Gomez et al. (in prep).

\section{The galaxy population}

A primary goal of galaxy formation models lies in explaining the stellar
content of the Universe. We will in the following demonstrate that our simulation 
yields a realistic galaxy and stellar population which is in reasonable
agreement with various recent observations.

\begin{figure}
\centering
\includegraphics[width=0.49\textwidth]{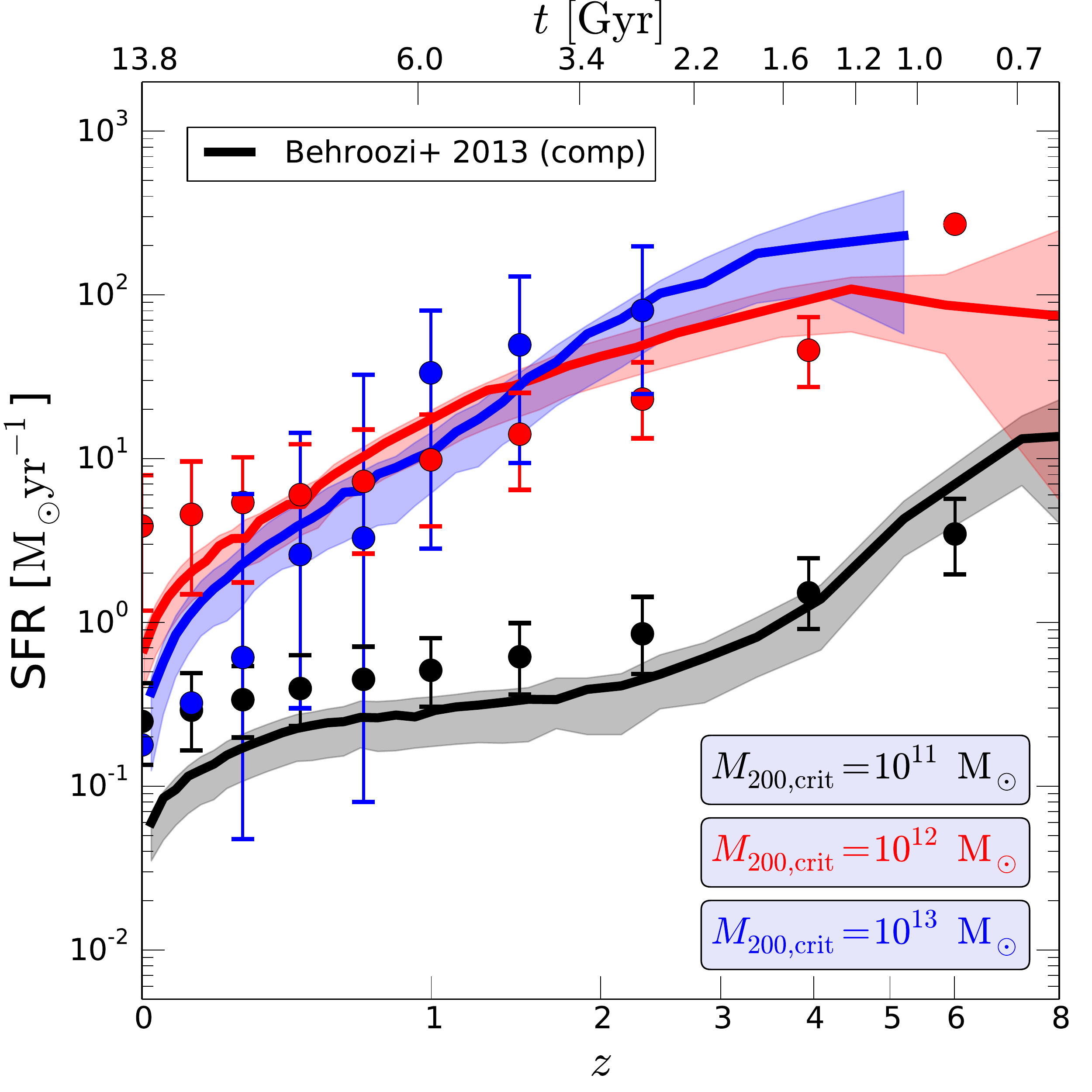}
\caption{Average star formation rates for galaxies in haloes at a given
instantaneous halo mass and redshift. Symbols with error bars show the
simulation results (median and $16$th and $84$th percentiles of the distribution), whereas observationally inferred results are shown by lines
with uncertainty bands. We recover the observed redshift trends. However, at
$z\lesssim 0.5$ less massive haloes below $M_{\rm 200,crit}\sim 10^{12}\msun$
tend to overproduce stars. Nevertheless our results are still consistent within
the uncertainties. Interestingly, our model tends to slightly underproduce
stars in the most massive haloes shown for a short period of time towards lower
redshifts, but the SFR at $z=0$ is consistent with observations for the most
massive haloes shown.}
\label{fig:sfr_halo}
\end{figure}

\subsection{The formation of galaxy stellar masses}

We start by looking at the global history of star formation over cosmic time.
Figure~\ref{fig:sfrd} shows the evolution of the cosmic SF rate density (SFRD),
a key prediction that can be tested against a plethora of measurements at
different epochs, extending back to just a few hundred million years after the
Big Bang.  The compiled data points in Figure~\ref{fig:sfrd} are based on
different types of surveys, including inferences from narrowband (e.g.,
$\mathrm{H}\alpha$), broadband (UV-IR), and radio ($1.4\,{\rm GHz}$)
observations.  We also include recent estimates for the very early SFRD beyond
$z=8$. These are based on the Cluster Lensing And Supernova survey with Hubble
(CLASH)~\citep[][]{Postman2012}, near-infrared Wide-Field Camera 3
(WFC3/IR)~\citep[][]{Kimble2008} images of the Hubble Ultra Deep Field (UDF),
and CANDELS WFC3/IR Great Observatories Origins Deep Survey
(GOODS)~\citep[][]{Giavalisco2004}-N imaging data.  Reliable SFRD estimates at
these high redshifts are extremely difficult to obtain, and it is still under
debate how this quantity evolves towards higher $z$.  Our predicted SF history
is in good agreement with the observations, including the recent lower
limits~\citep[][]{Ellis2013, Oesch2013} beyond $z \sim 8$. The dashed line
shows the extrapolated observational lower redshift trend ($\propto
(1+z)^{-3.6}$) and the yellow contour brackets the best-fit slopes for combined
$z\geq8$ measurements ($\propto (1+z)^{-10.9 \pm
2.5}$)~\protect\citep[][]{Oesch2013}. Interestingly, our model prediction
agrees best with the shallowest slope ($-10.9+2.5$).  Towards lower redshifts
we find some tension between the observational data and our predictions, where
our model tends to produce too many stars. It seems that our AGN feedback is
not efficient enough to quench SF sufficiently at these times. Specifically,
radio-mode AGN feedback still seems to be too weak to regulate SF sufficiently
in massive haloes.  This is despite the fact that our radio-mode feedback
parameters are chosen to result in more energetic AGN feedback compared to
previous studies \citep[see][for details]{Vogelsberger2013}. We note that
cosmic variance cannot account for the discrepancy found towards lower redshifts~\citep[see][for details]{Genel2014}.

Splitting the SFRD into contributions from galaxies in different stellar mass
bins (various coloured bold lines) reveals that the early SF is dominated by
low mass systems, whereas at late times more massive systems are most
important~\citep[see also][for a break down in halo mass]{Genel2014}.
However, in the most massive objects AGN feedback from central SMBHs has
quenched star formation activity, and so they do not contribute significantly
to present-day star formation.  The largest contribution to the present-day
SFRD comes from galaxies which have stellar masses of about $M_\star \sim
10^{10-10.5}\msun$.  We note that this is close to the regime where SF is most
efficient (see below) and it is the transition region below which SN feedback
is responsible for regulating SF. Above that mass scale radio-mode AGN feedback
is the main driver of SF quenching our simulation.

\begin{figure}
\centering
\includegraphics[width=0.49\textwidth]{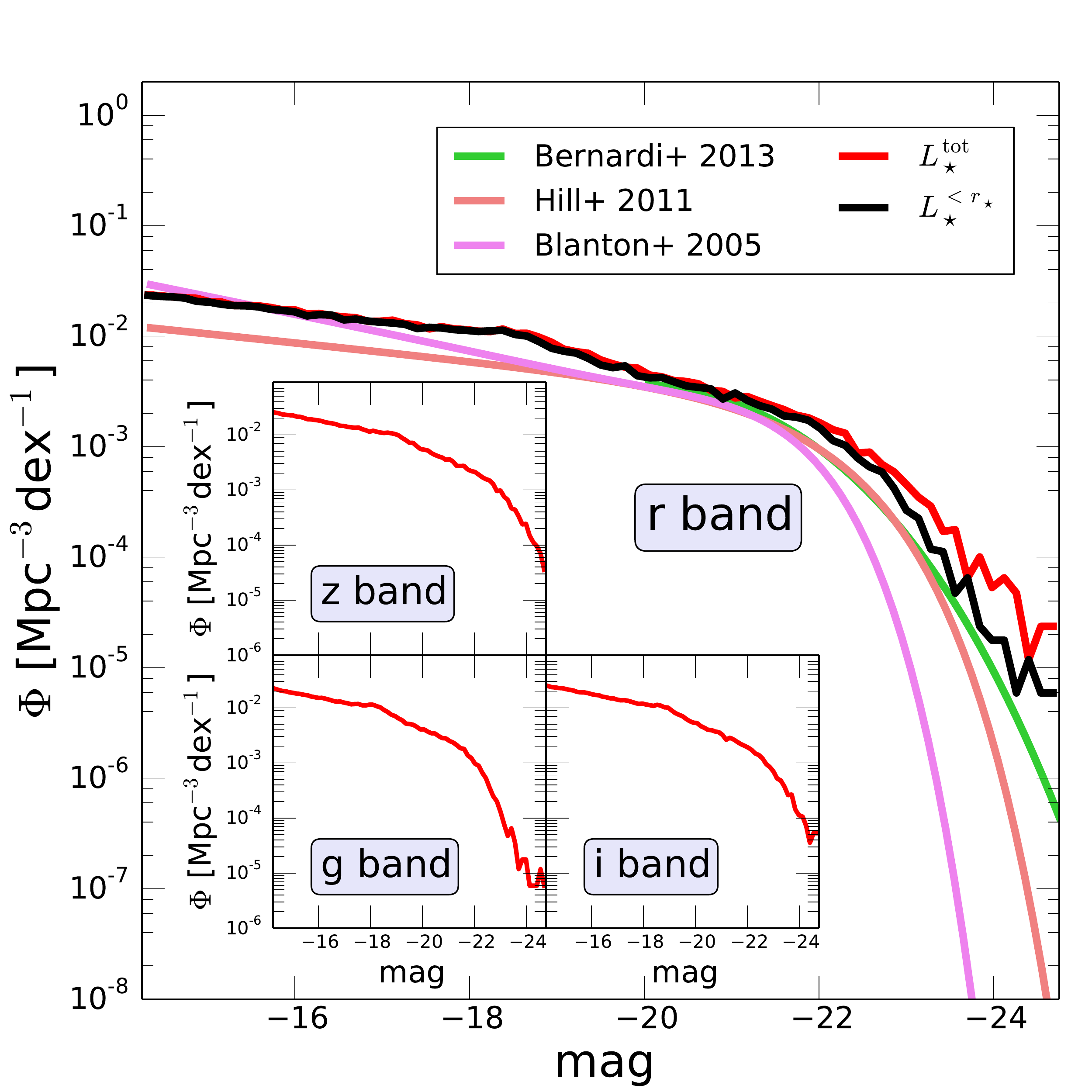}
\caption{Low
  redshift SDSS galaxy luminosity functions (GLFs) in g,r,i,z bands. The main panel compares the
  predicted r band GLF to recent observations~\protect\citep[][]{Blanton2005,
    Hill2011, Bernardi2013}, which probe the light distribution out to
  different radii. We present the predicted total light emission
  excluding satellites ($L^{\rm tot}_\star$) and the emission from the
  central galaxy ($L^{<\,r_\star}_\star$) to take into account observational
  uncertainties in the light assignment for massive galaxies~\protect\citep[see][for details]{Bernardi2013}. We note that simulation luminosities shown here do not include dust effects.}
\label{fig:GLF}
\end{figure}

The bottom panel of Figure~\ref{fig:sfrd} shows the ratios of the SFRD in the
individual subboxes (SUB-A to SUB-D) and the total SFRD. SUB-A has a high
overdensity, and therefore produces more stars compared to the total volume.
SUB-B, SUB-C, and SUB-D are less dense and produce fewer stars until $z ~\sim
0.7$. After that SF is also more efficient in these subboxes compared to the
total SFRD. The subboxes therefore show a significant variation in SFRD compared
to the global SFRD, sampling significantly different regions of the simulation
volume.

In addition to inspecting the global SFRD of the total simulation volume, we
can also compare star formation rates within haloes of a given mass and compare
to observationally derived results. This is shown in Figure~\ref{fig:sfr_halo},
where we compare star formation rates for three different instantaneous halo
masses $M_{\rm 200,crit}=10^{11-11.2},10^{12-12.2},10^{13-13.2}\msun$. The
symbols with error bars show the simulation results, whereas observational
estimates are shown by lines with uncertainty bands.  We recover the general
trends of the observations although not all data points agree. At late times
our model tends to produce slightly too many stars compared to observations,
especially in the $M_{\rm 200,crit} \sim 10^{12}\msun$ halo mass range. We saw
this already in Figure~\ref{fig:sfrd}, where the late time SFRD was slightly
too high for the corresponding stellar mass range. As shown in \cite{Genel2014} this halo mass range is also exactly the one that contributes the most to
the overestimated global SFRD.

\begin{figure}
\centering
\includegraphics[width=0.49\textwidth]{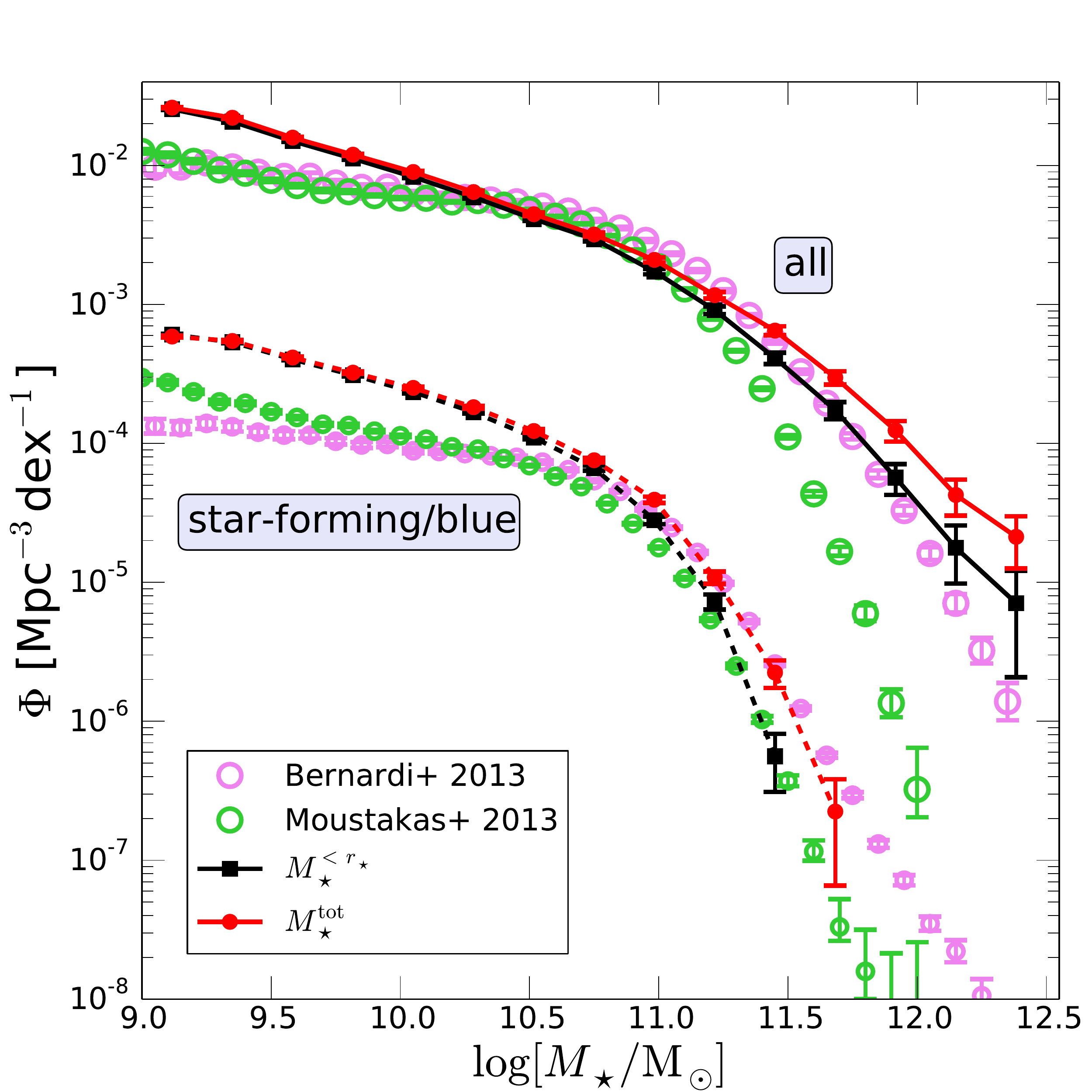}
\caption{Galaxy stellar mass function (GSMF)
  compared to observationally derived values~\protect\citep[][]{Moustakas2013, Bernardi2013}.  The star-forming GSMFs (${\log[{\rm SFR}/(\msun {\rm yr}^{-1})]>-0.6 +
0.65\,\log[M_\star/10^{10}\msun]}$)
  are shifted down by $1.5\,{\rm dex}$ for graphical clarity.
  We present the simulated GSMF
  using two measurements of stellar mass: total stellar mass excluding
  satellite contributions ($M^{\rm tot}_\star$) and the central
  stellar mass ($M^{<\,r_\star}_\star$) to compare with
  observations which are uncertain towards the massive end~\protect\citep[see][for details]{Bernardi2013}. Error bars show Poisson errors.}
\label{fig:GSMF}
\end{figure}

\begin{figure}
\centering
\includegraphics[width=0.49\textwidth]{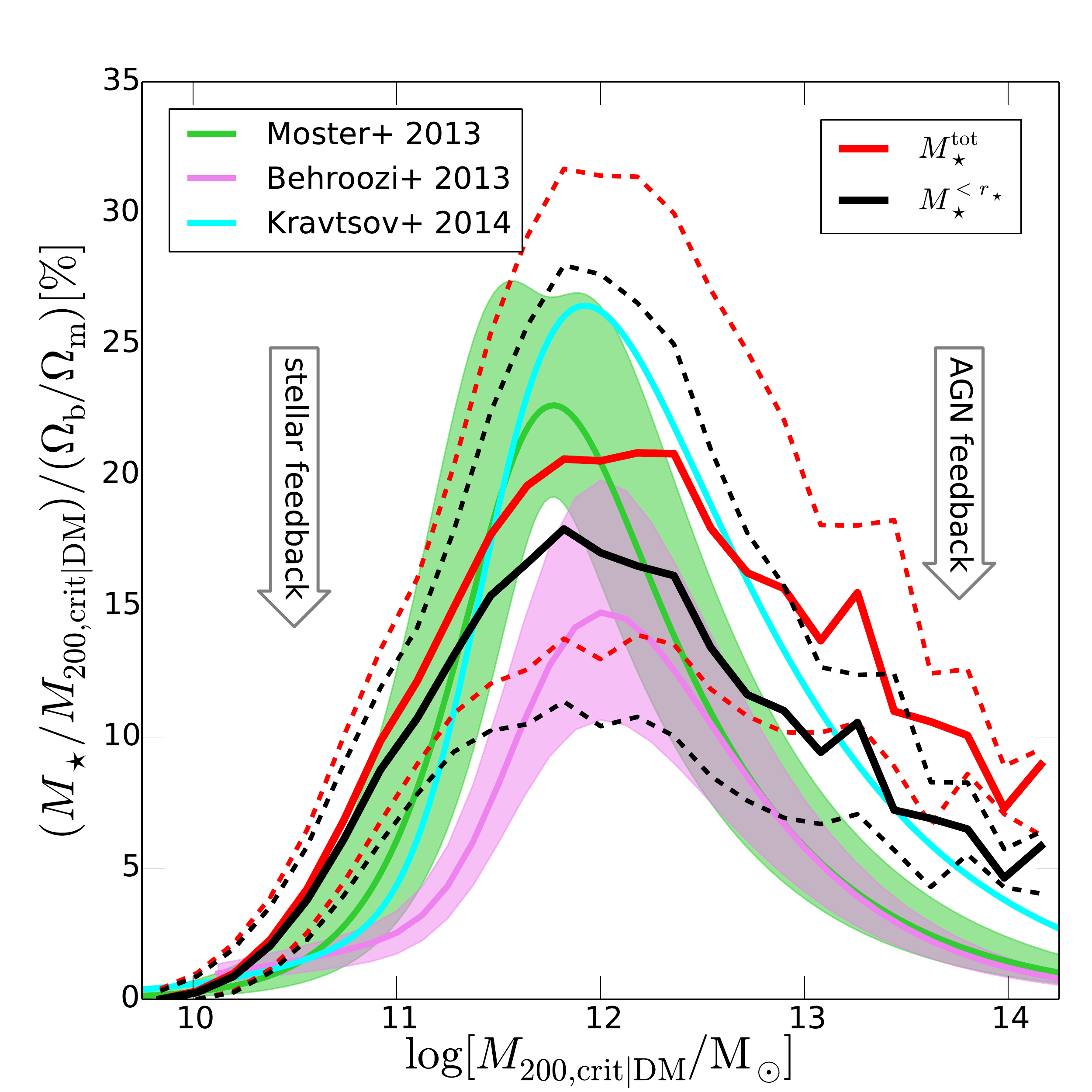}
\caption{Stellar mass to halo mass ratio as a function of halo mass compared to
observationally derived abundance matching
results~\protect\citep[][]{Moster2013, Behroozi2013, Kravtsov2014}. The thick
line shows the median simulation relation. The shaded areas show the $1\sigma$
regions around the abundance matching relations, and dashed lines around the
predicted ratio from the simulation.  $M^{\rm tot}_\star$ includes all stars
except satellites, whereas $M^{<\,r_\star}_\star$ includes only the stellar
mass within $r_\star$.}
\label{fig:shmr}
\end{figure}

The galaxy population can be quantified through the galaxy luminosity function
(GLF).  In Figure~\ref{fig:GLF} we examine  the
present-day GLF for four broadband SDSS filters (g,r,i,z) and compare to
observations~\citep[][]{Blanton2005, Hill2011, Bernardi2013} for the r-band
based on SDSS data. Different observations probe the stellar distribution out
to various radii depending on the assumed light profile.  Particularly for the
bright end of the luminosity function this adds measurement uncertainties,
because the assigned light can differ significantly depending on whether, for
example, Petrosian-like or Sersic-fit based magnitudes are
extracted~\citep[see][for details]{Bernardi2013}.  To crudely mimic these uncertainties, we measure
the predicted stellar light in the simulation also within different radii, and
present the predicted total bound stellar light of each galaxy ($L^{\rm
tot}_\star$) as well as the light associated with the central part
($L^{<\,r_\star}_\star$), where $r_\star$ is our fiducial galaxy radius defined
as twice the stellar half mass radius. Taking this into account we find
reasonable agreement with the shape and normalisation of the observed
luminosity function if we compare to recent Sersic-fit based GLFs~\citep[][]{Bernardi2013}. 

The abundance of faint galaxies follows a power-law
whereas brighter galaxies are exponentially suppressed in their
abundance~\citep[][]{Schechter1976}. This shape is caused by scale-dependent
feedback processes: Low mass systems are affected mostly by stellar SN
feedback, whereas star formation in massive systems is suppressed through
SMBH-related AGN feedback~\citep[][]{Bower2006}.  Our model slightly
overpredicts the number of massive and bright galaxies. This is related to the
disagreement we found for the SFRD, and potentially caused by slightly insufficient AGN feedback.
Measurements of the luminosity function hence provide important constraints on
galaxy formation, and the agreement found here demonstrates that a
self-consistent treatment of the physics that is thought to be relevant can
indeed reconcile the rather different shapes of the CDM halo mass function and
the galaxy luminosity function. We note that we find a similar level of agreement in the other bands (g,i,z), and therefore only show the simulation results in the inset panels.

\begin{figure*}
\centering
\includegraphics[width=0.48\textwidth]{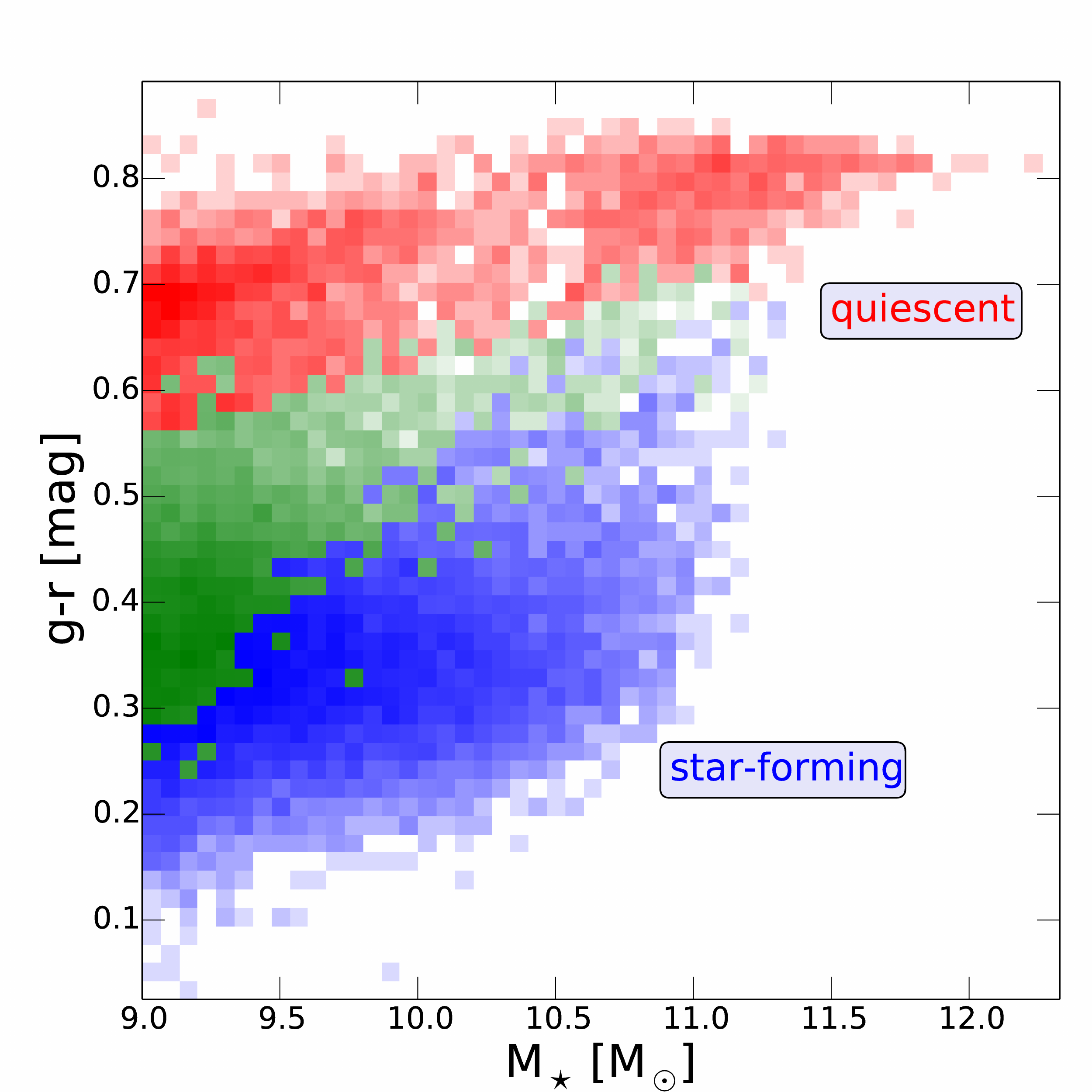}
\includegraphics[width=0.48\textwidth]{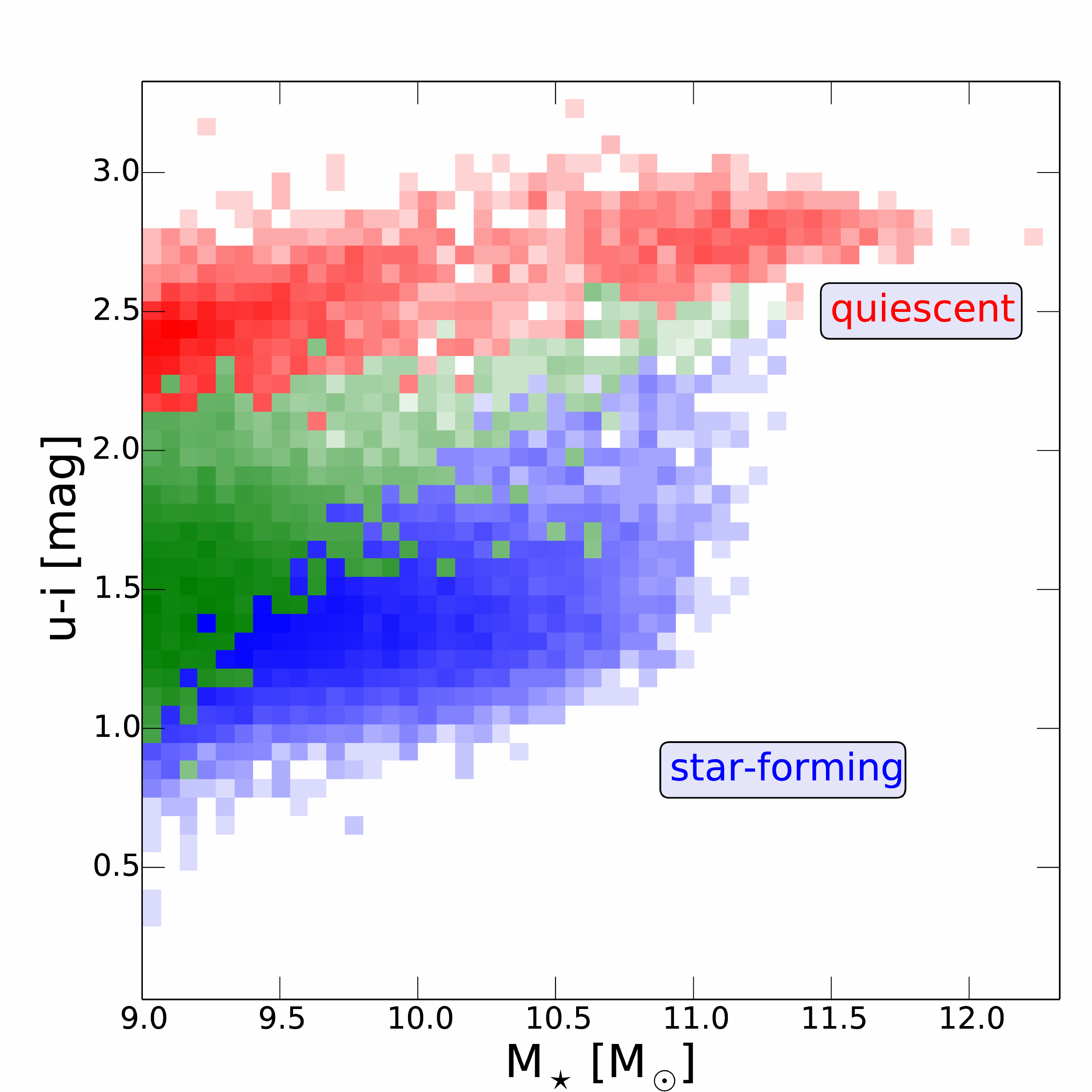}
\includegraphics[width=0.48\textwidth]{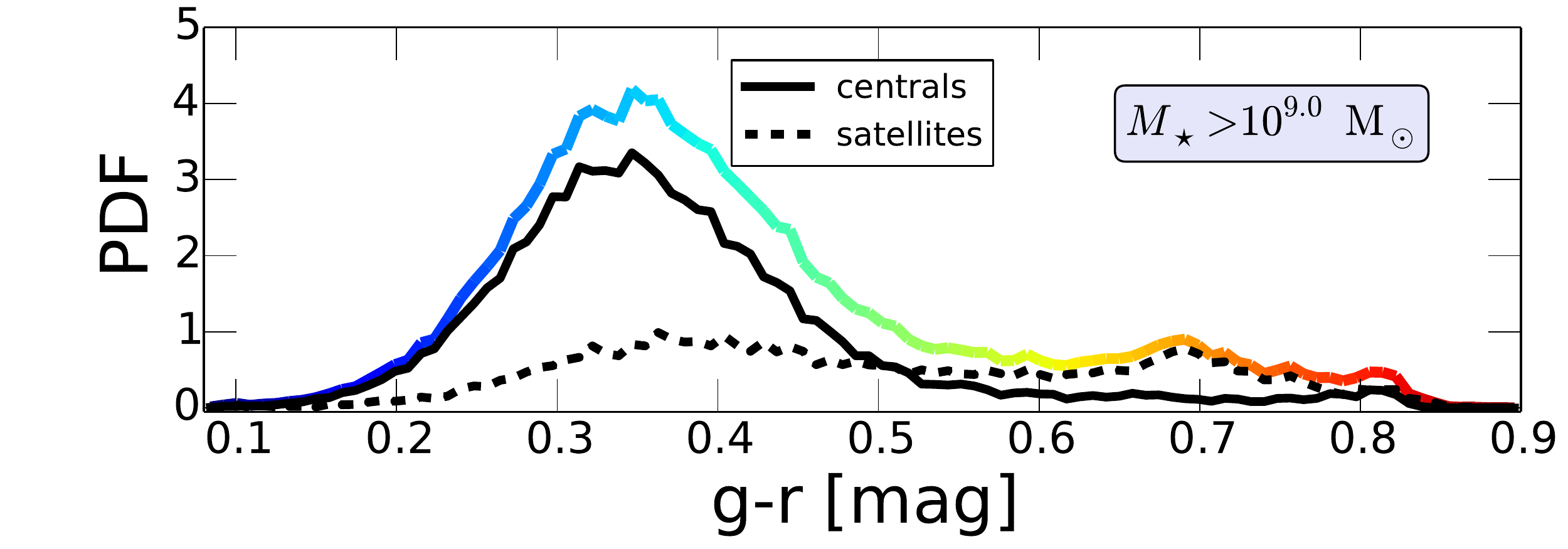}
\includegraphics[width=0.48\textwidth]{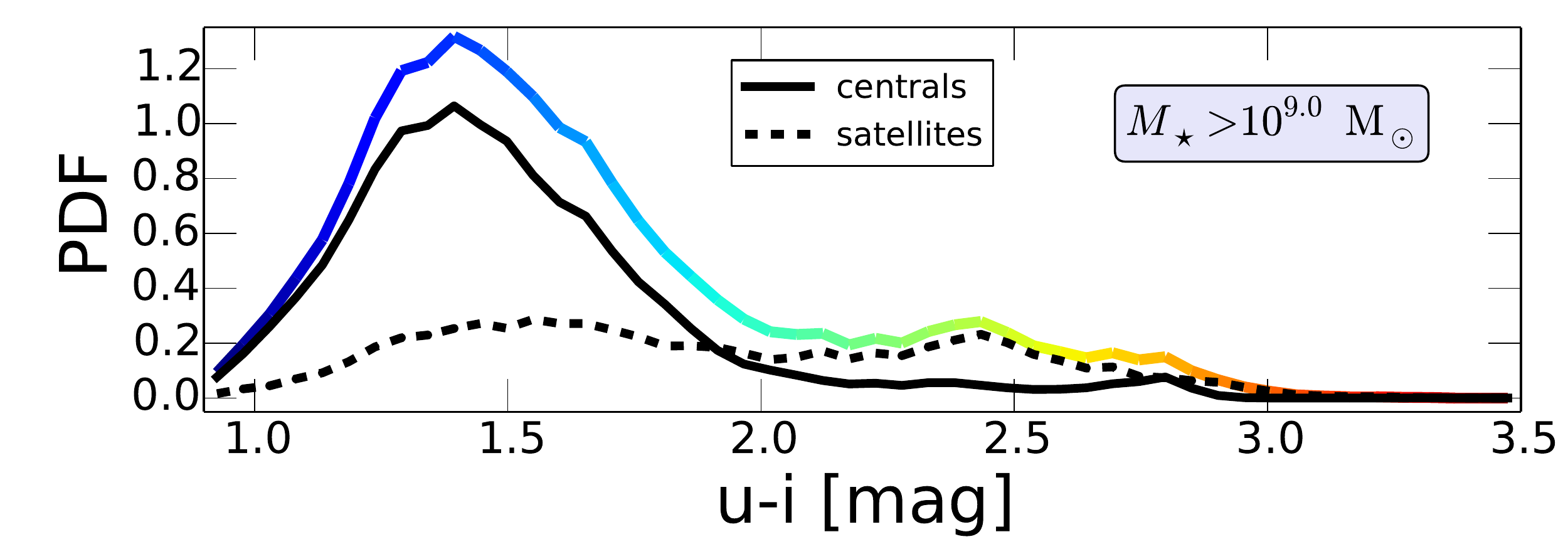}
\caption{Distribution of {g-r} and {u-i} colours as a function of stellar mass.
Top panels: Two-dimensional histograms showing galaxy {g-r} (left) and {u-i}
(right) colour versus stellar mass. We do not take dust effects into account and only show intrinsic galaxy colours.
The histograms are divided into three different galaxy populations:
star-forming, quiescent, and a mixed region, where both types of galaxies
occur. The distinction between star-forming and quiescent is made based on the
star formation rate threshold employed above for the GSMF (see
Figure~\ref{fig:GSMF}). Bottom panels: Distribution of {g-r} (left) and {u-i}
(right) colours for all galaxies with $M_\star > 10^9\msun$ (coloured lines). The integrals over
the histograms are normalised to one. The black dashed and solid lines split the total sample in centrals and satellites.}
\label{fig:color_mass}
\end{figure*}

Observationally, the GLF can be ``directly'' obtained. However, simulations
have to make certain assumptions and use stellar population synthesis models to
assign luminosities to stellar particles. Furthermore, the modelling of dust,
which we neglected in our analysis, adds further uncertainties.  Alternatively,
stellar population synthesis models fitted to broadband spectral energy
distributions can be used to transform an observed luminosity function into a
galaxy stellar mass function (GSMF). This quantity can be straightforwardly
compared to simulations, where the stellar masses of galaxies are a direct
output.

Figure~\ref{fig:GSMF} adopts this complimentary view and compares the predicted
GSMF to current observations~\citep[][]{Moustakas2013,
Bernardi2013} based on data from SDSS, and the PRism MUlti-object Survey
(PRIMUS)~\citep[][]{Coil2011, Cool2013}.  To bracket uncertainties in the
observational estimates for bright and massive
galaxies~\citep[][]{Bernardi2013} we calculate the total stellar mass excluding
satellites ($M^{\rm tot}_\star$) and for just the central stellar mass
($M^{<\,r_\star}_\star$) for the simulated galaxies following the approach used
above for the GLF.  Given these uncertainties, the theoretically estimated
stellar mass function follows the observational trends reasonably well,
even though the predicted faint end is slightly overpopulated around $M_\star
\sim 10^{9}\msun$.  We note, however, that these galaxies are relatively poorly
resolved in the simulation (with only $\sim 1000$ stellar resolution elements).
The simulation volume of Illustris is too small to sample the massive end
very well, which is why there is still some statistical uncertainty in the
predicted GSMF towards larger masses. Interestingly, the stellar mass
function of star-forming galaxies with ${\log[{\rm SFR}/(\msun {\rm
yr}^{-1})]>-0.6 + 0.65\,\log[M_\star/10^{10}\msun]}$ matches the
observations~\citep[][]{Moustakas2013} better than the total sample.
Furthermore, although the total GSMF does not perfectly agree with the
observational data, we recover the finding that the total galaxy stellar mass
function at the massive end is almost entirely comprised of non-star forming
galaxies. For $M_\star>10^9\msun$ we estimate that about $52\%$ of all stellar
mass at $z=0$ is in galaxies with a low SF rate, while
observations~\citep[][]{Baldry2004, Moustakas2013} report $54\%-60\%$. However,
here we also find that the low mass end is even more pronounced for the SF
population than for the total GSMF, indicating that we specifically
over-produce star-forming galaxies at those masses.

\begin{figure*}
\centering
\includegraphics[width=0.49\textwidth]{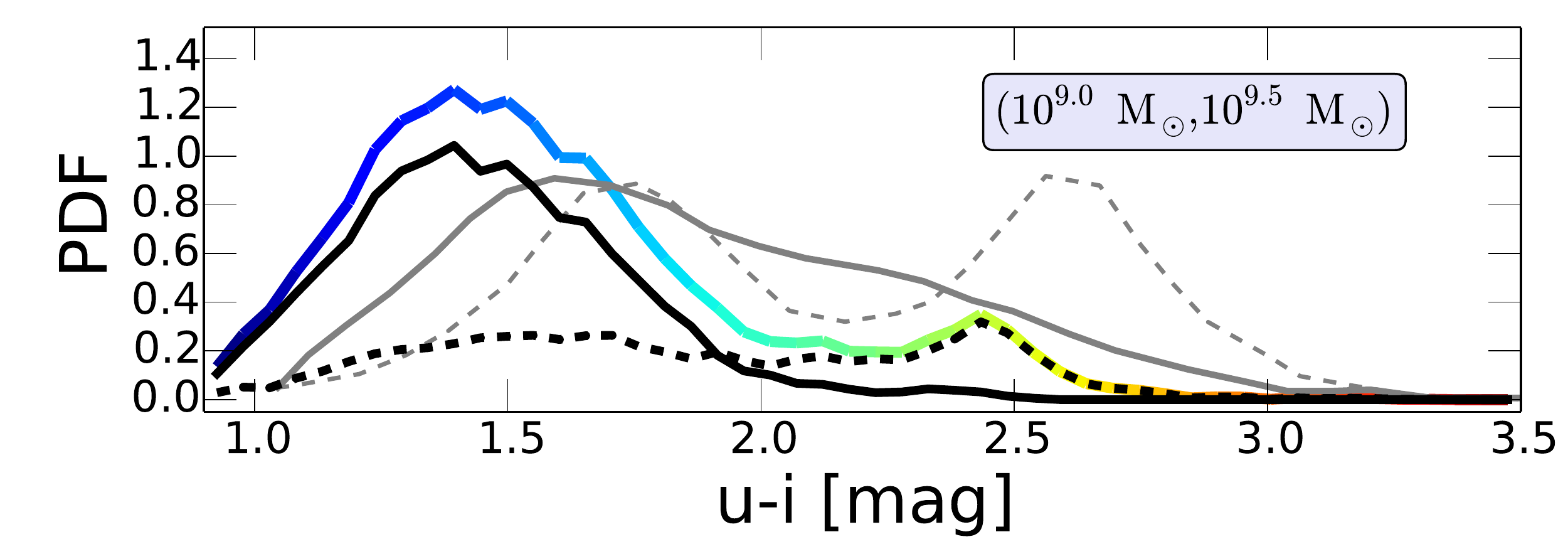}
\includegraphics[width=0.49\textwidth]{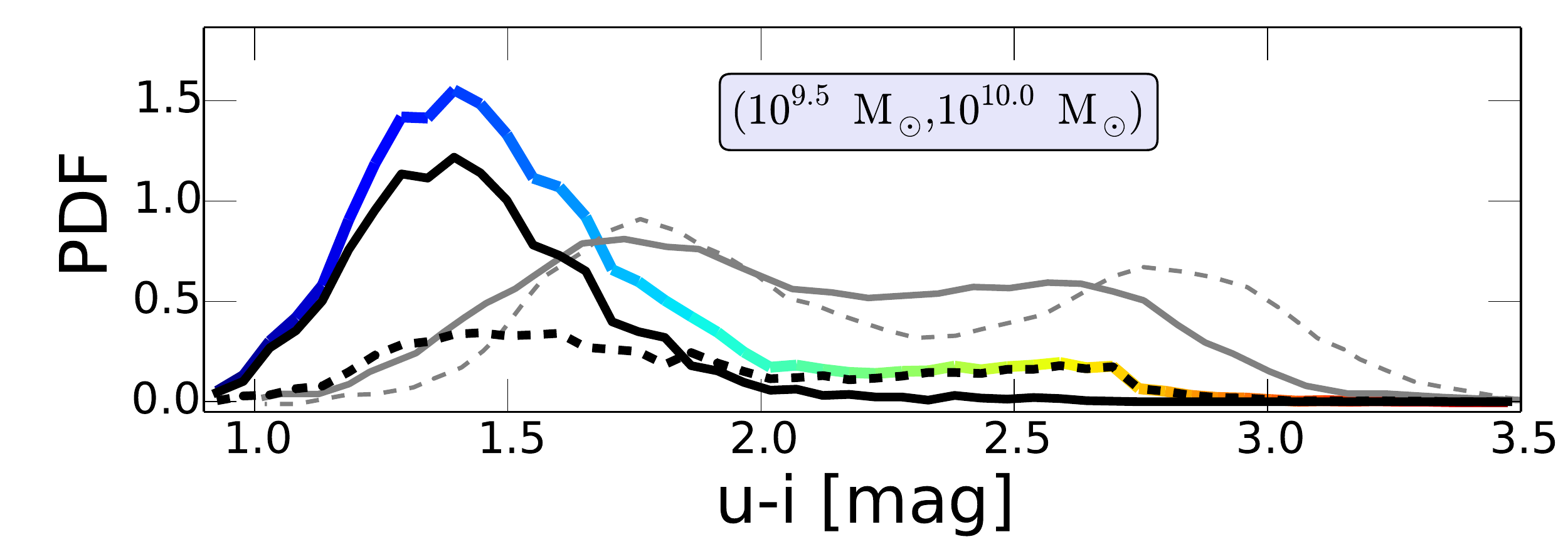}
\includegraphics[width=0.49\textwidth]{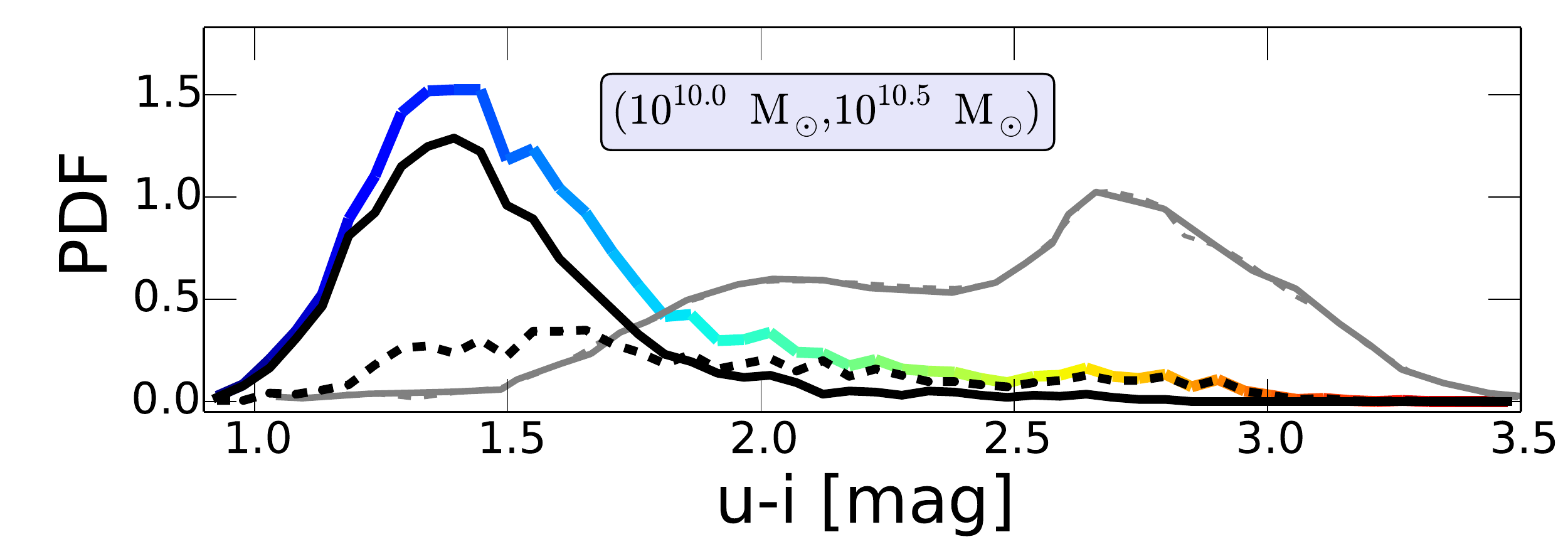}
\includegraphics[width=0.49\textwidth]{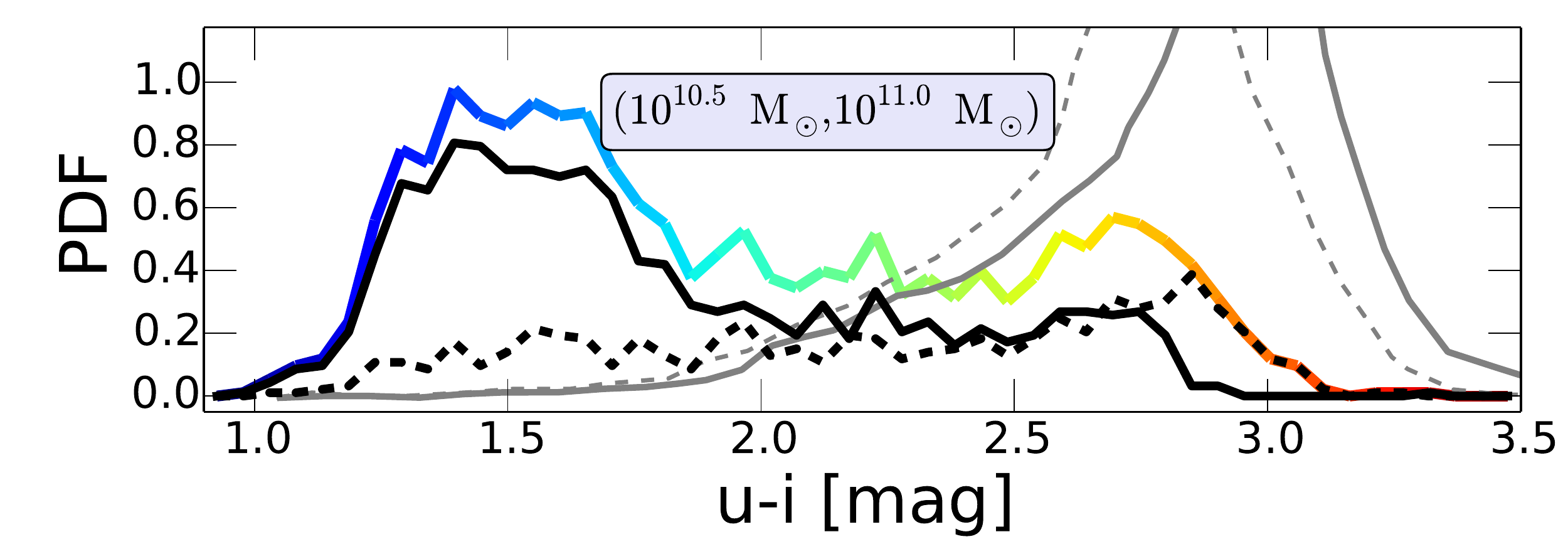}
\includegraphics[width=0.49\textwidth]{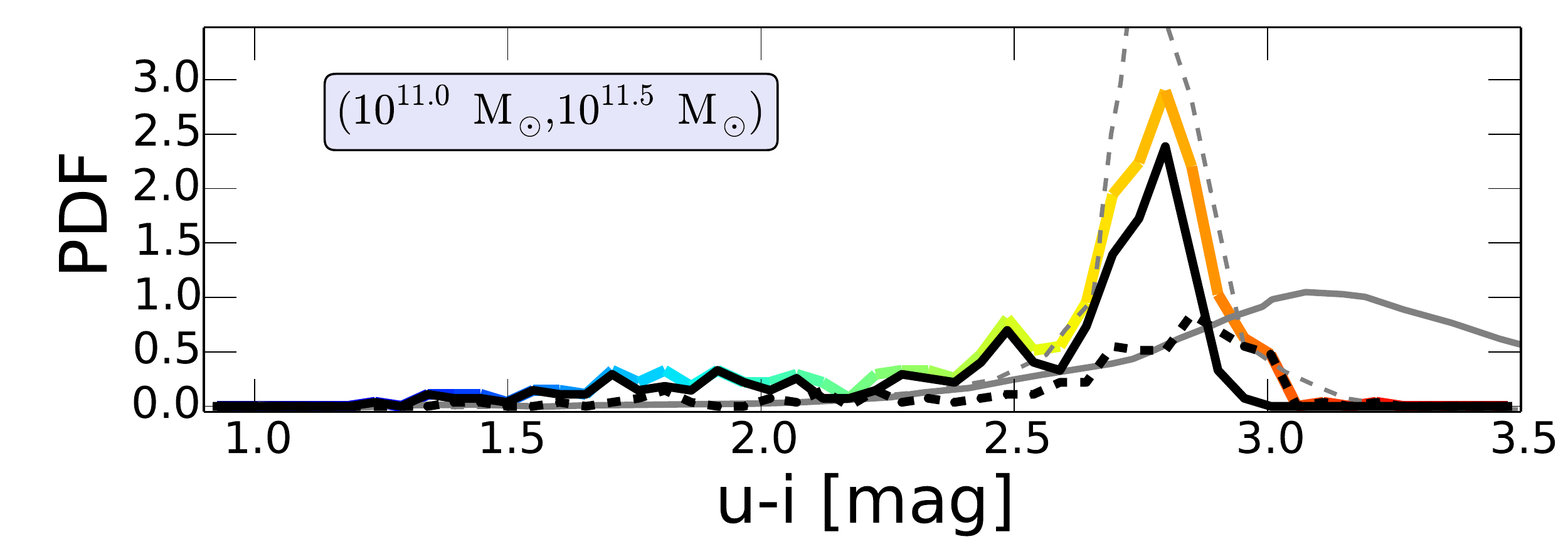}
\includegraphics[width=0.49\textwidth]{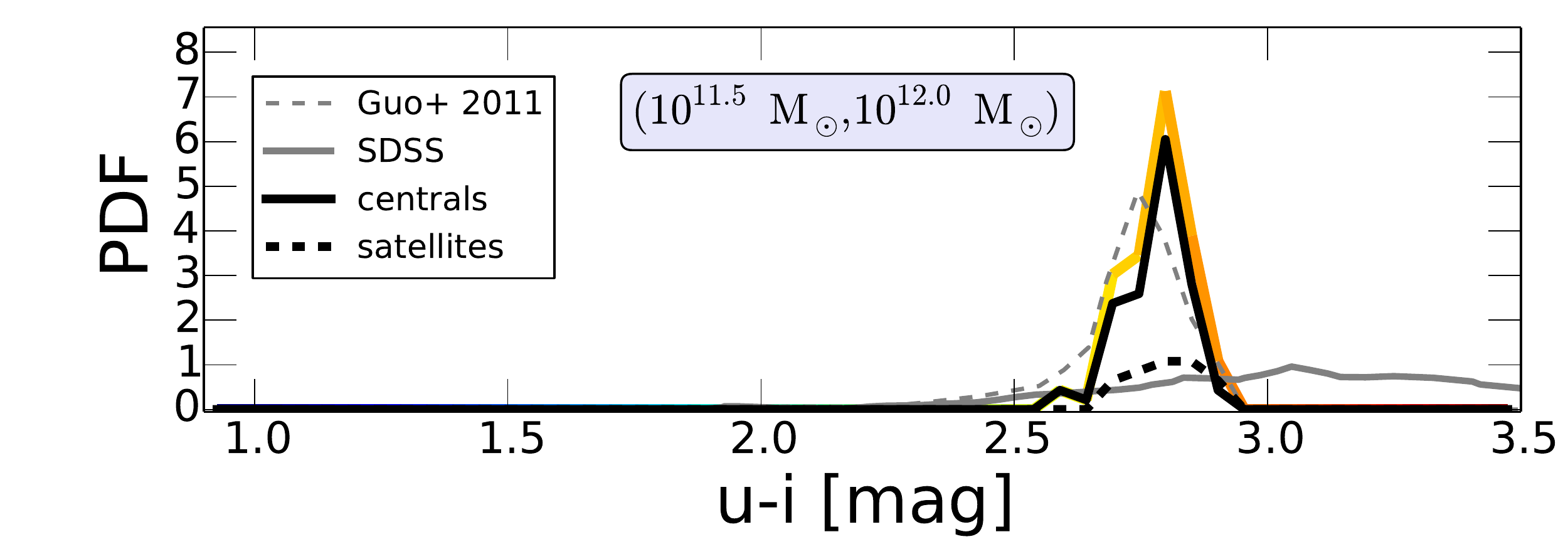}
\caption{Distribution of {u-i} colours for galaxies split into bins of stellar mass ($M_\star$) as indicated.
The integral over the histograms is normalised to one, and the coloured lines
indicate the colour value. The distributions demonstrate that
the simulated galaxy population clearly reddens towards the massive end. We note that the colours
do not include the effects of dust reddening. We split the sample into centrals (black solid) and
satellites (black dashed). We also show observational results based on SDSS (gray solid), and results
from the semi-analytic model of \protect\cite{Guo2011} (gray dashed).}
\label{fig:color_hist}
\end{figure*}

Feedback and cooling processes control the baryon conversion efficiency from
the gas phase to the stellar phase, and are required to match the GLF and GSMF.
Regulating SF is therefore a key ingredient in modern galaxy formation
simulations, since without feedback galaxies would have significantly too high
stellar masses.  Abundance matching techniques have over the last decade used
DM-only N-body simulations combined with large galaxy surveys like SDSS to
constrain the expected baryon conversion efficiency as a function of halo mass.
We present a comparison of our simulation to recent abundance matching
estimates~\citep[][]{Moster2013,Behroozi2013,Kravtsov2014} in
Figure~\ref{fig:shmr}.  Here we show the stellar mass to halo mass ratio as a
function of halo mass.  Abundance matching results are derived by
simultaneously ranking DM haloes obtained through DM-only N-body simulations
and observed stellar masses. This implies, however, that a proper comparison to
hydrodynamical simulations should be based on the relation of halo masses
obtained from DM-only simulations and the associated stellar masses as
predicted by the corresponding hydrodynamical
simulation~\citep[][]{Sawala2013,Munshi2013}. We follow this procedure using
the matched halo sample discussed above.  

In Figure~\ref{fig:shmr}, we present therefore the stellar mass content of
central galaxies in units of the universal baryon fraction ($\Omega_{\rm
b}/\Omega_{\rm m}$) as a function of the matched DM-only halo mass ($M_{\rm
200,crit|DM}$) for our galaxy population at $=0$; the redshift evolution of
this relation is presented in \cite{Genel2014}. Here, we use two stellar
mass estimates: total ($M_\star^{\rm tot}$) and the stellar mass contained
within our fiducial galaxy radius $r_\star$ ($M_\star^{<r_\star}$). This is to
take again into account uncertainties in the stellar mass estimates for more
massive systems~\citep[see][for more details]{Kravtsov2014}.
Figure~\ref{fig:shmr} demonstrates that we find a reasonable agreement with
the abundance matching results; SF is most efficient close to the mass-scale of
the Milky way ($\sim 10^{12}\msun$), where the observationally inferred baryon
conversion efficiency reaches $\sim 20-30\%$ over a large redshift range. Lower
and higher mass haloes have orders of magnitude smaller baryon conversion
efficiencies so that most of the stellar mass is found in haloes around this
mass scale. Reproducing this result is crucial, for example, for predicting the
correct total amount of stellar mass in the Universe.  Our simulation results
are well within the $1\sigma$ observational uncertainties (shaded regions)
demonstrating that our feedback implementation leads to a stellar mass growth
consistent with observations. The $1\sigma$ regions of the simulation data are
indicated through thin dashed lines.  The abundance matching result
of~\cite{Kravtsov2014} agrees reasonably well with our results taking into
account all stellar mass of haloes excluding the mass residing in satellites of
the host. 

We conclude that our feedback models both for SN and AGN feedback are
sufficient to reduce SF roughly to the observed level. However, we stress that
uncertainties in the stellar light assignment affect the GLF function and GSMF
as discussed above.  Since abundance matching techniques rely on
observationally derived stellar mass estimates, we have to consider this effect
also when comparing the stellar mass content of haloes to that derived through
abundance matching. Taking these effects into account we find reasonable
agreement for the GLF, the GSMF, and the amount of stellar mass for a given
halo mass. Most importantly our model shows a maximum SF efficiency at the
observationally inferred mass scale, which is achieved through the interplay of
SN and AGN feedback. 

We stress that it is dangerous to construct and tune galaxy formation models
such that they exactly reproduce observations like the galaxy stellar mass
function. For example, the systematic errors at the bright end of the galaxy
stellar mass function have to be considered when constructing models. The same
is true for matching the stellar to halo mass relationship.  These relations
are also uncertain at higher mass and different abundance matching results
differ also towards lower halo masses. Most recently, \cite{Sawala2014}
demonstrated that classical abundance matching even breaks down below halo
masses of $3\times 10^9\msun$. It is therefore not advisable to construct
large-scale galaxy formation models such, that they are tuned to reproduce
these observations exactly.

\subsection{Distribution of colours}

Our simulation also predicts galaxy colours in different bands. In this first
analysis we do not take into account dust effects on the emitted galaxy light
and colours, but rather focus on the intrinsic color distributions.
However, we note that dust can substantially affect colours of
galaxies.  The distribution of colours against stellar mass is shown in
two-dimensional histograms in the two top panels of
Figure~\ref{fig:color_mass}, where we present the {g-r} colours (left) and
{u-i} colours (right) as a function of stellar mass. We divide the histograms
further in three different galaxy populations: star-forming (blue), quiescent
(red), and a mixed region (green), where both types of galaxies occur. The
distinction between star-forming and quiescent is made based on the star
formation rate threshold employed for the GSMF above (see
Figure~\ref{fig:GSMF}). We stress that the three regions are strictly disjoint,
e.g., the quiescent region (red) does not include any star-forming galaxies
(blue) according to our cut in the specific star formation rate.  Green bins in
Figure~\ref{fig:color_mass} contain both star-forming and quiescent galaxies.
The histogram therefore illustrates that the reddest galaxies are typically
quiescent, especially at the massive end, where all red galaxies are quiescent.
However, there is also a substantial fraction of quiescent low mass galaxies,
which are also red. On the other hand, star-forming galaxies are typically much
bluer, and more biased towards less massive systems. In between these two
regimes we identify a narrow region where we find both types of galaxies:
quiescent and star-forming. 

\begin{figure}
\centering
\includegraphics[width=0.49\textwidth]{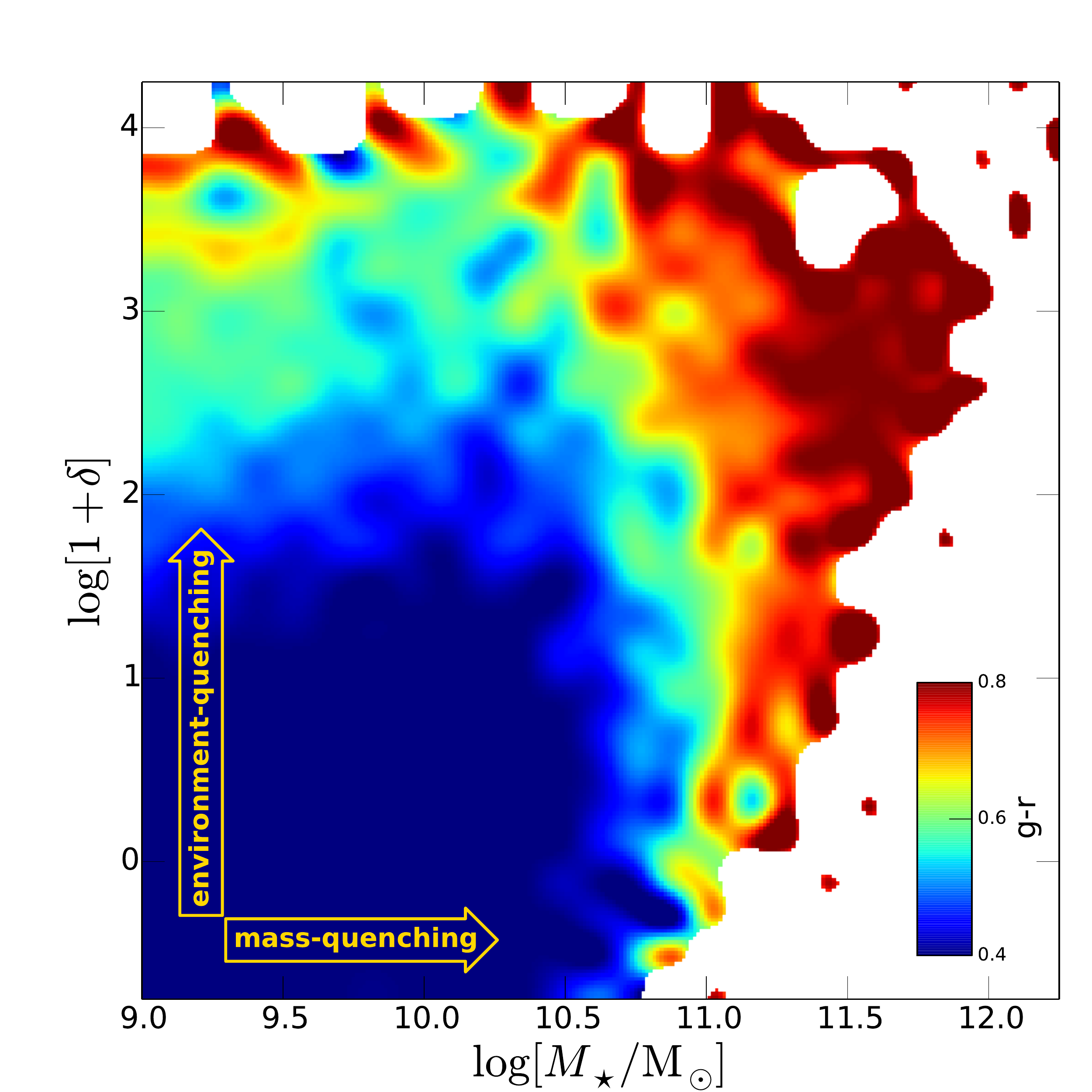}
\caption{Average {(g-r)} colour of galaxies
  as a function of stellar mass and galaxy overdensity $1+\delta$. We find mass-quenching
  along the horizontal axis and environment-quenching along the vertical axis. Along both axes
  galaxies tend to become redder. The overdensity field ($1+\delta$) is based
on the fifth nearest galaxy considering only systems with $r<-19.5$.}
\label{fig:colours}
\end{figure}

The coloured lines in the bottom panels of Figure~\ref{fig:color_mass} show
one-dimensional distributions of {g-r} colours (left) and {u-i} colours (right)
for all galaxies with $M_\star > 10^9\msun$,~i.e. galaxies resolved with more
than $\sim 1000$ stellar resolution elements. All histograms are normalised
such that the integral over the distribution sums up to unity.  The simulation
predicts a slightly bimodal distribution both in {g-r} and {u-i} colours.
However, the second maximum for red galaxies in both distributions is not as
pronounced as found in observations~\citep[e.g.,][]{Blanton2005b}.  This can be
caused by different effects probably related to quenching in massive haloes.

Firstly, it seems that our radio-mode AGN feedback is not efficient enough to
suppress SF sufficiently in massive haloes as mentioned above.  Too little
quenching can lead to stellar populations which are too young and therefore
biased towards bluer colours. A second reason for the less pronounced second
maximum lies potentially in the rather small sample of massive objects in our
simulation volume. For example, we have only ten haloes with masses above
$10^{14}\msun$ (C-1 to C-10, see above). The massive part of the galaxy
population is therefore not captured well, and could potentially be biased due
to the initial density field that is sampled by the simulation. A final reason
is related to the effects of dust, which renders galaxies redder than
intrinsically expected. Since we do not include this effect here, we naturally
find a smaller number of red galaxies. Also the maximum ``redness'' of galaxies
is expected to be lower compared to real galaxies.  Only the first of these
three reasons is inherent to our model, and it remains to be seen whether other
mechanisms or a different AGN radio-mode model need to be introduced to quench
SF more efficiently in massive systems like groups and clusters. Assuming that
the second and third reasons are not the main driver of the discrepancy, our
results seem to indicate that radio-mode AGN feedback alone may not be
sufficient to quench massive haloes enough. However, the exact duty cycle of
the radio-mode will also affect this behaviour. Below, we will demonstrate a
fourth possibility, namely that the quenching and reddening of intermediate
mass haloes is also not sufficiently strong in our model. 

The bottom panels of Figure~\ref{fig:color_mass} also split the galaxy
population into centrals (solid black line) and satellites (dashed black line).
Here we find that the red part of the distribution is actually dominated by
satellite galaxies, and not by centrals. Centrals clearly dominate the blue
part of the distribution.  To inspect this behaviour further we show in
Figure~\ref{fig:color_hist} {u-i} colour distributions for different stellar
mass cuts going from low to higher mass systems (coloured lines). Our results
agree qualitatively with observations in the sense that more massive systems
tend to have significantly redder colours distributions compared to low mass
systems. In fact, the most massive systems included in
Figure~\ref{fig:color_hist} peak around ${\rm u-i}\sim 2.8$. The
$10^{10.5}\msun$ to $10^{11.0}\msun$ mass range shows a clear bimodal
distribution with broad maxima at ${\rm u-i}\sim 1.5$ and ${\rm u-i}\sim 2.75$.
We split all distributions again into centrals (solid black) and satellites
(dashed black), and find that the general trend is that the red population at
low stellar masses is made up by satellite galaxies, whereas centrals dominate
the red population towards higher stellar masses. Centrals start to become more
relevant in the red part of the distribution above $M_\star \gtrsim
10^{10.5}\msun$.

Although we do not include dust effects, we also show
observational SDSS results (thin gray lines) in the different panels of
Figure~\ref{fig:color_hist}. This reveals another qualitative agreement of our
model with observations: only intermediate mass galaxies show a bimodal
distribution.  Despite the fact that we do not include dust effects, we find
for some mass ranges reasonable agreement with the observed colour
distribution.  However, in the intermediate mass regime the predicted
quantitative colour distribution clearly deviates from the observed samples.
Here we miss a significant fraction of red and quenched galaxies, which are
observed, but not present in our model.  This relates to the  fourth reason for
why our total colour distribution is not as bimodal as found observationally:
we miss a significant fraction of red and quenched galaxies in the intermediate
mass range.  A similar conclusion is reached by looking at the quenched
fraction as a function of galaxy mass~\citep[see][]{Genel2014}.  

We have
also added model predictions from~\cite{Guo2011} based on a semi-analytic model
run on the Millennium-I and -II simulations N-body simulations.  At higher
masses our simulation results agree well with the semi-analytic predictions
from~\cite{Guo2011}.  However, both models tend to produce too few red galaxies
at the massive end, and the maximum {u-i} values agree well between the two
models, but are too low compared to observations.  Furthermore, both models
predict a rather narrow and peaked distribution, which is not observed. Instead
in the massive mass bins observations favor a broad distribution extending to
{u-i} colours larger than $3.5$. At lower masses we find galaxies which are
bluer than observationally found, but dust effects can in principle compensate
for this. 

\begin{figure*}
\centering
\includegraphics[width=0.49\textwidth]{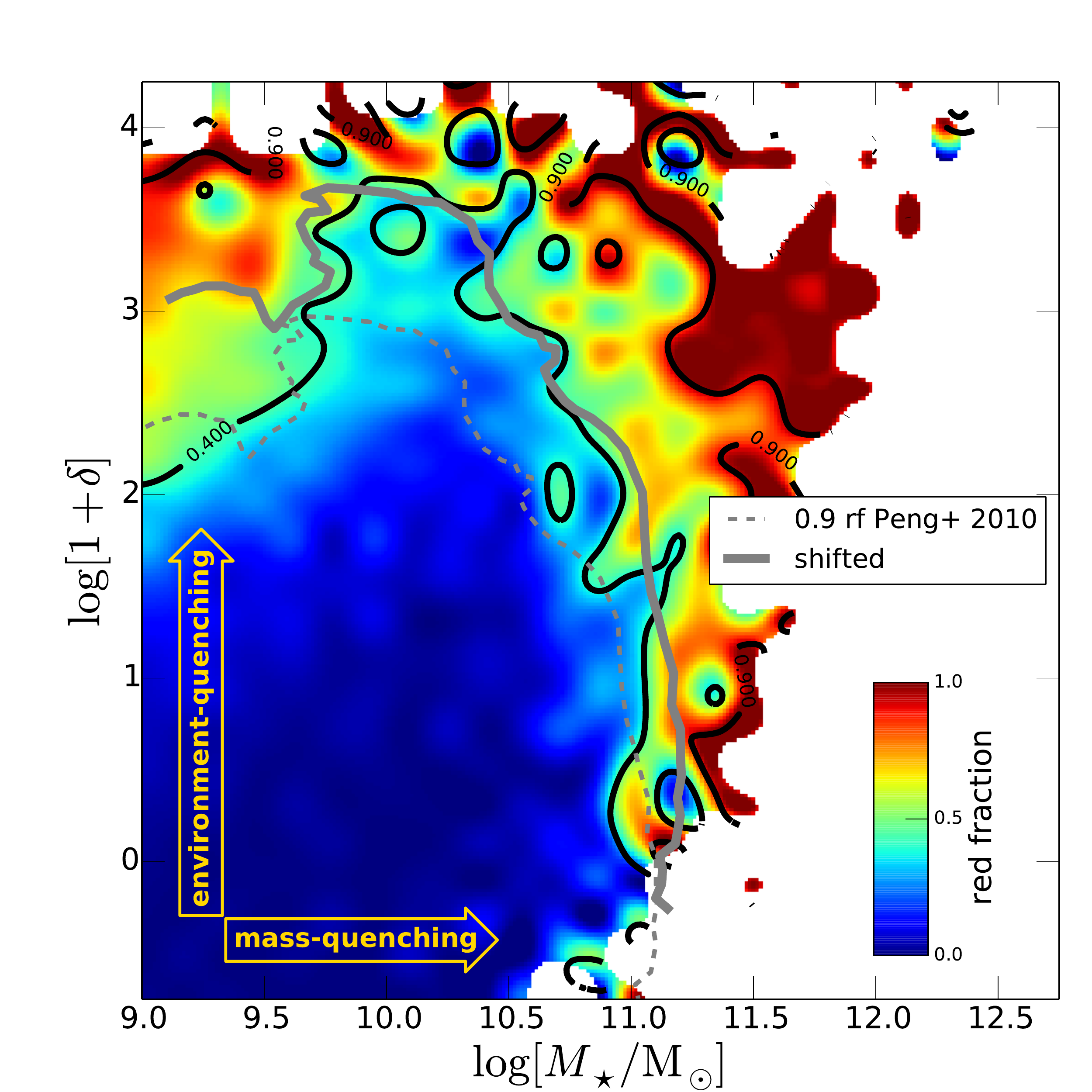}
\includegraphics[width=0.49\textwidth]{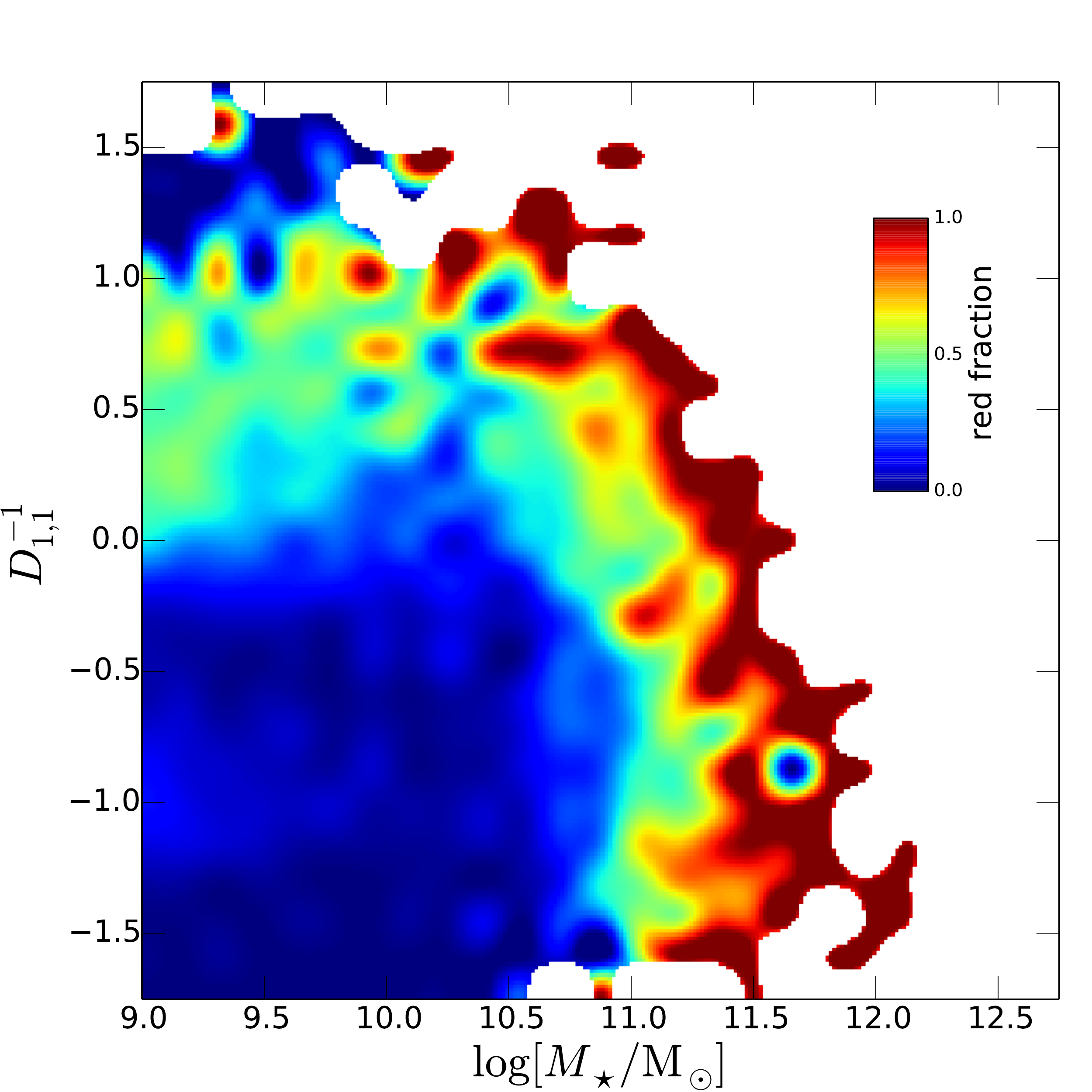}
\includegraphics[width=0.49\textwidth]{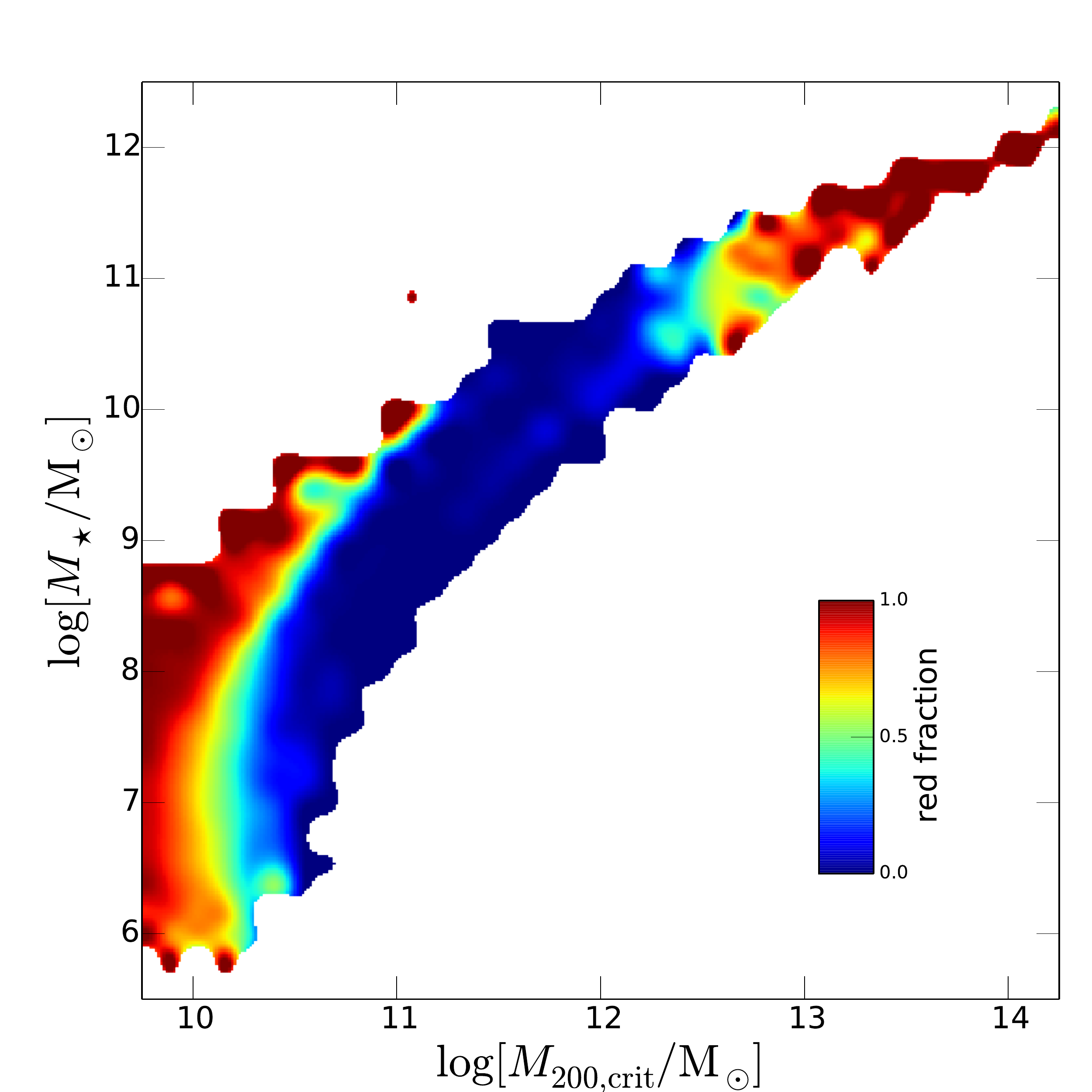}
\includegraphics[width=0.49\textwidth]{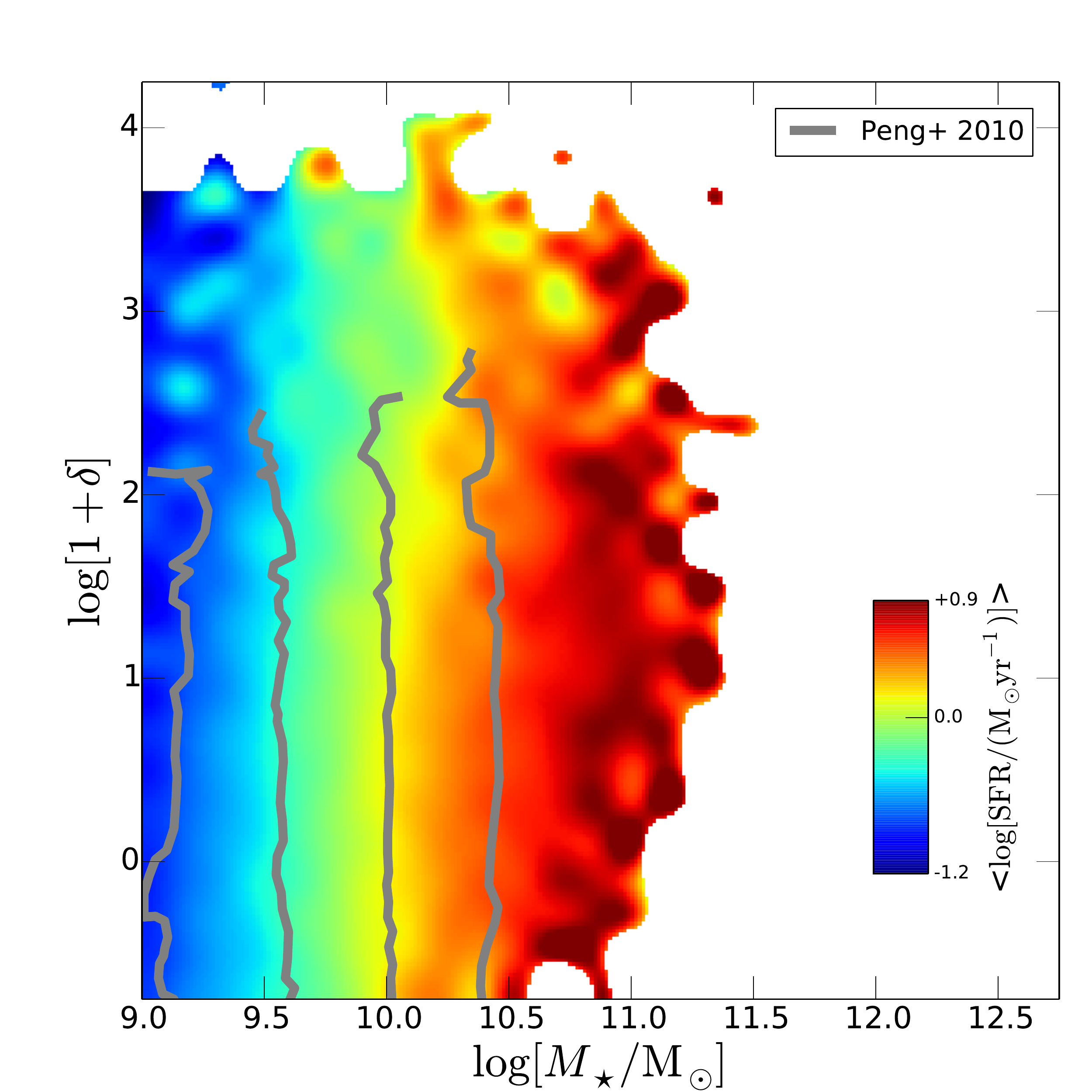}
\caption{Environmental dependence of the galaxy population.  Upper left panel:
Red fraction of galaxies as a function of stellar mass and galaxy overdensity.
Reddening occurs along the mass and overdensity axes. We also include
observational contour lines for a red fraction of $0.9$
from~\protect\cite{Peng2010} (dashed). The shifted solid line represents a mass
and overdensity shift to better fit our data. This shifted contour line
coincides with the simulated contour line for a red fraction of $0.4$ (black
line), but not with that of $0.9$. Upper right panel: Red fraction of galaxies
as a function of stellar mass and galaxy overdensity, but now using a
theoretically motivated environment estimator described
in~\protect\cite{Haas2012}, which has only little halo mass dependence. Lower
left panel: Stellar mass as a function of halo mass colour coded according to
red fraction. SF is most efficient around $M_{\rm 200,crit} \sim 10^{12}\msun$
where we find blue star forming galaxies sitting on the SF main sequence.
Lower right panel: Mean SF rate as a function of stellar mass and galaxy
overdensity.  We include here only star forming galaxies. For these, the SF
does not depend on $\delta$, but instead is a pure function of stellar mass
through the SF main sequence.  Observational contour lines from
\protect\cite{Peng2010} are also shown, and those tend to show no dependence on
the overdensity either. Combined with the findings in the other panels this
implies that the rate of SF does not depend on environment, but the probability
of a galaxy being quenched and becoming redder does.}
\label{fig:environment}
\end{figure*}

\subsection{The impact of environment}

As demonstrated above, SN and AGN feedback are key processes in regulating SF
and in determining the stellar mass content of haloes.  However, many studies have
also established the view that the evolution of galaxies is a strong function
of environment, both observationally~\citep[e.g.,][]{Hogg2003, Kauffmann2004,
Blanton2005b, Baldry2006, Peng2010} and theoretically~\citep[e.g.,][]{Crain2009},
based on various measures used to quantify galaxy environments~\citep[see][for
an overview]{Haas2012}.  Galaxies entering high density regions like clusters
are prone to processes like ram pressure stripping, starvation, strangulation, and tidal
stripping, which are captured self-consistently by hydrodynamical simulations, a major
advantage over ad hoc treatments employed in semi-analytic models.  These
physical processes are significant because they influence, for example, the SF
rates and colours of galaxies as a function of environment.  

We demonstrate this in Figure~\ref{fig:colours}, where we present the average
{(g-r)} colours of galaxies as a function of stellar mass and local galaxy
overdensity $1+\delta$ at $z=0$. The overdensity field is calculated based on
the fifth nearest galaxy considering only systems with $r$-band magnitude
$<-19.5$ for the construction of the galaxy density field at $z=0$. This leaves
us with $11,112$ galaxies from which we calculate the local overdensity field
($1+\delta$) for each other galaxy that we can correlate with other galaxy
properties. We note that this procedure is commonly adopted in observational
studies~\citep[e.g.,][]{Peng2010}.  Figure~\ref{fig:colours} demonstrates that
galaxies tend to be redder at larger stellar masses and at higher galaxy
overdensities, in accord with observations.  These two ``reddening axes'' are
however related to entirely different physical processes.  Quenching of massive
galaxies proceeds mostly through AGN feedback in large haloes, which is
associated with the heating and removal of large amounts of interstellar gas so
that star formation ceases and the galaxy starts to redden as its stellar
population ages.  Environmental quenching due to denser surroundings primarily
affects lower mass satellite galaxies, which are stripped of their gas as they
experience ram pressure when falling into larger haloes.

Next we explore more quantitatively the dependence of the red fraction on
stellar mass and $1+\delta$ in the upper left panel of
Figure~\ref{fig:environment}. Here we present the un-weighted average red
fraction in each bin with a threshold of ${(g-r) = 0.65
- 0.03\,(r+20)}$~\citep[see][]{Blanton2005}. The results again demonstrate that
  galaxies tend to be redder for larger stellar masses and higher galaxy
overdensities as defined through $1+\delta$.  We stress that the trends we
recover in this panel are qualitatively in good agreement with recent
observations~\citep[e.g.,][]{Peng2010}.  To make this clearer we also include
observational contour lines for a red fraction of $0.9$ from~\cite{Peng2010}.
The dashed line is the original result from~\cite{Peng2010}, whereas the
shifted line represents a mass and overdensity shift to fit better to our data.
The observational contour line was defined based on a different galaxy
overdensity $1+\delta$, and a different definition of red fraction. We have
therefore shifted this line by $+0.1\,{\rm dex}$ in mass, and $+0.7\,{\rm dex}$
in overdensity. Interestingly, this brings the general shape of the contour
line into good agreement with our results.  Most importantly, we recover the
drop in the red fraction around $M_\star \sim 10^{10}\msun$ at higher
overdensities. However, the shifted contour line coincides with the
simulated contour line for a red fraction of $0.4$ (black line), but not with
that of $0.9$. 

\begin{figure}
\centering
\includegraphics[width=0.49\textwidth]{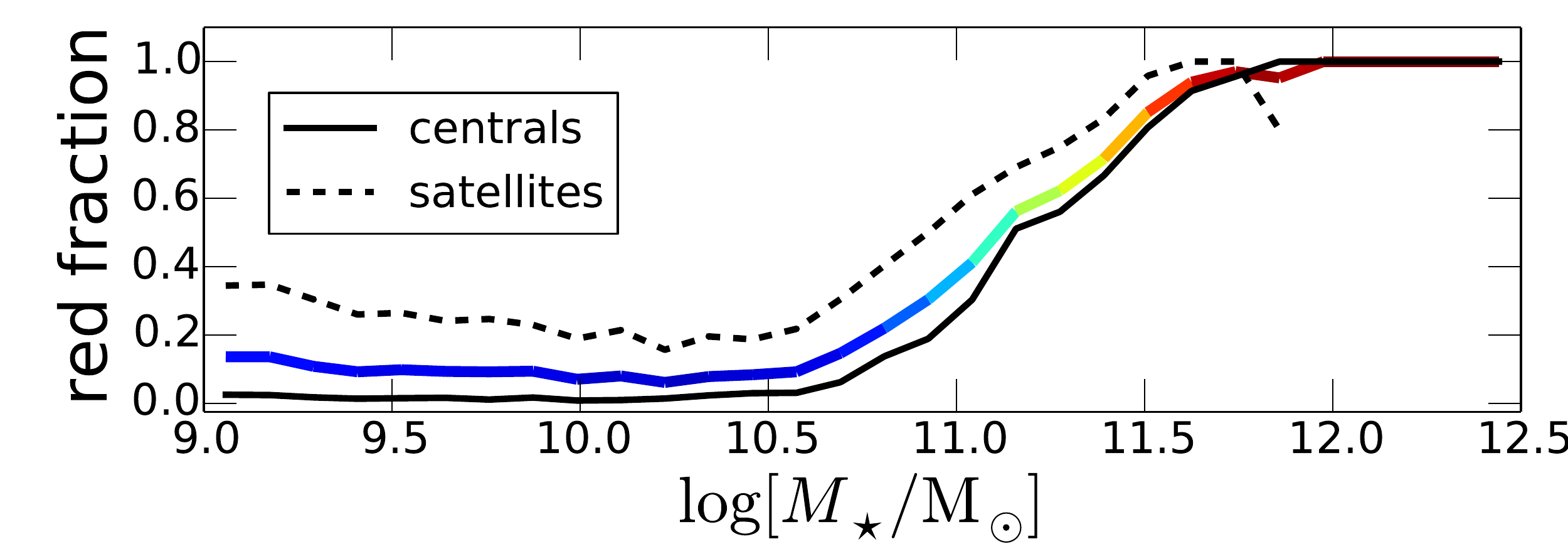}
\includegraphics[width=0.49\textwidth]{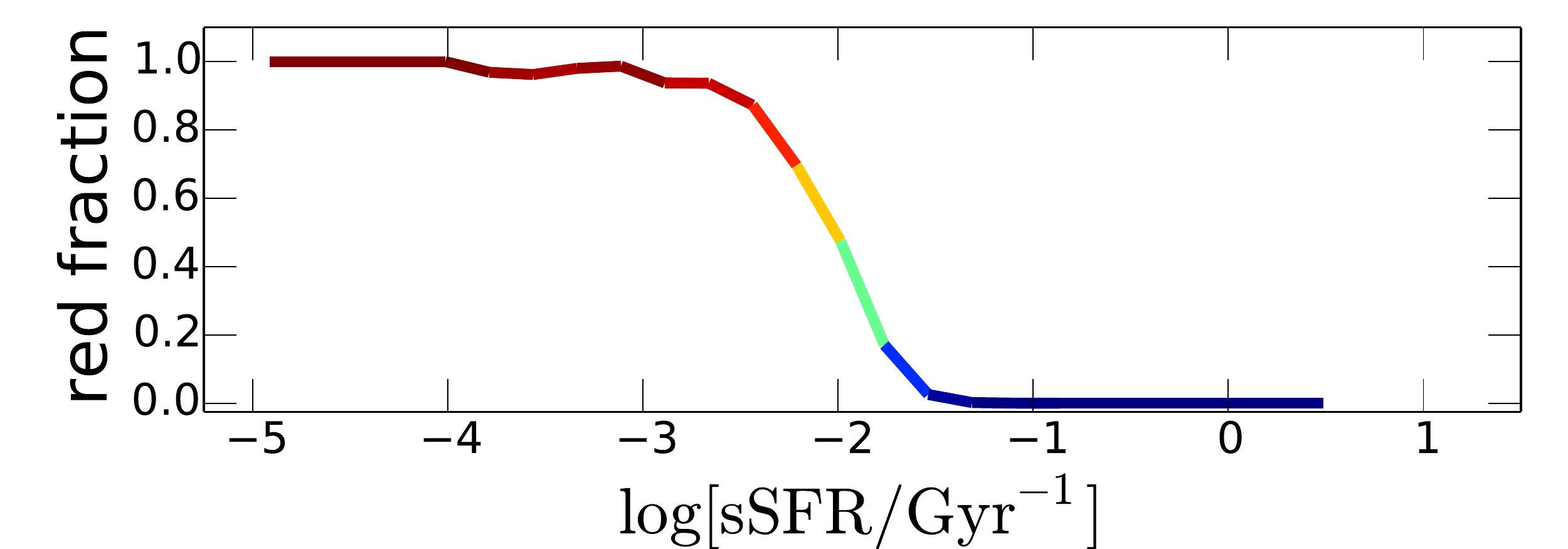}
\includegraphics[width=0.49\textwidth]{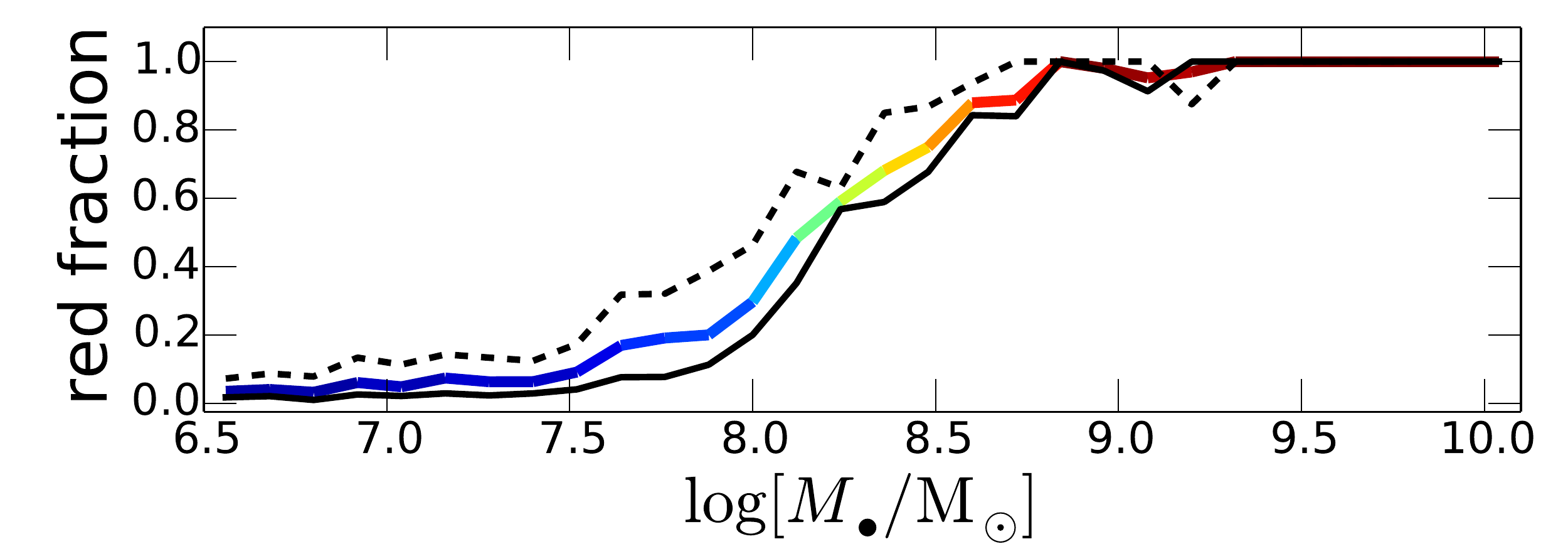}
\includegraphics[width=0.49\textwidth]{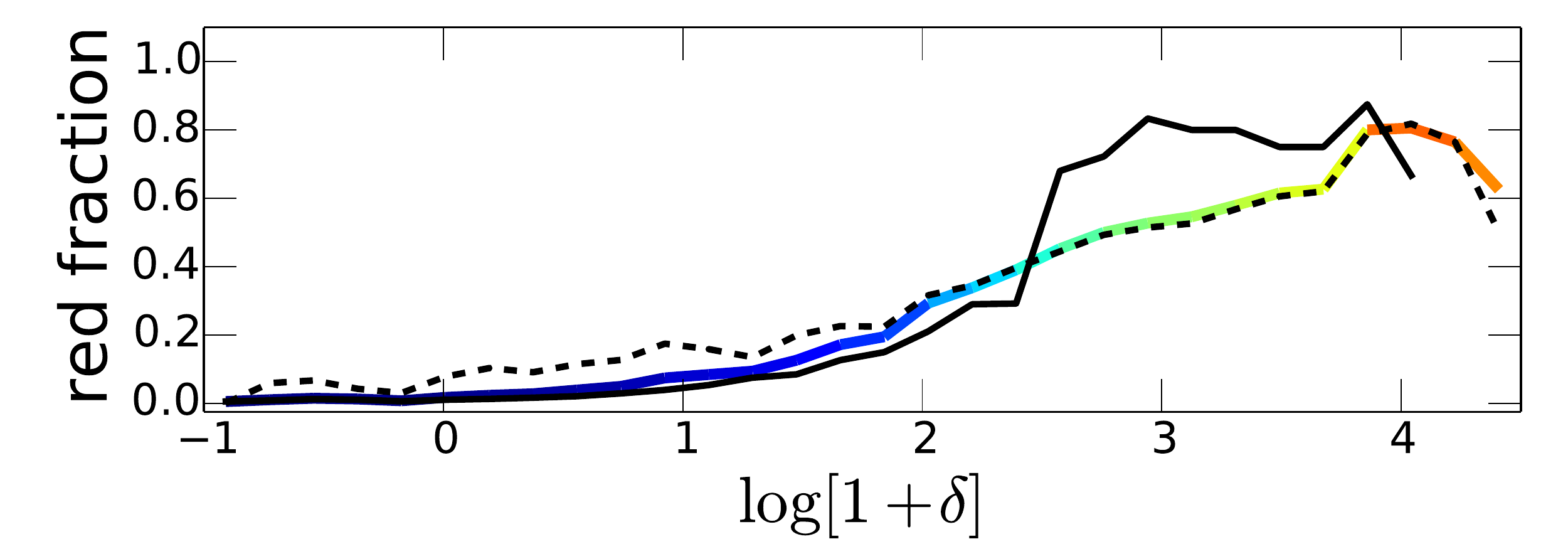}
\caption{Red fraction as (from top to bottom) a function of stellar mass,
specific SFR, BH mass, and galaxy overdensity for all galaxies more massive
than $10^{9}\msun$. The coloured line represents the total galaxy population,
whereas the solid and dashed black lines denote centrals and satellites,
respectively. The colour of the line simply reflects the red fraction on the
y-axis. The dependence of the red fraction on sSFR does not show any strong
dependence on galaxy type (centrals and satellites), and we do therefore not
include the central and satellite lines.}
\label{fig:redfraction}
\end{figure}

Many of the observationally employed overdensity estimators do not strictly
probe the galaxy environment.  In fact, it has been pointed out that most of
the common environment indicators, among them $1+\delta$, are to some degree
correlated with halo mass~\citep[see][]{Haas2012}.  They are therefore not
strictly probing environmental effects, but are ``contaminated'' also by the
halo mass dependence.  Other characteristics of probing the impact of the
environment on galaxy properties have therefore been proposed
by~\cite{Haas2012}. 

We consider this in the upper right panel of Figure~\ref{fig:environment},
where we use $D_{1,1}$, the three-dimensional distance to the nearest neighbour
with a virial mass that is at least that of the halo under consideration,
divided by the virial radius of the nearest neighbour, to probe the
environment~\citep[see][for more details]{Haas2012}.  This estimator aims to
minimise any halo mass dependence in the direction of the vertical axis. Using
this estimator we still recover two different quenching mechanisms:
mass-quenching along the horizontal axis and environment-quenching along the
vertical axis.  Although the upper left and upper right panels of
Figure~\ref{fig:environment} look similar, they differ in the dependence of the
red fraction as a function of $1+\delta$ and $D_{1,1}$ for fixed stellar masses
around $M_\star \sim 10^{10}\msun$. The estimator $D_{1,1}$ does not show an
equally strong drop in the red fraction towards higher overdensities compared
to $1+\delta$. This might be related to the fact that $1+\delta$ is still
contaminated by a remaining halo mass dependence as pointed out
by~\cite{Haas2012}.  Interestingly, the two upper panels of
Figure~\ref{fig:environment} also differ at the highest overdensities and
lowest masses, but in the opposite direction: the $D_{1,1}$ measure has its
upper-left corner very blue again, which is not the case for $1+\delta$.

The lower left panel of Figure~\ref{fig:environment} combines the stellar mass
halo mass relation and the fraction of red galaxies. SF is most efficient
around $M_{\rm 200,crit} \sim 10^{12}\msun$ where the galaxy population is
dominated by blue star forming galaxies which populate the SF main sequence
(see below).  Towards lower and higher masses, SF becomes less efficient and
the fraction of red galaxies increases in both directions due to quenching,
which is consistent with observations~\citep[][]{Woo2013}.  Although the red
fraction depends on environment, it is observationally found that the SF rate
of star forming galaxies does not depend on $1+\delta$, but only varies as a
function of mass in the $M_\star-\delta$ plane~\citep[][]{Peng2010}. The lower
right panel of Figure~\ref{fig:environment} demonstrates that we recover the
same trend in the simulation. Here we present the average SF rate as a function
of stellar mass and galaxy overdensity $1+\delta$ for star-forming galaxies
using the same cut as discussed above (see discussion of the GSMF, and
colours).  Gray lines in the lower right panel represent contour lines from
\cite{Peng2010}. These lines are essentially parallel to the contour lines
predicted by the simulation demonstrating that our model describes the
observational finding correctly.  We note that this result is not in conflict
with those presented in the upper panels. The fact that the red fraction of
galaxies increases with $1+\delta$ combined with a $1+\delta$-independent SF
rate only implies that the SF rate itself does not depend on environment,
whereas the quenching and reddening probability is a strong function of
environment. Both observationally and theoretically this points towards a
rather rapid process that quenches and reddens galaxies in higher density
environments. 

\begin{figure*}
\centering
\includegraphics[width=1\textwidth]{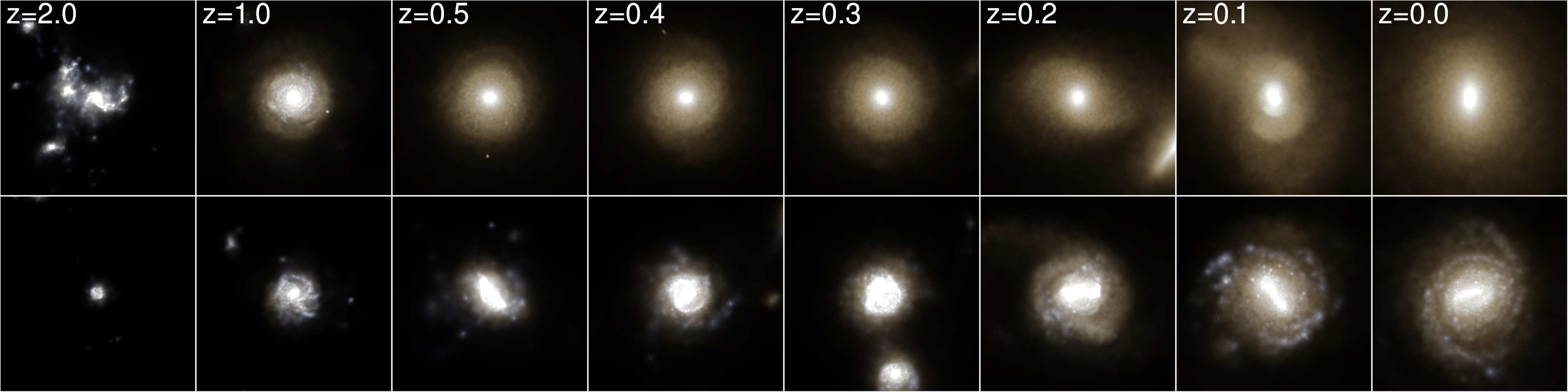}
\caption{Evolution of two selected galaxies across cosmic time. The images show
rest-frame g,r,i band composite images of the stellar light for the two
galaxies at various redshifts as indicated. The top galaxy evolves into a red
elliptical galaxy, which has only very little remaining star formation at
$z=0$. The bottom galaxy evolves into a blue star-forming spiral with a
bar-like structure in the center. Both galaxies experienced recent mergers,
which leads to the formation of shells in the stellar light distribution of the
elliptical galaxy. The merger of the blue galaxy leads to the formation of a
central bar. Both galaxies are distinct at $z=0$ in their morphology, their
colour, and their present star formation rate. At higher $z$ both galaxies are
significantly bluer.}
\label{fig:galevolve_image}
\end{figure*}

We summarise the dependence of the galaxy red fraction on the stellar mass, the
specific star formation rate, the BH mass, and galaxy overdensity in the four
panels of Figure~\ref{fig:redfraction}. The coloured line represents the total
galaxy population, whereas the solid and dashed black lines denote centrals and
satellites, respectively. The dependence of the red fraction on sSFR does not
show any strong dependence on the galaxy type (centrals and satellites), and
therefore we do not include the central and satellite lines.
Figure~\ref{fig:redfraction} reveals some clear trends of the red fraction. 

The first panel emphasises the strong increase of the red fraction with
increasing stellar mass.  As emphasised above red galaxies have typically only
little ongoing SF which can also be seen from the second panel.  Here we find a
rather steep drop of the red fraction around a specific SFR of about $\sim
10^{-2}\Gyr^{-1}$. Below (above) this threshold, we find only a low number of
blue (red) galaxies. Also as discussed above AGN feedback of SMBHs plays an
important role in quenching star formation towards more massive halo masses. It
is therefore not surprising that the red fraction increases with black hole
mass as shown in the third panel. The galaxy environment also shapes the galaxy
colours and affects the red fraction as demonstrated above. The last panel of
Figure~\ref{fig:redfraction} demonstrates this clearly. Interestingly, the red
fraction in centrals and satellites differs significantly towards higher
overdensities. For an observationally employed density estimator like
$1+\delta$ we find that around $1+\delta \sim 10^3$ centrals tend to be
significantly redder than satellites of the same overdensity. 

The results in this section demonstrate that SF formation, quenching, and
reddening are deeply connected through feedback processes and environment
effects, most of which are recovered by our model reasonably well.

\begin{figure*}
\centering
\includegraphics[width=0.49\textwidth]{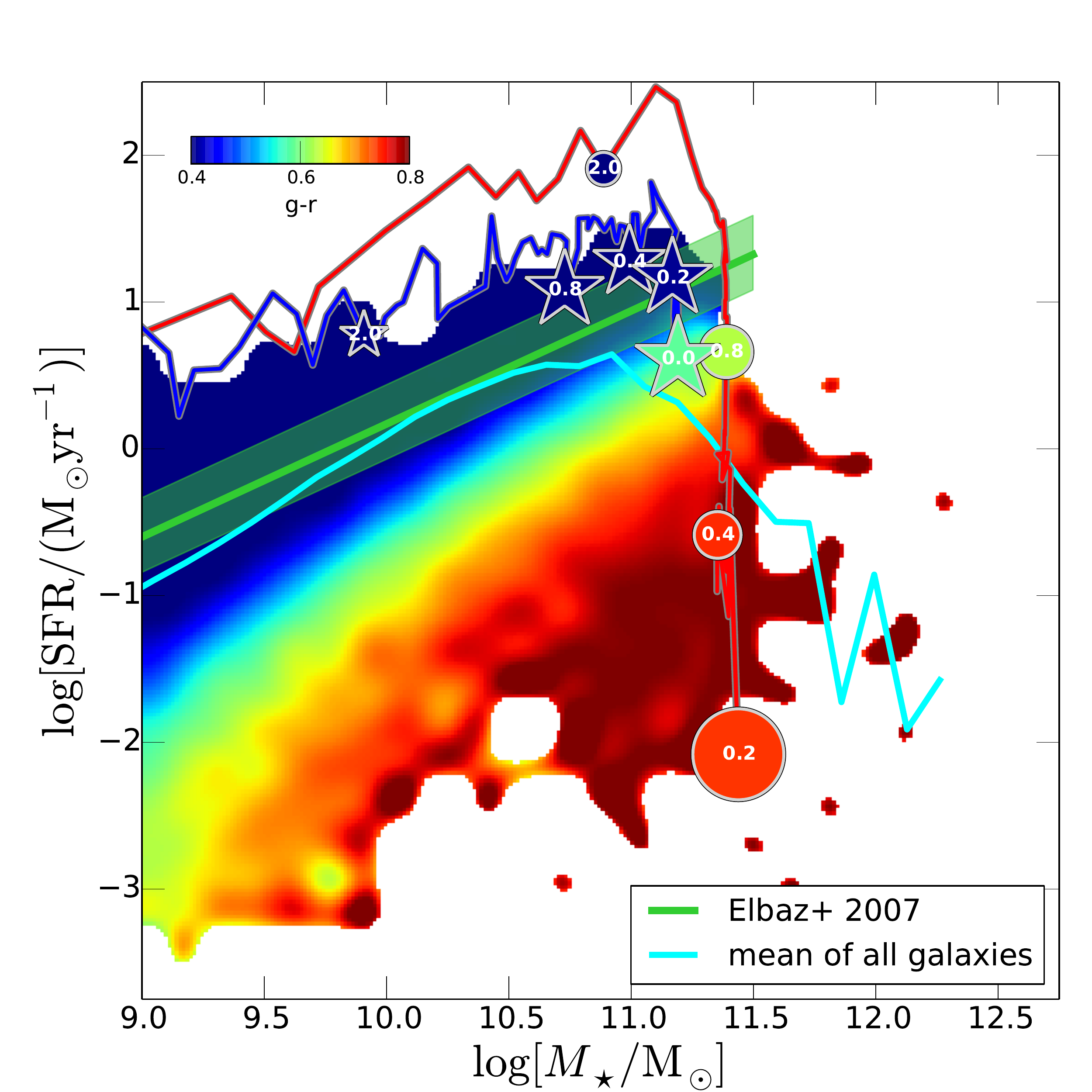}
\includegraphics[width=0.49\textwidth]{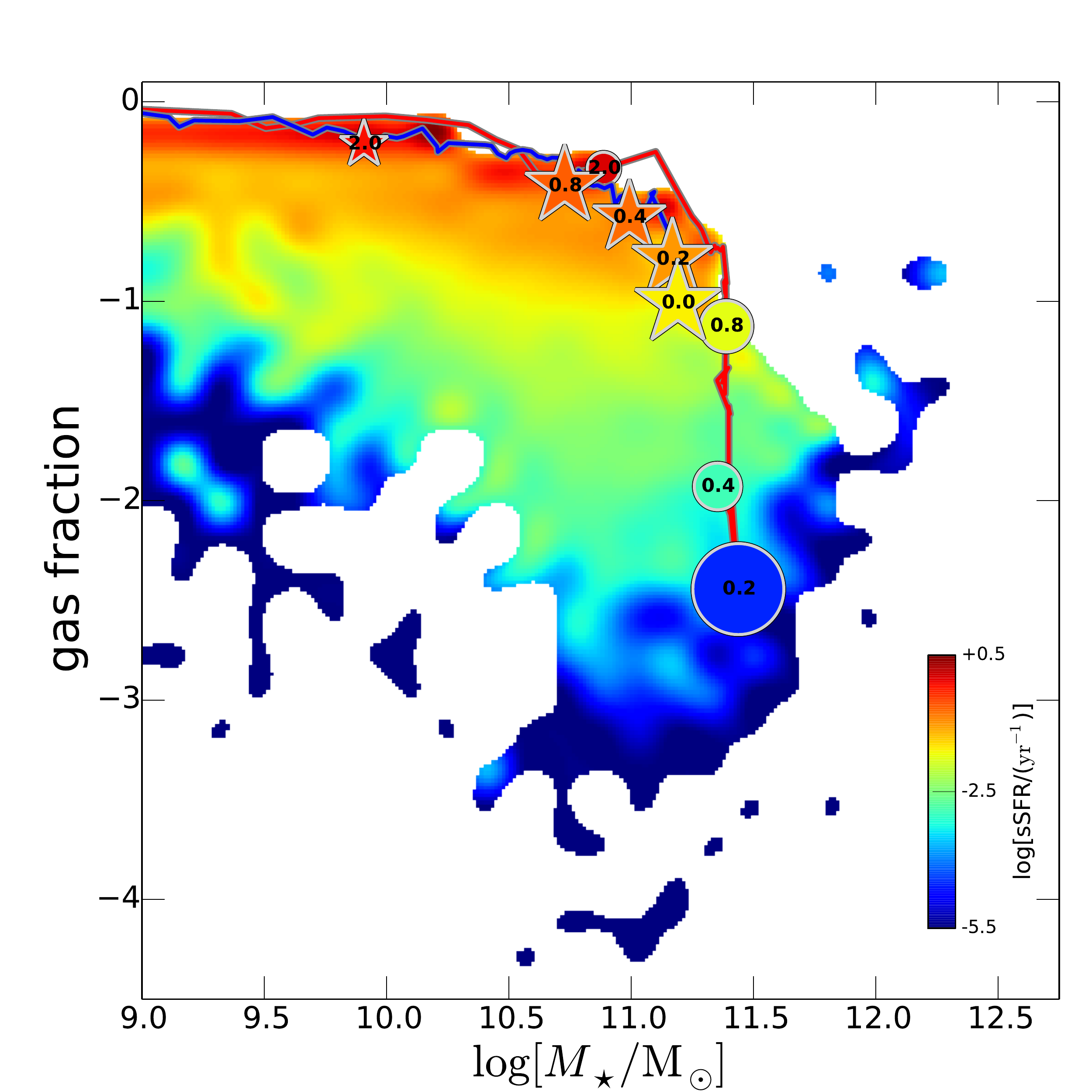}\\
\caption{Galaxy evolution across cosmic time. Left panel: Star formation rate
versus stellar mass at $z=0$ ({g-r} colour coded). The green band shows the
observed SF main sequence~\protect\citep[][]{Elbaz2007}.  The cyan line shows
the mean relation between stellar mass and SF predicted by the simulation. Symbols connected by
lines trace the evolution of the two sample galaxies shown in
Figure~\ref{fig:galevolve_image}: one evolving into an elliptical (circles /
red line) and the other one ending as a disk galaxy (stars / blue line).
Numbers in the symbols show the redshift and the symbol colour represent the
galaxy's {g-r} colour. Symbol sizes are proportional to the stellar half mass
radius normalised for each galaxy evolution track individually. Right panel:
Central gas fraction versus stellar mass at $z=0$ (specific SF rate colour
coded).  Symbol colours are chosen according to the specific SF rates. We do
not show the elliptical galaxy beyond $z=0.2$   since the SFR is essentially
zero after that time. Symbol sizes in the right panel are the same as in the
left panel.}
\label{fig:galevolve}
\end{figure*}

\section{Characteristics of individual galaxies}

Although our effective ISM model prevents us from making detailed predictions
about the internal gas structure of galaxies, we can still estimate coarse
grained quantities, and test those predictions against observations. We will
demonstrate this in what follows for a few observables, and we will also show
that our model produces a reasonable mix of early- and late-type galaxies.
Furthermore, the more massive galaxies in our simulation volume are resolved
well enough to study the detailed distribution of stars in them.

Before studying specific details of the galaxy population we present in
Figure~\ref{fig:galevolve_image} the redshift evolution of two galaxies based
on stellar images in rest frame g,r,i bands. One of the galaxies evolves into a
red elliptical galaxy (top), and the other one (bottom) into a blue
star-forming spiral galaxy. We trace these galaxies back in time for more than
$10\Gyr$ until $z=2$ as indicated in the figure. The images reveal that both
galaxies undergo some mergers. For example the red galaxy experienced a dry
merger around $z \sim 0.2$ resulting in shell like structures in the stellar
image at $z \sim 0.1$. The shells are not prominent anymore at $z=0$, where we
find a smooth and featureless light distribution. The blue late-type galaxy in
the bottom of the figure also undergoes a merger around $z \sim 0.3$ resulting
in the formation of a dominant bar, which is clearly visible around $z \sim
0.2-0.1$ and still noticeable at $z=0$. Some of the blue galaxy's images also
show star-forming regions as knots in the light distribution. By $z=0$ these
systems have evolved into very different galaxies, which differ strongly in
their morphology, colour, and specific star formation rates.

\begin{figure*}
\centering
\includegraphics[width=0.95\textwidth]{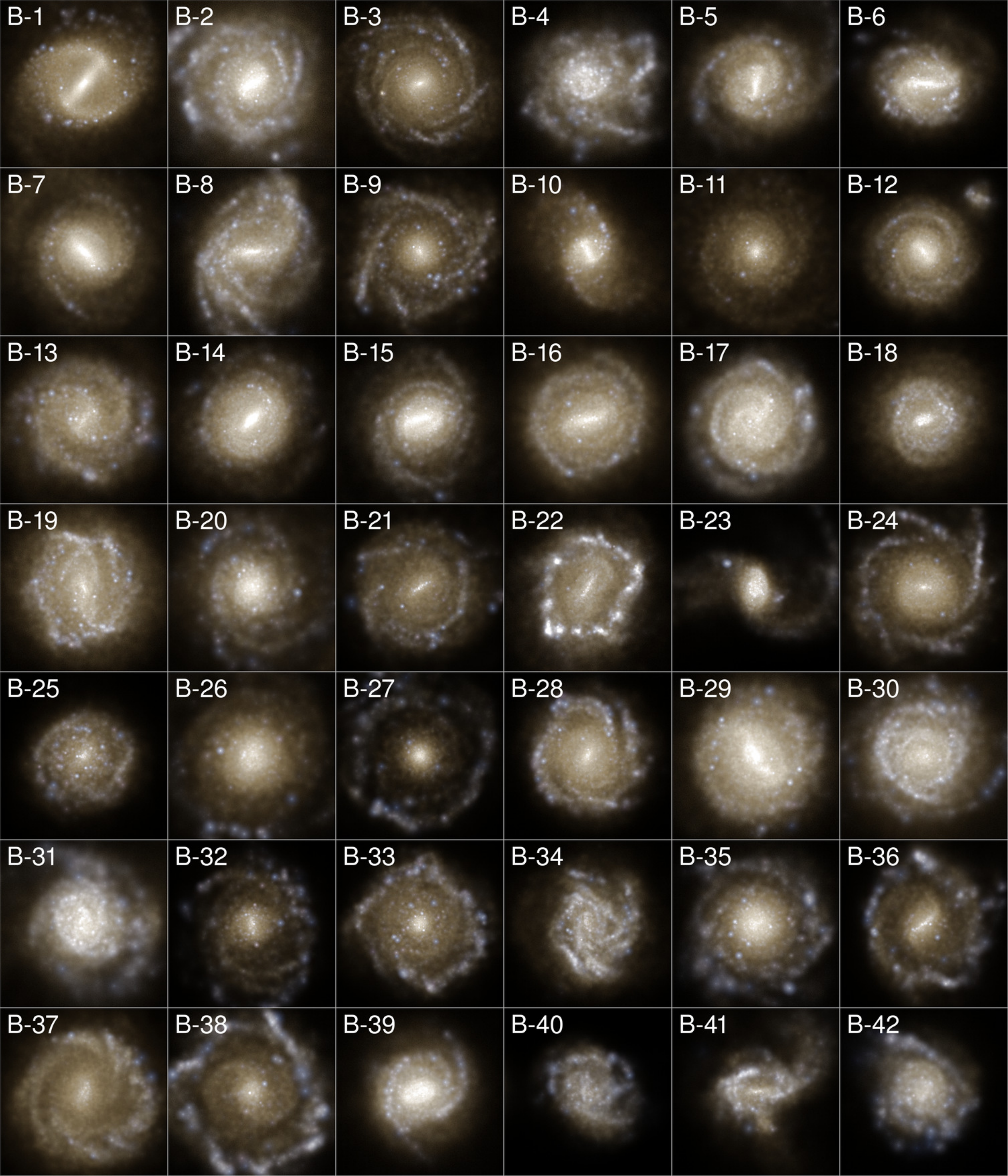}
\caption{Blue galaxies: Stellar light distribution (g,r,i SDSS band composites) of a
  few selected blue galaxies.
  These blue galaxies have typically high
  present-day SF rates and most of them are disk-like. Basic
  characteristics of these galaxies are listed in
  Table~\ref{table:galaxies}. The blue star-forming `rings' in some galaxies are star-forming regions, which are exaggerated in significance due to the employed colour scale.}
\label{fig:blue_galaxies}
\end{figure*}

\begin{figure*}
\centering
\includegraphics[width=0.95\textwidth]{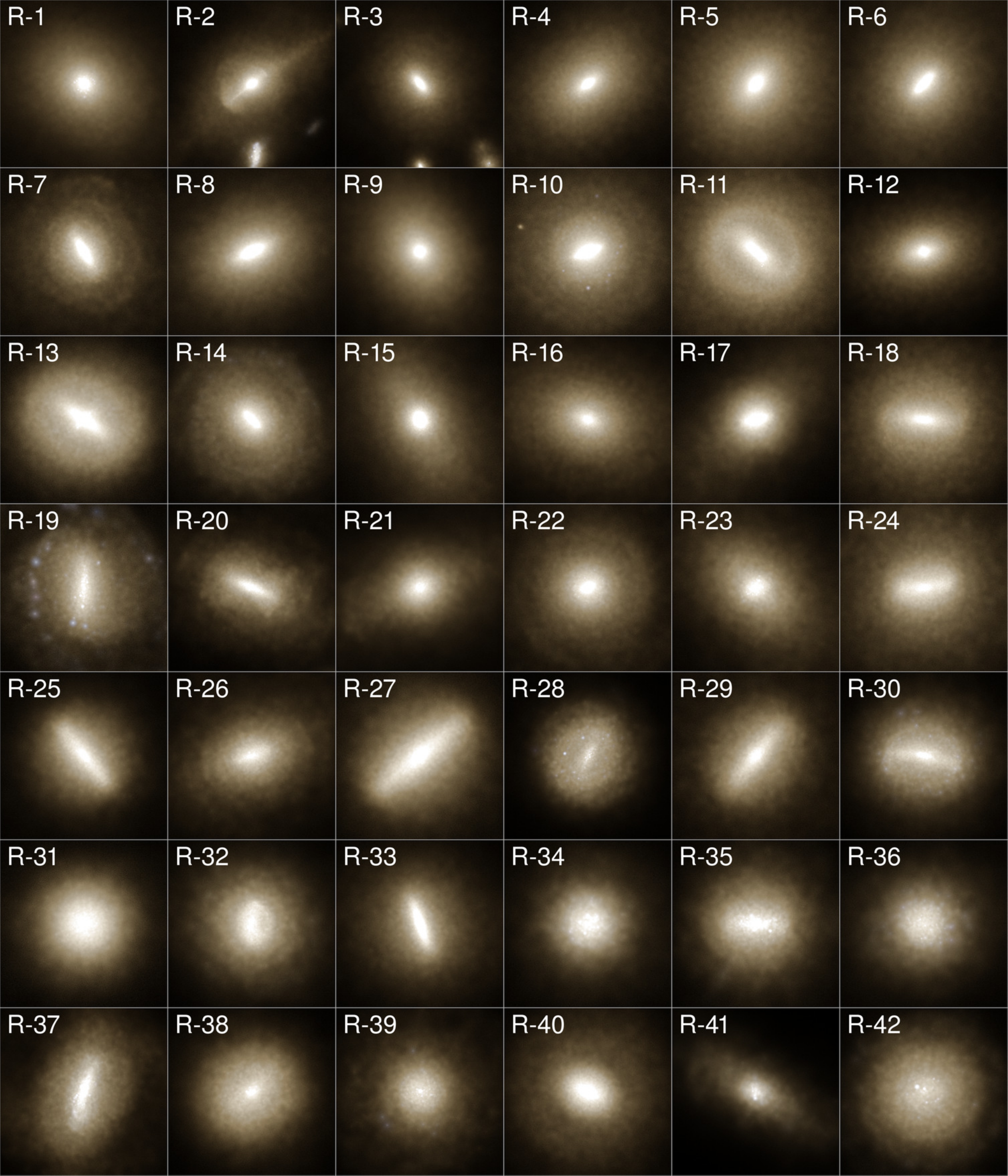}
\caption{Red galaxies: Stellar light distribution (g,r,i SDSS band composites) of a
  sample of red galaxies.
  Most of the red galaxies have little on-going SF and are
  ellipticals.  Basic
  characteristics of these galaxies are listed in
  Table~\ref{table:galaxies}. Some of the galaxies (e.g., R-2, R-7, and R-11) show shell-like structures in the stellar light distribution.}
\label{fig:red_galaxies}
\end{figure*}

We trace the history of these two galaxies more quantitatively in
Figure~\ref{fig:galevolve}, where we show their evolutionary paths together
with the distributions of all galaxies at $z=0$. In the left panel of
Figure~\ref{fig:galevolve} we present the distribution of all galaxies as a
function of stellar mass and SF rate at $z=0$.  The colour coding in the
histogram shows the average {(g-r)} colour of galaxies in each bin.  The green
band shows the observed SF main sequence~\citep[][]{Elbaz2007} which is in good
agreement with the predictions of our model.  We also include the evolutionary
paths of the two galaxies shown in Figure~\ref{fig:galevolve_image}.  The
early-type red galaxy (circle symbol) shows a rapid late-time drop in SF, which
is absent for the blue galaxy (star symbol).  This is caused by quenching
through radio-mode AGN feedback.  The star formation rate decreases by more
than one order of magnitude from $z=0.8$ to $z=0.4$ ($\Delta t \sim 2.6\Gyr$)
and subsequently by a similar factor from $z=0.4$ to $z=0.2$ ($\Delta t \sim
1.9\Gyr$) at which point SF of the central galaxy shuts off completely, which
is why we show no data point for the early-type galaxy at $z=0$.  The colour of
the symbols indicates the actual {g-r} colour of the galaxy at the particular
time. One can clearly see how the elliptical galaxy reddens while its SF is
reduced by feedback. On the other hand, the blue galaxy has ${\rm g-r} \sim
0.4$ until about $z\sim 0.2$. Then SF is slightly reduced and the galaxy turns
slightly redder, but is still significantly bluer than the red elliptical at
late times. The symbol sizes for the two galaxies are proportional to the
stellar half mass radius normalised for each galaxy evolution track
individually. The red galaxy seems to grow substantially from $z=0.4$ to
$z=0.2$. This is due to the dry merger happening around that time leading to a
more extended stellar distribution. The growth of the blue galaxy is less
dramatic, but it also grows continuously until $z=0$.

\begin{table*}
\centering
\begin{tabular}{lllllllll}
\hline
\noalign{\vskip 0.5mm}
galaxy                  & $M_{\rm 200,crit}$  & $M_\star$          & $M_\bullet$         & SFR                &$\log[{\rm sSFR}/\Gyr^{-1}]$         & g-r    & r       & $\log[1+\delta]$\\
\noalign{\vskip 0.5mm}
                        & [$10^{12}\msun$]    & [$10^{11}\msun$]   & [$10^{8}\msun$]     & [$\msun\yr^{-1}$]  &                                     & [mag]  & [mag]   &              \\
\noalign{\vskip 0.5mm}
\hline
\hline
\noalign{\vskip 0.5mm}
B/R-1   	&	11.07 / 59.21   &	2.92 / 6.19   	&	10.53 / 25.16   &	2.3718 / 1.0746   	&	-1.09 / -1.76   &	0.72 / 0.75   	&	-23.38 / -24.30   	&	2.41 / 2.83   \\
B/R-2   	&	10.51 / 41.60  	&	1.97 / 4.60   	&	2.68 / 17.42   	&	10.8637 / 0 	   	&	-0.26 / $--$   	&	0.53 / 0.76   	&	-23.58 / -24.00   	&	3.65 / 3.47   \\
B/R-3   	&	7.69 / 36.50   	&	2.69 / 3.80   	&	2.20 / 32.63   	&	4.5586 / 0	   	&	-0.77 / $--$   	&	0.65 / 0.75   	&	-23.56 / -23.74   	&	1.95 / 3.13   \\
B/R-4   	&	5.46 / 26.69   	&	0.73 / 4.02   	&	0.15 / 41.81   	&	11.9022 / 0	   	&	0.21 /  $--$   	&	0.35 / 0.77   	&	-22.82 / -23.67   	&	0.63 / 2.25   \\
B/R-5   	&	5.13 / 24.87   	&	1.17 / 5.41   	&	1.48 / 36.30   	&	2.1489 / 0.0235   	&	-0.74 / -3.36   &	0.60 / 0.79   	&	-22.87 / -23.95   	&	0.89 / 2.46   \\
B/R-6   	&	4.60 / 24.62   	&	1.57 / 3.97   	&	1.25 / 19.64   	&	7.7917 / 0	   	&	-0.30 /  $--$  	&	0.54 / 0.77   	&	-23.33 / -23.77   	&	1.22 / 3.21   \\
B/R-7   	&	4.57 / 19.79   	&	1.87 / 2.23   	&	3.03 / 18.86   	&	1.8975 / 0.0393   	&	-0.99 / -2.75   &	0.68 / 0.76   	&	-23.35 / -23.13   	&	0.51 / 2.24   \\
B/R-8   	&	4.45 / 16.50   	&	1.40 / 3.13   	&	0.66 / 16.82   	&	9.9469 / 0	   	&	-0.15 /  $--$  	&	0.50 / 0.77   	&	-23.26 / -23.51   	&	3.00 / 2.58   \\
B/R-9   	&	4.24 / 14.19   	&	1.09 / 3.40   	&	1.69 / 22.86   	&	4.0928 / 0.0162   	&	-0.43 / -3.32   &	0.56 / 0.79   	&	-22.51 / -23.45   	&	0.70 / 2.20   \\
B/R-10   	&	4.13 / 14.03   	&	0.93 / 3.85   	&	2.07 / 23.22   	&	1.2217 / 0.2478   	&	-0.88 / -2.19   &	0.69 / 0.78   	&	-22.28 / -23.78   	&	2.58 / 1.80   \\
B/R-11   	&	3.81 / 12.40   	&	1.10 / 4.92   	&	2.50 / 11.94   	&	0.6194 / 0.1505   	&	-1.25 / -2.51   &	0.67 / 0.79   	&	-22.44 / -23.82   	&	1.50 / 1.89   \\
B/R-12   	&	3.71 / 12.23   	&	1.43 / 3.30   	&	1.15 / 11.37   	&	2.3570 / 0.0015   	&	-0.78 / -4.36   &	0.67 / 0.78   	&	-22.92 / -23.51   	&	1.23 / 2.24   \\
B/R-13   	&	3.65 / 11.72   	&	0.95 / 3.02   	&	1.02 / 11.06   	&	3.5973 / 0.5288   	&	-0.42 / -1.76   &	0.58 / 0.77   	&	-22.42 / -23.32   	&	0.90 / 2.55   \\
B/R-14   	&	3.24 / 10.66   	&	1.94 / 2.56   	&	1.27 / 11.99   	&	3.7626 / 0.2000   	&	-0.71 / -2.11   &	0.63 / 0.76   	&	-23.32 / -23.25   	&	0.91 / 2.09   \\
B/R-15   	&	3.19 / 9.80   	&	1.55 / 2.53   	&	1.49 / 21.56   	&	4.0788 / 0	   	&	-0.58 /  $--$  	&	0.58 / 0.77   	&	-23.25 / -23.34   	&	1.80 / 1.49   \\
B/R-16   	&	3.01 / 9.08   	&	1.08 / 2.65   	&	1.51 / 8.02   	&	1.5833 / 0.1344   	&	-0.83 / -2.29   &	0.64 / 0.78   	&	-22.53 / -23.17   	&	1.46 / 1.36   \\
B/R-17   	&	2.99 / 7.45   	&	1.30 / 1.14   	&	1.11 / 12.61   	&	3.3859 / 0	  	&	-0.58 /  $--$  	&	0.56 / 0.77   	&	-22.94 / -22.35   	&	0.85 / 0.90   \\
B/R-18   	&	2.99 / 5.66   	&	1.05 / 2.17   	&	1.09 / 6.61   	&	5.8264 / 0	  	&	-0.26 /  $--$  	&	0.53 / 0.80   	&	-22.83 / -22.85   	&	0.66 / 1.08   \\
B/R-19   	&	2.94 / 5.56   	&	1.81 / 1.54   	&	0.75 / 2.05   	&	7.9209 / 1.1771   	&	-0.36 / -1.12   &	0.56 / 0.70   	&	-23.18 / -22.90   	&	0.45 / 1.67   \\
B/R-20   	&	2.73 / 5.53   	&	0.78 / 1.84   	&	1.16 / 3.22   	&	2.9106 / 0.2241   	&	-0.43 / -1.91   &	0.58 / 0.74   	&	-22.41 / -23.12   	&	0.04 / 2.87   \\
B/R-21   	&	2.64 / 5.52   	&	0.97 / 1.05   	&	0.82 / 4.36   	&	4.4156 / 0.2280   	&	-0.34 / -1.66   &	0.55 / 0.74   	&	-22.62 / -22.39   	&	0.70 / 0.67   \\
B/R-22   	&	2.59 / 5.36   	&	1.73 / 2.55   	&	0.51 / 5.78   	&	9.4646 / 0.4769   	&	-0.26 / -1.73   &	0.51 / 0.76   	&	-23.33 / -23.38   	&	2.55 / 1.84   \\
B/R-23   	&	2.53 / 5.22   	&	0.66 / 1.09   	&	0.67 / 4.91   	&	3.7511 / 0.2894   	&	-0.25 / -1.57   &	0.50 / 0.75   	&	-22.58 / -22.28   	&	1.22 / 0.15   \\
B/R-24   	&	2.51 / 4.71   	&	1.26 / 1.85   	&	0.70 / 3.41   	&	2.8558 / 0.0054   	&	-0.65 / -3.53   &	0.59 / 0.77   	&	-22.80 / -23.03   	&	1.00 / 1.16   \\
B/R-25   	&	2.46 / 4.13   	&	0.74 / 0.93   	&	0.49 / 5.69   	&	4.2743 / 0	   	&	-0.24 /  $--$  	&	0.50 / 0.78   	&	-22.25 / -21.93   	&	2.12 / 0.76   \\
B/R-26   	&	2.41 / 4.09   	&	0.65 / 1.28   	&	0.66 / 2.70   	&	1.0882 / 0.0813   	&	-0.77 / -2.20   &	0.60 / 0.77   	&	-22.02 / -22.55   	&	0.86 / 1.47   \\
B/R-27   	&	2.40 / 3.08   	&	0.42 / 1.22   	&	1.50 / 4.63   	&	0.5720 / 0	   	&	-0.86 /  $--$  	&	0.47 / 0.80   	&	-21.88 / -22.20   	&	0.07 / 0      \\
B/R-28   	&	2.38 / 2.89   	&	1.43 / 1.18   	&	0.67 / 1.18   	&	4.9461 / 0.7463   	&	-0.46 / -1.20   &	0.55 / 0.73   	&	-22.82 / -22.47   	&	0.78 / 1.39   \\
B/R-29   	&	2.28 / 2.65   	&	1.45 / 1.10   	&	0.84 / 2.38   	&	2.9813 / 0.0131   	&	-0.69 / -2.92   &	0.63 / 0.78   	&	-22.91 / -22.32   	&	0.86 / 0.61   \\
B/R-30   	&	2.12 / 2.65   	&	1.13 / 1.34   	&	0.72 / 1.89   	&	7.5327 / 0.7895   	&	-0.18 / -1.23   &	0.48 / 0.74   	&	-22.98 / -22.61   	&	0.80 / 1.36   \\
B/R-31   	&	2.04 / 2.29   	&	0.65 / 0.67   	&	0.49 / 3.55   	&	8.5263 / 0	  	&	0.12 /  $--$   	&	0.42 / 0.79   	&	-22.74 / -21.55   	&	0.33 / 0.19   \\
B/R-32   	&	2.03 / 2.23   	&	0.59 / 0.73   	&	0.88 / 1.81   	&	3.0271 / 0.0984   	&	-0.29 / -1.87   &	0.49 / 0.77   	&	-22.15 / -21.81   	&	1.54 / 1.35   \\
B/R-33   	&	1.92 / 2.12   	&	0.97 / 1.41   	&	1.08 / 3.51   	&	5.5084 / 0	   	&	-0.25 /  $--$  	&	0.48 / 0.79   	&	-22.52 / -22.61   	&	1.01 / 0.72   \\
B/R-34   	&	1.89 / 2.09   	&	0.88 / 0.40   	&	0.58 / 2.10   	&	11.4449 / 0.3732   	&	0.12 / -1.03   	&	0.39 / 0.69   	&	-22.86 / -21.34   	&	1.84 / 1.64   \\
B/R-35   	&	1.74 / 1.97   	&	0.80 / 0.84   	&	0.76 / 0.99   	&	1.7241 / 0.6094   	&	-0.67 / -1.14   &	0.51 / 0.70   	&	-22.38 / -22.10   	&	0.91 / 0.72   \\
B/R-36   	&	1.72 / 1.93   	&	0.66 / 0.52   	&	0.37 / 1.08   	&	4.7982 / 0.7674   	&	-0.14 / -0.83   &	0.44 / 0.71   	&	-22.32 / -21.67   	&	-0.54 / 0.06  \\
B/R-37   	&	1.68 / 1.87   	&	1.11 / 1.44   	&	0.50 / 1.15   	&	1.0564 / 1.3607   	&	-1.02 / -1.03   &	0.62 / 0.67   	&	-22.55 / -22.95   	&	1.03 / 1.32   \\
B/R-38   	&	1.65 / 1.84   	&	0.54 / 0.64   	&	0.40 / 0.94   	&	3.8845 / 0.1060   	&	-0.14 / -1.78   &	0.46 / 0.78   	&	-22.07 / -21.47   	&	0.25 / 1.05   \\
B/R-39   	&	1.55 / 1.83   	&	1.08 / 0.51   	&	1.19 / 2.43   	&	5.4434 / 0.1860   	&	-0.30 / -1.43   &	0.54 / 0.75   	&	-22.92 / -21.38   	&	1.85 / 0.59   \\
B/R-40   	&	1.50 / 1.80   	&	0.33 / 1.26   	&	0.09 / 1.81   	&	5.6719 / 0.2119   	&	0.23 / -1.77   	&	0.29 / 0.75   	&	-22.11 / -22.83   	&	0.91 / 1.60   \\
B/R-41   	&	1.48 / 1.68   	&	0.49 / 0.28   	&	0.27 / 0.80   	&	8.0608 / 0.3356   	&	0.22 / -0.93   	&	0.39 / 0.65   	&	-22.41 / -21.26   	&	0.22 / 0.43   \\
B/R-42   	&	0.84 / 1.65   	&	0.27 / 0.70   	&	0.13 / 1.20   	&	4.7164 / 0.1817   	&	0.24 / -1.59   	&	0.34 / 0.74   	&	-21.94 / -21.76   	&	0.76 / 0.25   \\

\noalign{\vskip 0.5mm}
\hline
\end{tabular}
\caption{Basic structural parameters of the galaxy sample presented in
Figure~\ref{fig:blue_galaxies} (blue galaxies, B) and
Figure~\ref{fig:red_galaxies} (red galaxies, R).  The different columns list:
virial mass ($200$ critical), stellar mass, SMBH mass, star formation rate,
specific star formation rate,  {(g-r)} color, r band magnitude, and local
galaxy overdensity. Colours do not include the effects of dust. The local galaxy overdensity
is the same density as used in Figure~\ref{fig:environment}. Some of the red galaxies are quenched
so strongly that they have no ongoing star formation, e.g., R-2 or R-3. The stellar component of the galaxies presented here is typically resolved with $\sim 10^5$ stellar particles, and the DM halo with about $\sim 10$ times more DM particles. Galaxies are ranked according to their virial mass.}
\label{table:galaxies}
\end{table*}

The right panel of Figure~\ref{fig:galevolve} shows the central gas fraction
(${M_{\rm gas}/(M_{\rm \star}+M_{\rm gas})}$) as a function of stellar mass.
The colour coding indicates the average specific SF rate in each bin. As
galaxies become more massive, their gas content is reduced by AGN feedback and,
as a result, star formation becomes less efficient. The left panel demonstrates
that this is accompanied by a reddening of the galaxy (increasing {(g-r)}
values).  This effect can also be seen visually in
Figure~\ref{fig:galevolve_image}: both galaxies are blue at earlier times owing
to their young stellar populations and ongoing star formation.  Feedback
quenches star formation in the more massive system and leads to reddening
towards $z=0$. Both the gas fraction and the central gas content drop quickly
from $z=0.8$ to $z=0.2$ along with a significant reduction of the specific star
formation rate. At early times both galaxies have an approximately constant gas
fraction (right panel) along with a slowly increasing star formation rate (left
panel).

Figure~\ref{fig:galevolve_image} and Figure~\ref{fig:galevolve} demonstrate two
different evolutionary tracks of galaxies, which lead to a very distinct 
galaxy type at $z=0$. We stress that these differences in the evolutionary
history are largely driven by the feedback processes included in our model.
Specifically, the late time evolution of the early-type galaxy is strongly
influenced by energetic radio-mode AGN feedback. Such processes are therefore
responsible for producing a diverse galaxy population at $z=0$ in our simulation.

\begin{figure*}
\centering
\includegraphics[width=0.245\textwidth]{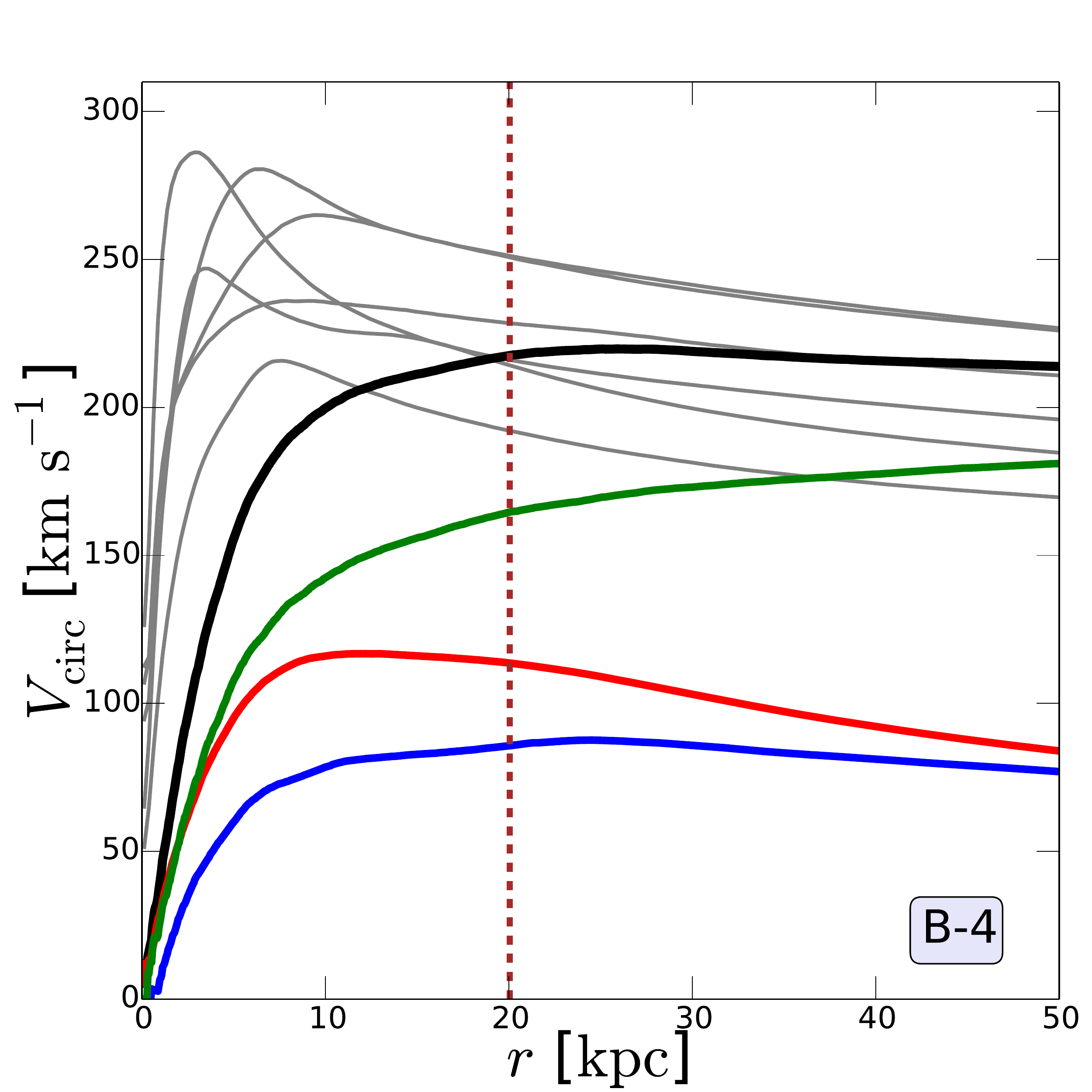}
\includegraphics[width=0.245\textwidth]{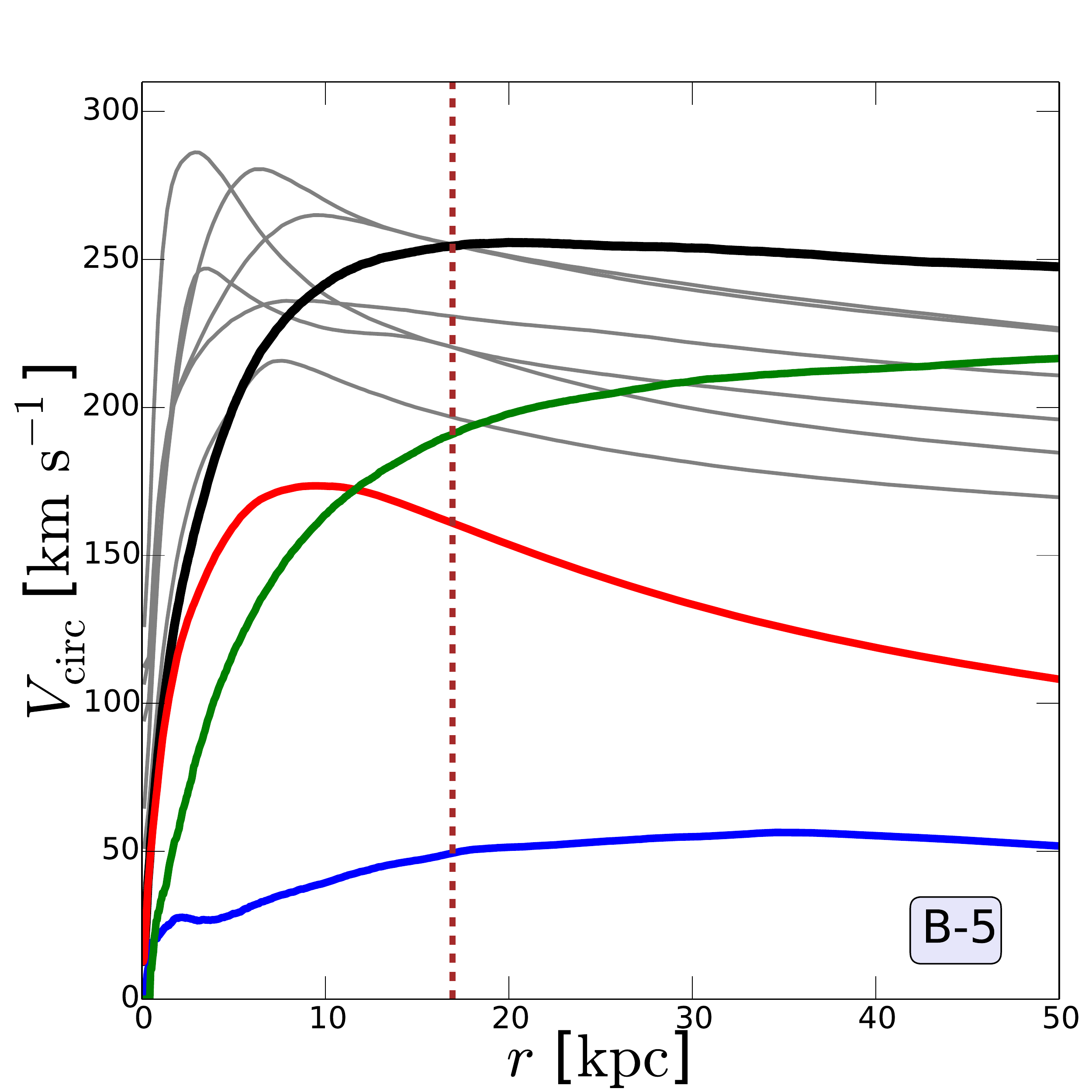}
\includegraphics[width=0.245\textwidth]{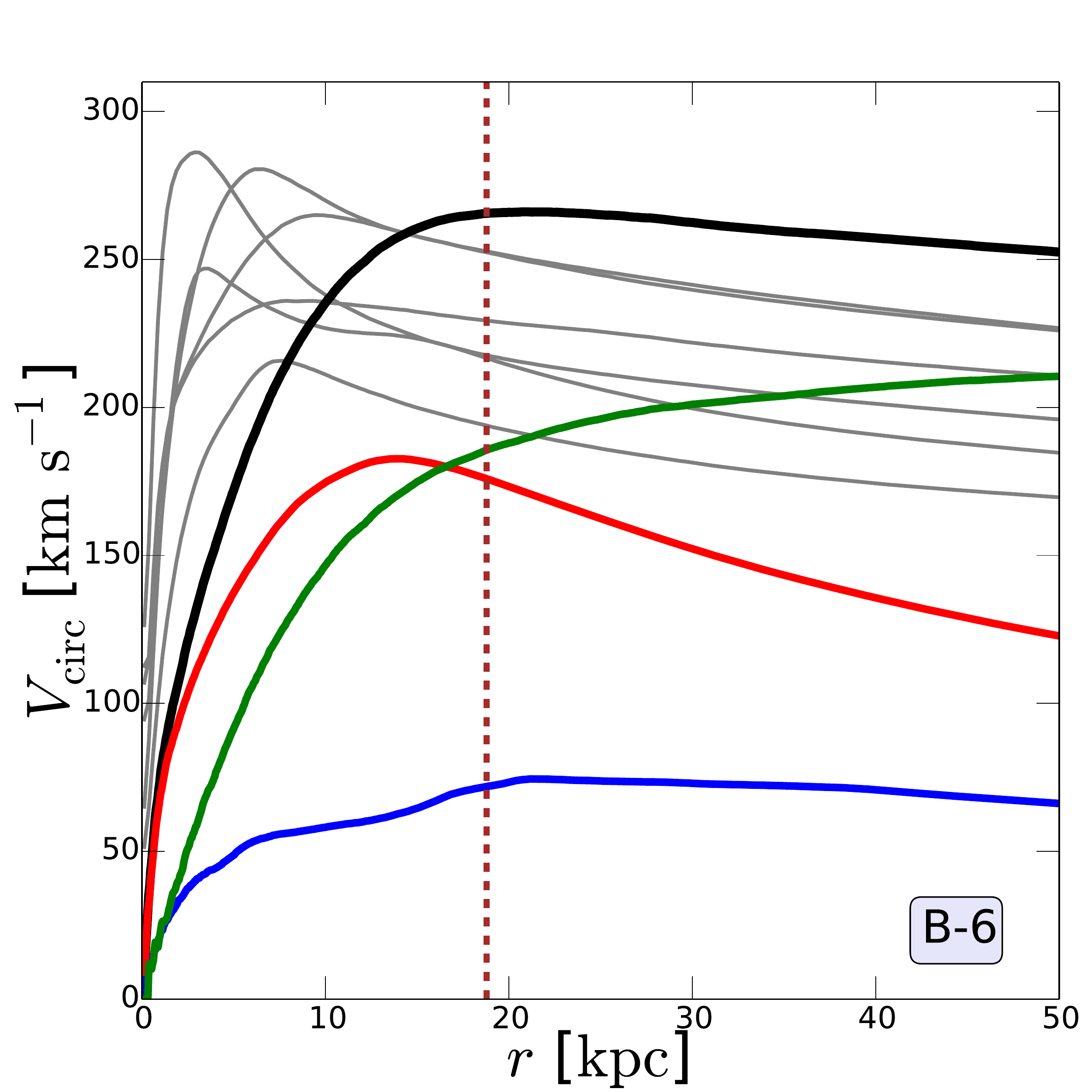}
\includegraphics[width=0.245\textwidth]{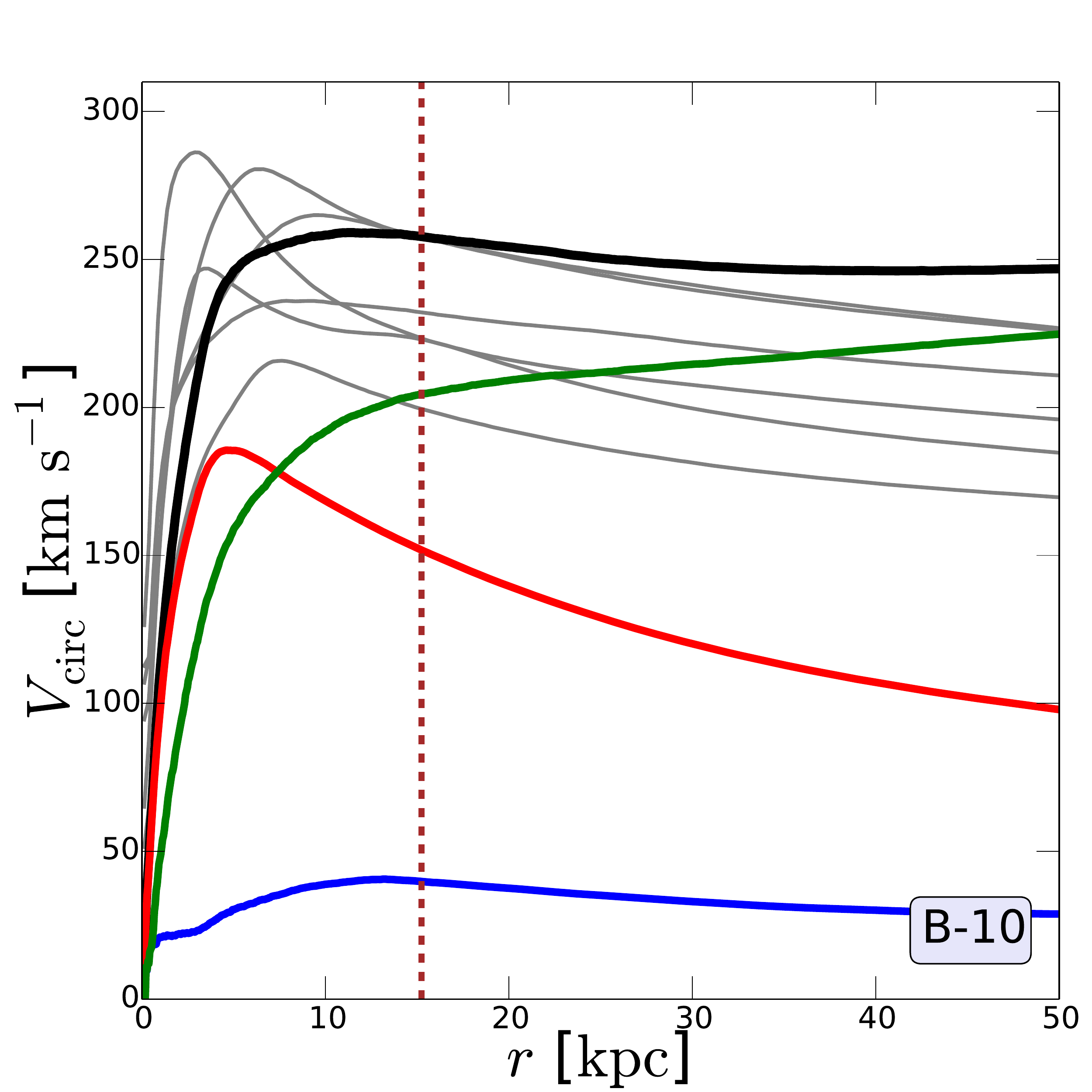}
\includegraphics[width=0.245\textwidth]{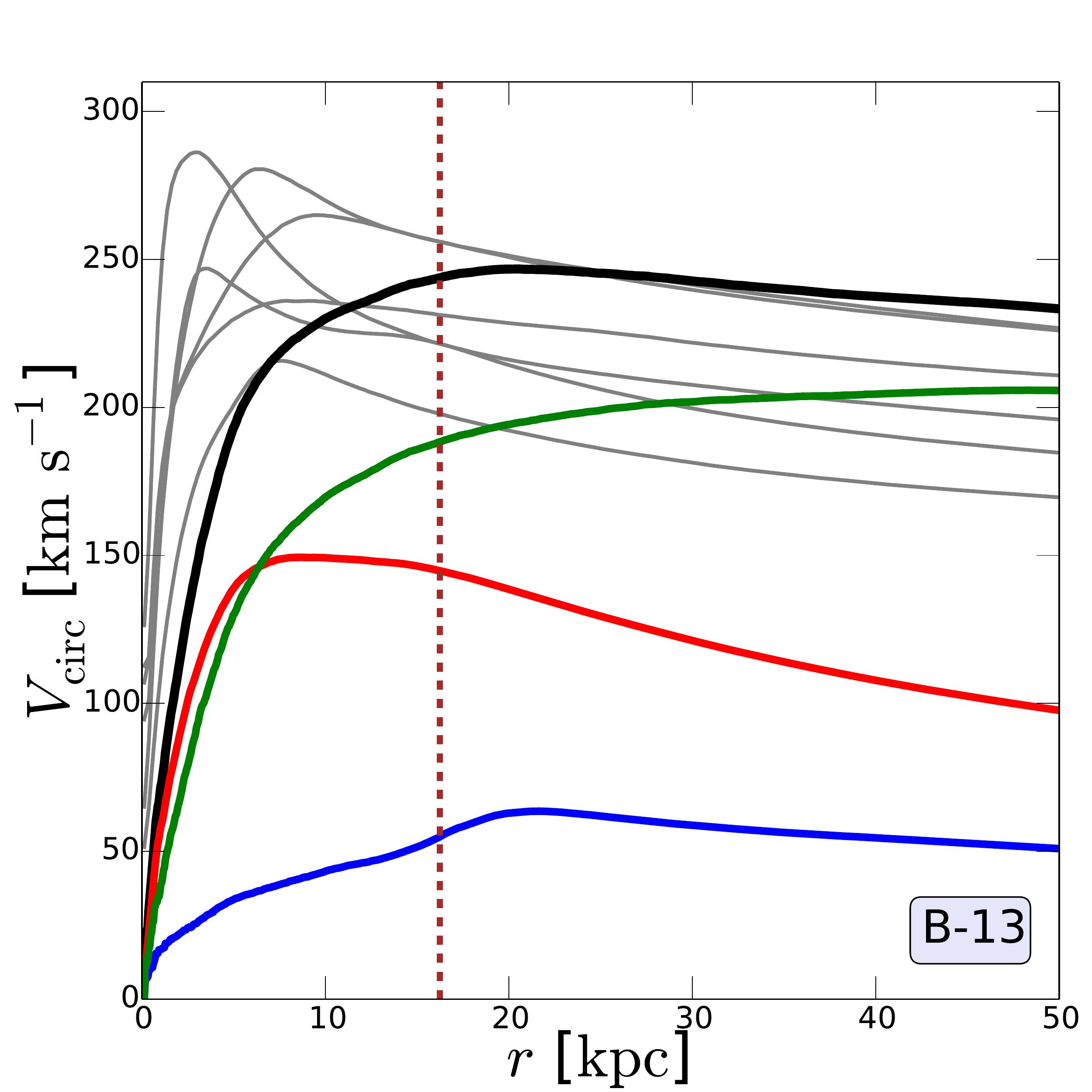}
\includegraphics[width=0.245\textwidth]{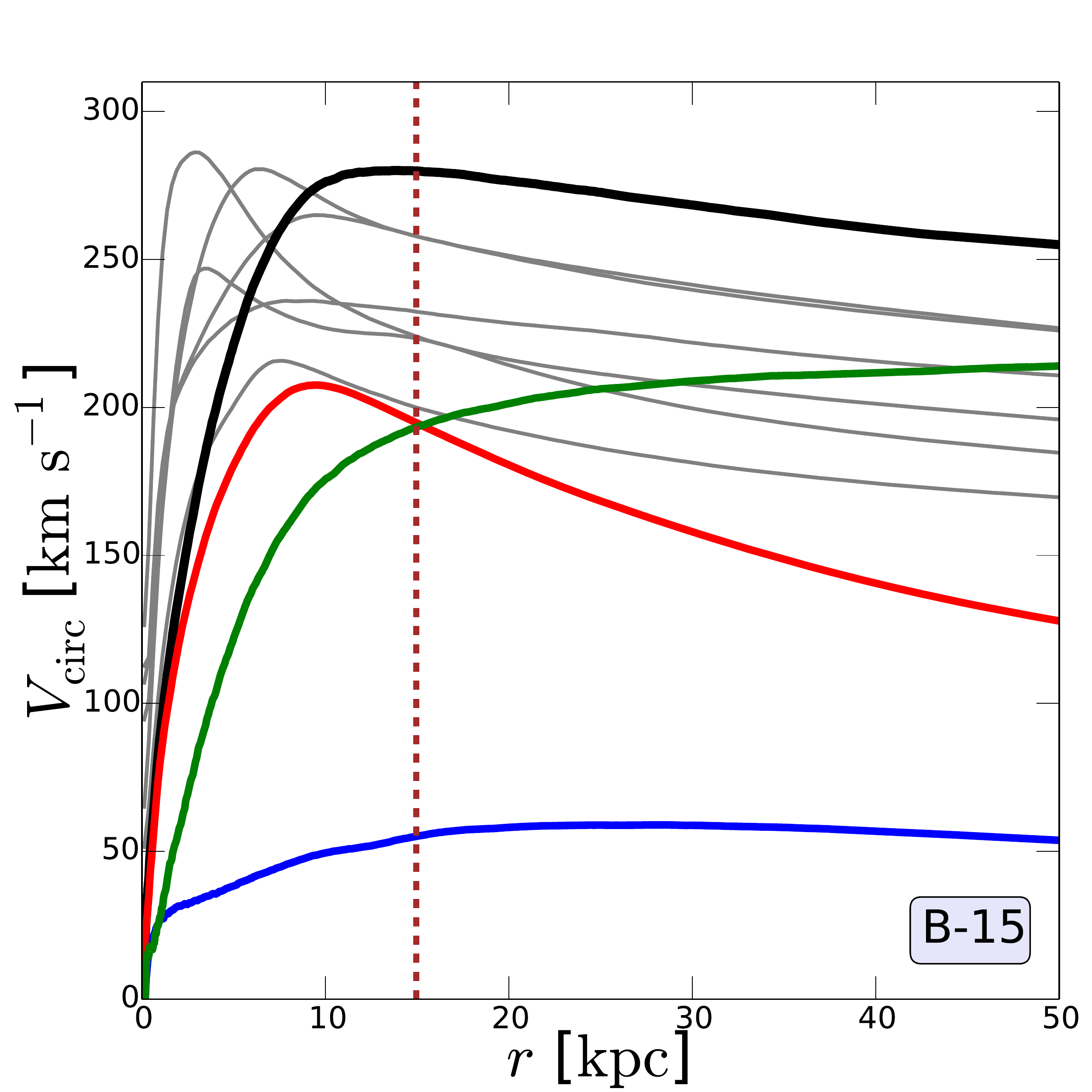}
\includegraphics[width=0.245\textwidth]{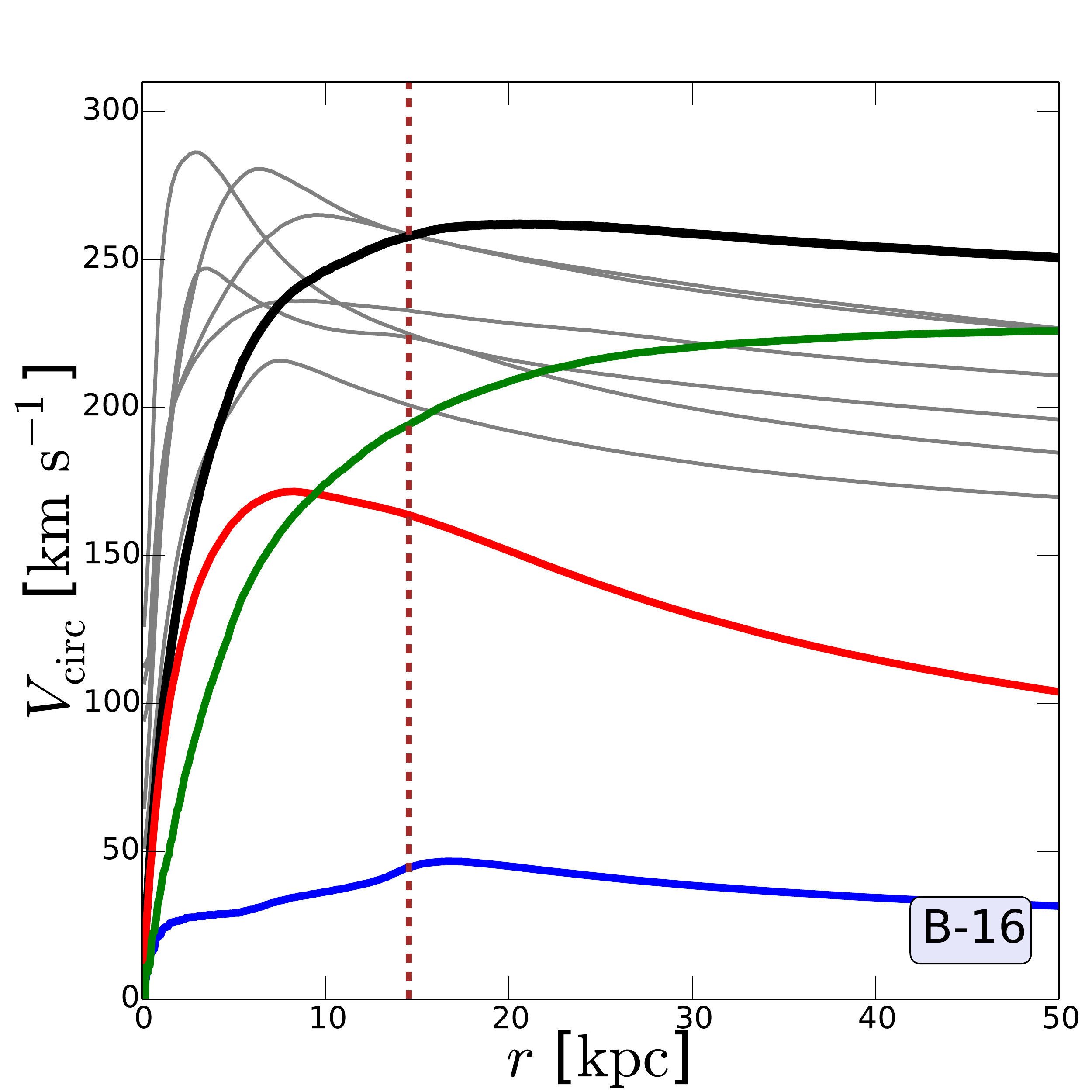}
\includegraphics[width=0.245\textwidth]{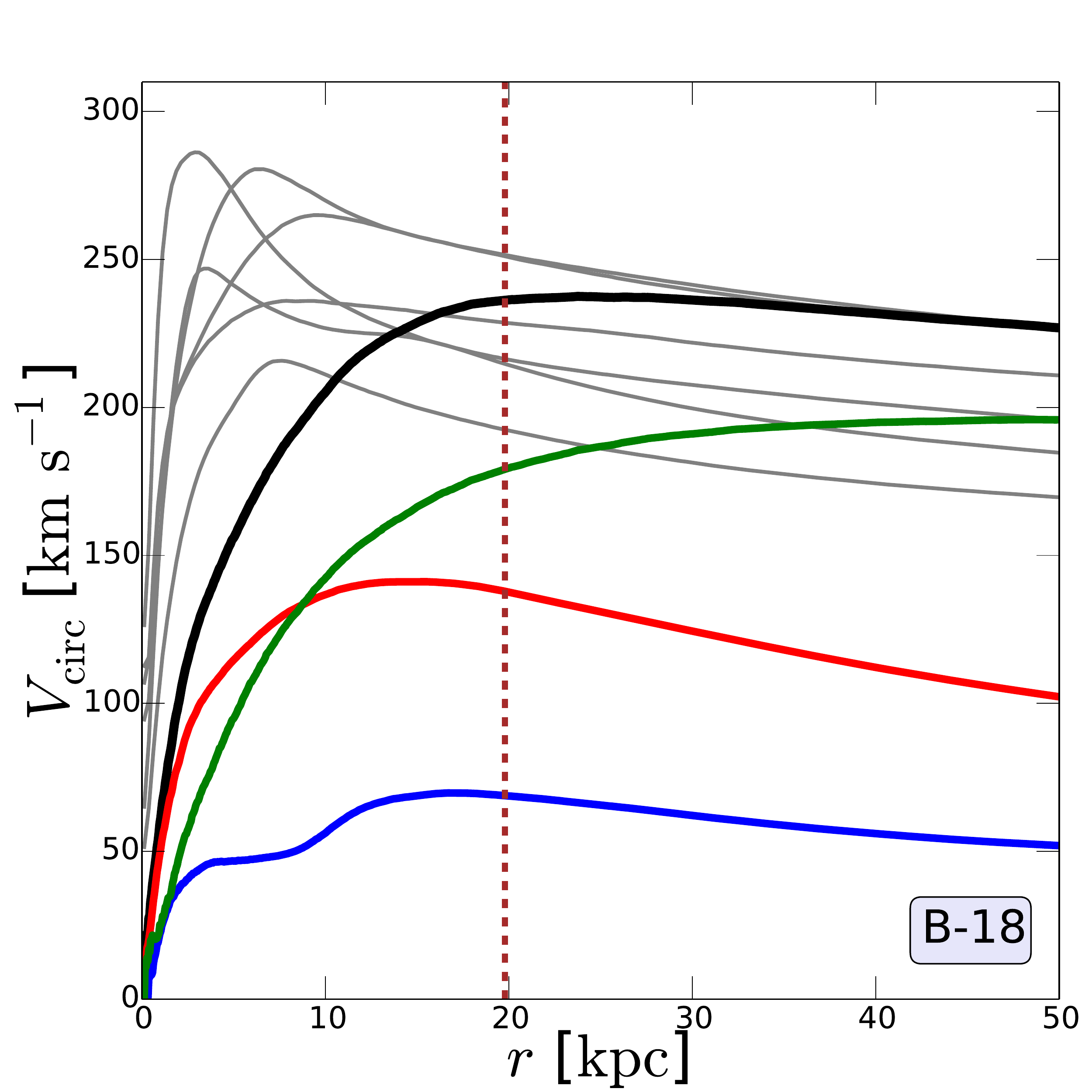}
\includegraphics[width=0.245\textwidth]{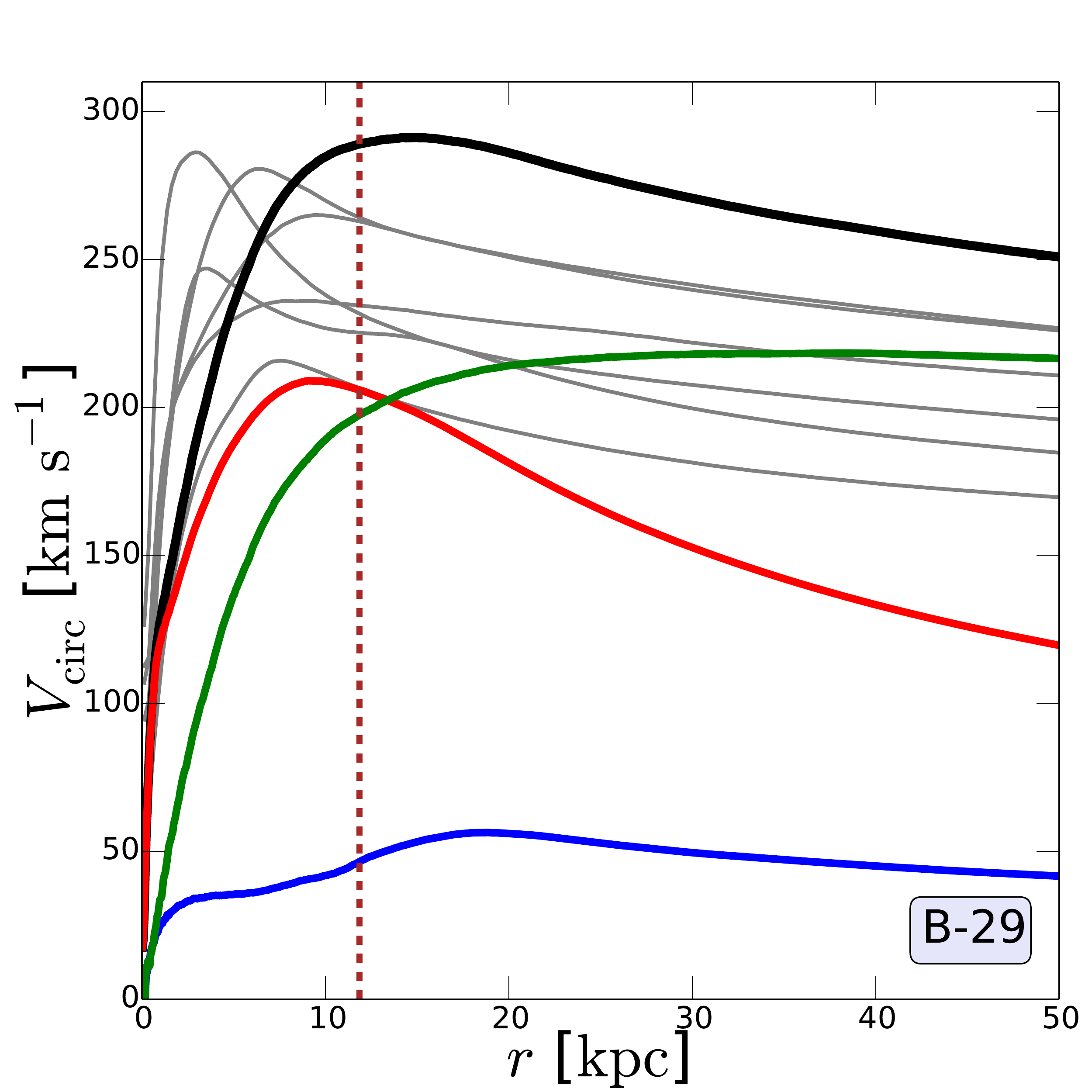}
\includegraphics[width=0.245\textwidth]{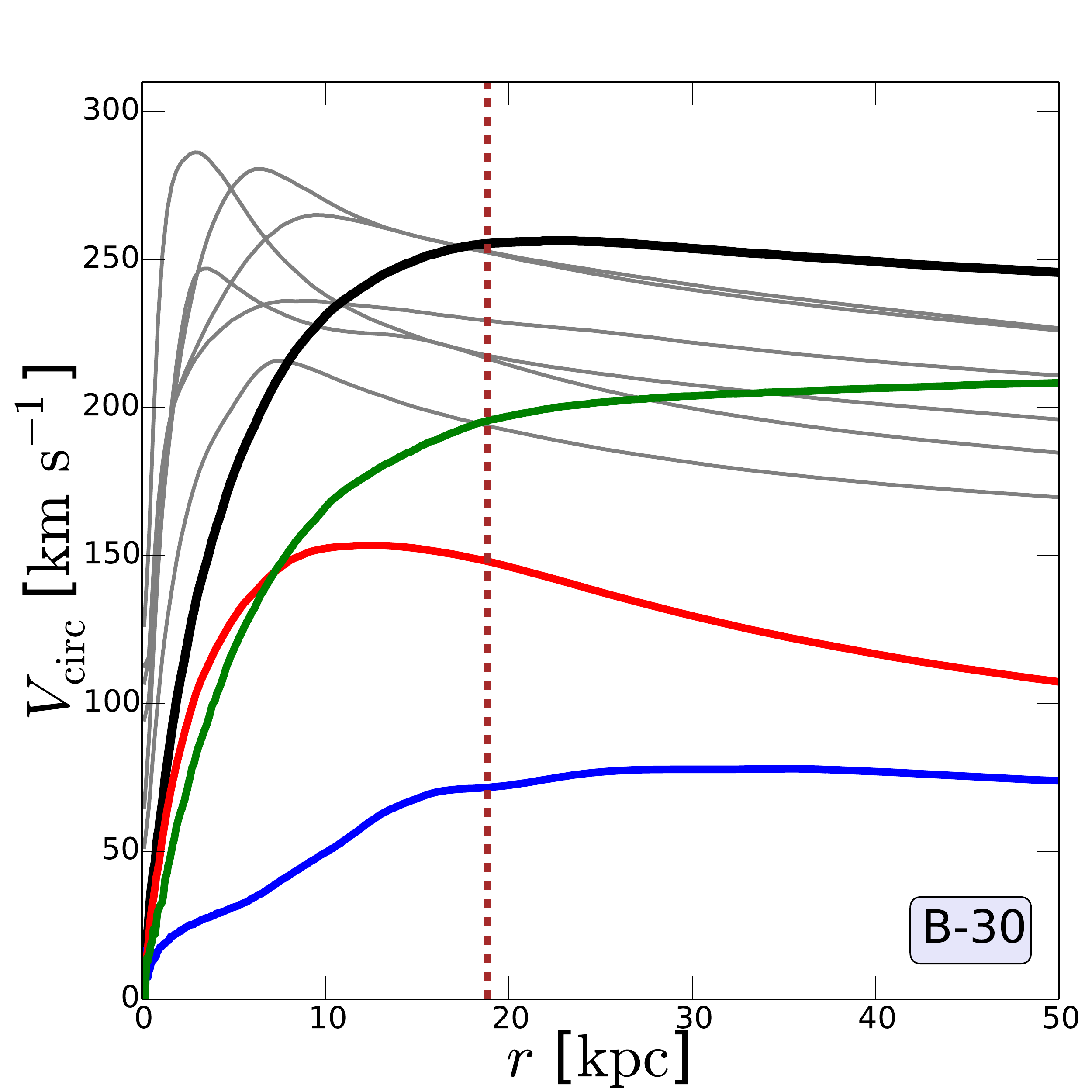}
\includegraphics[width=0.245\textwidth]{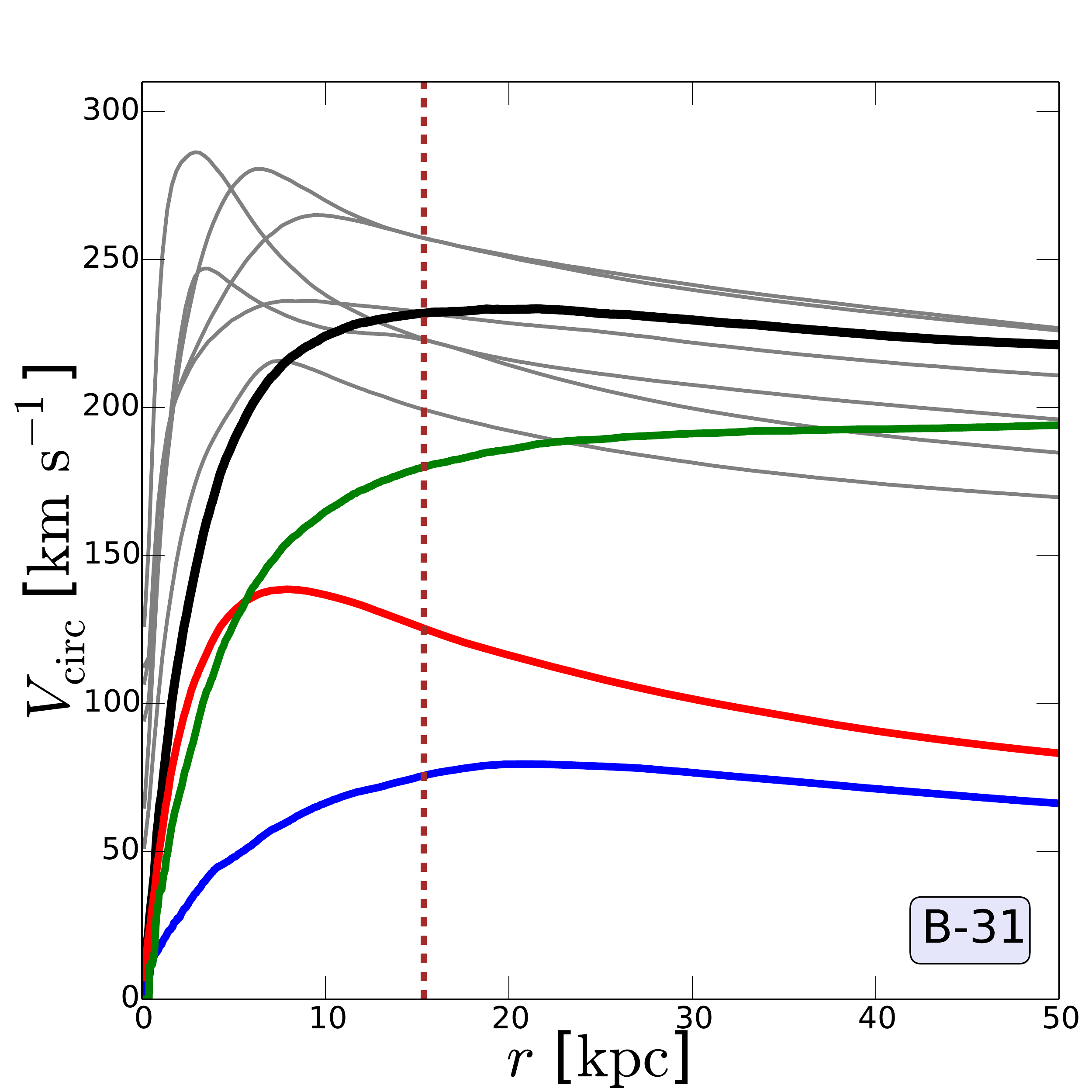}
\includegraphics[width=0.245\textwidth]{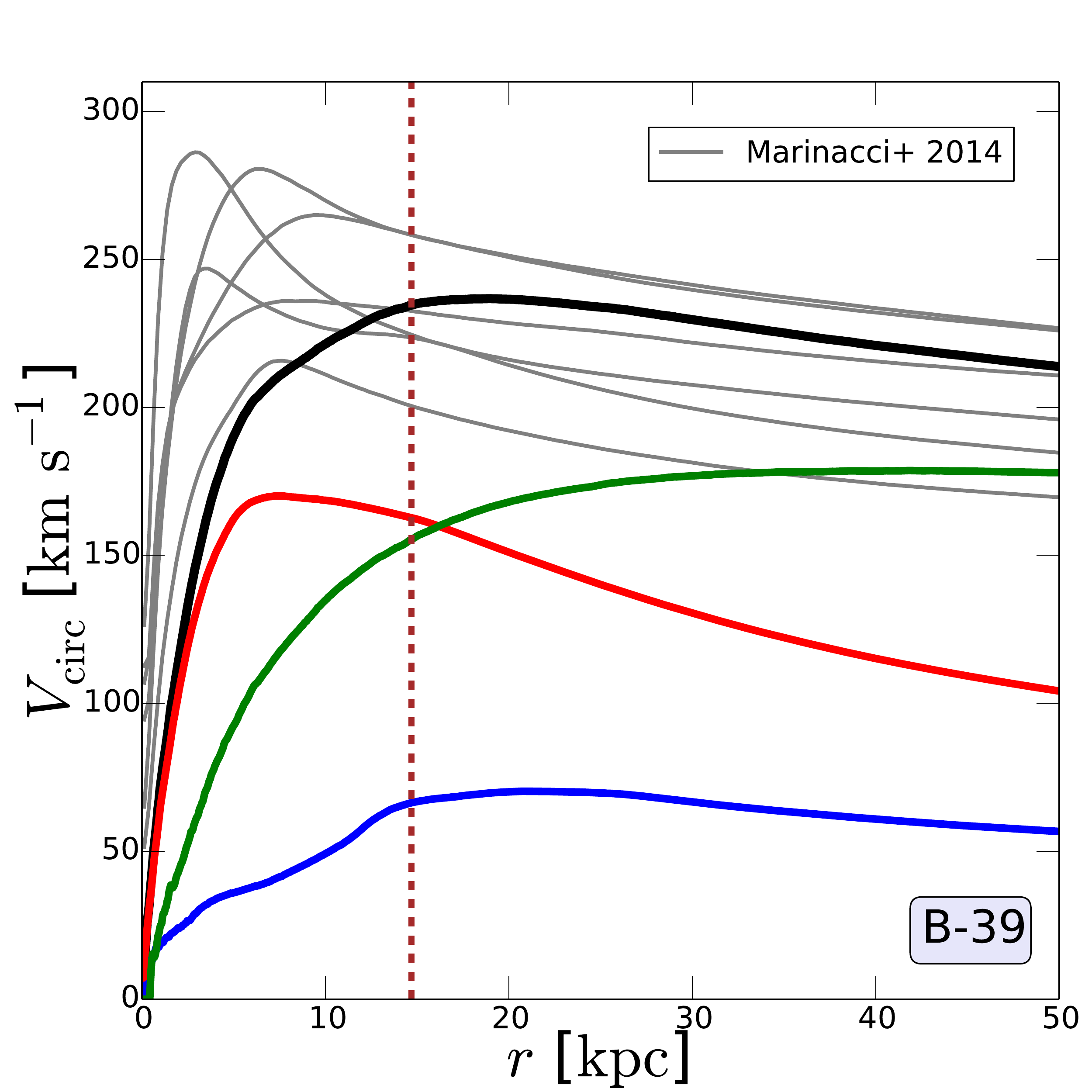}
\caption{Circular velocity curves of a few selected disk galaxies from the sample
presented in Figure~\ref{fig:blue_galaxies}. All of these galaxies show a
steeply rising and then nearly flat circular velocity curve (black line), which is
characteristic of late-type spiral galaxies. The different lines show the
contributions from different mass components: gas (blue), stars (red), and DM
(green). The gas contribution is typically small with maximum circular
velocities around $\sim70\kms$ for some cases, and for most systems around
$\sim50\kms$. The dashed brown vertical lines shows our fiducial galaxy radius
$r_\star$,~i.e. twice the stellar half-mass radius. This is the radius where we
measure the circular velocity for the construction of the baryonic Tully-Fisher
relation (see below).  We also show in each panel the eight circular velocity curves for
the Aquila haloes taken from \protect\cite{Marinacci2014a} (gray thin lines).
We note that those are based on the level-5 Aquarius haloes;~i.e. sampled at
higher spatial and mass resolution than the samples selected from the Illustris
volume.}
\label{fig:rotcurve}
\end{figure*}

At $z=0$ Illustris-1 contains $41,416$ galaxies which are resolved with more
than $500$ stellar particles. For many coarse-grained galaxy properties like
total metal content \cite[see][]{Nature2014} or total luminosity $\sim 1000$
particles are typically enough to get reasonable estimates. For example, we can
reliably probe the GLF down to this mass limit.  However, quantifying the
internal structure of the stellar and gas distribution requires significantly
more resolution elements to achieve accurate results. For example, probing the
internal velocity structure, morphology or detailed sizes of galaxies requires
substantially more resolution elements than a few thousand. The resolution of
Illustris-1 is therefore only sufficient to roughly describe the morphology and
internal kinematics of the more massive galaxies $M_\star \gtrsim
10^{10-11}\msun$ where the stellar population is sampled by about $\sim 10^4 -
10^5$ stellar particles. Although this resolution is still rather low compared
to state-of-the-art zoom-in simulations, we will demonstrate below that
internal characteristics of galaxies like the circular velocity curves are described
reasonably well with this resolution. Specifically, we will inspect in the
following the structure of a few well-resolved galaxies.  

\begin{figure*}
\centering
\includegraphics[width=0.49\textwidth]{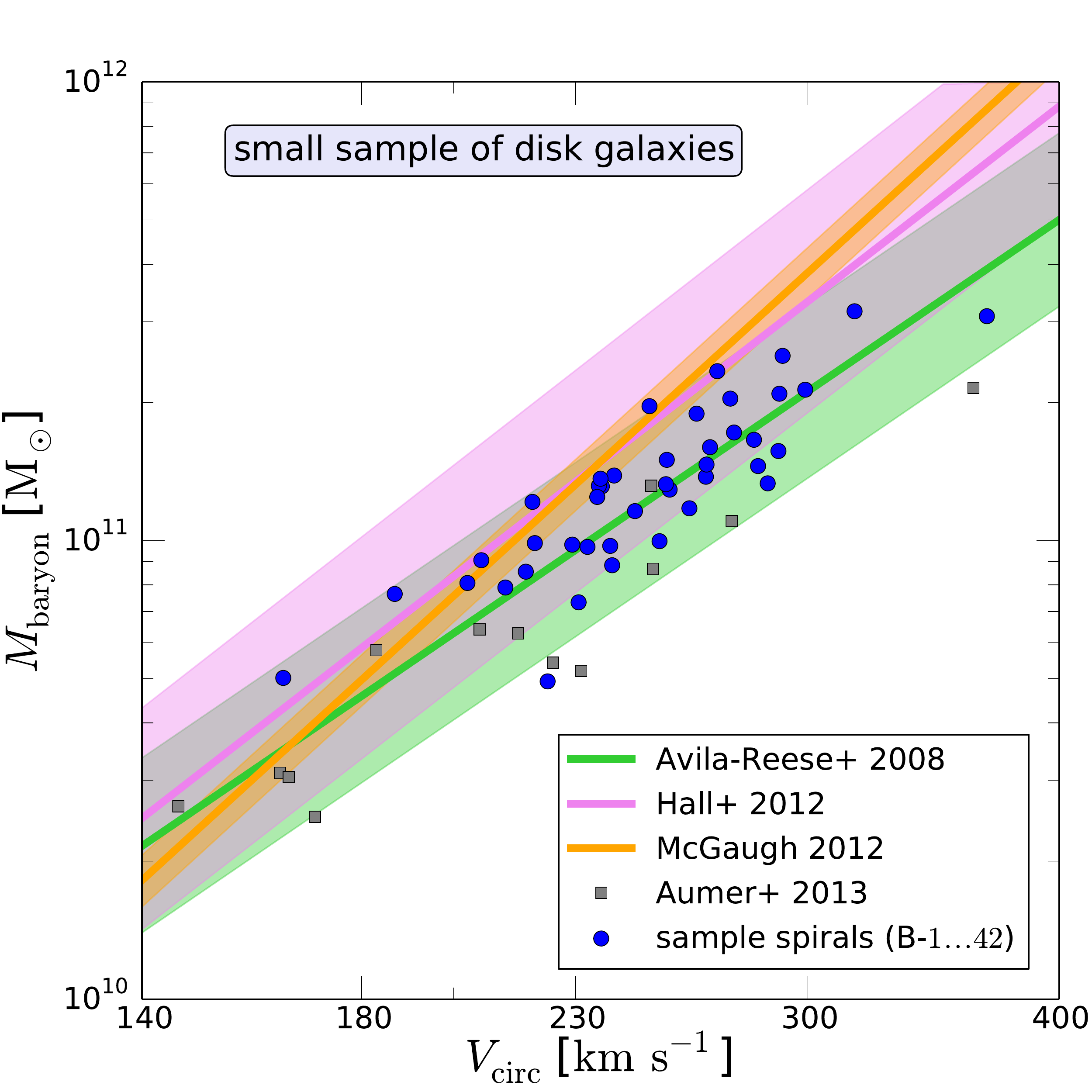}
\includegraphics[width=0.49\textwidth]{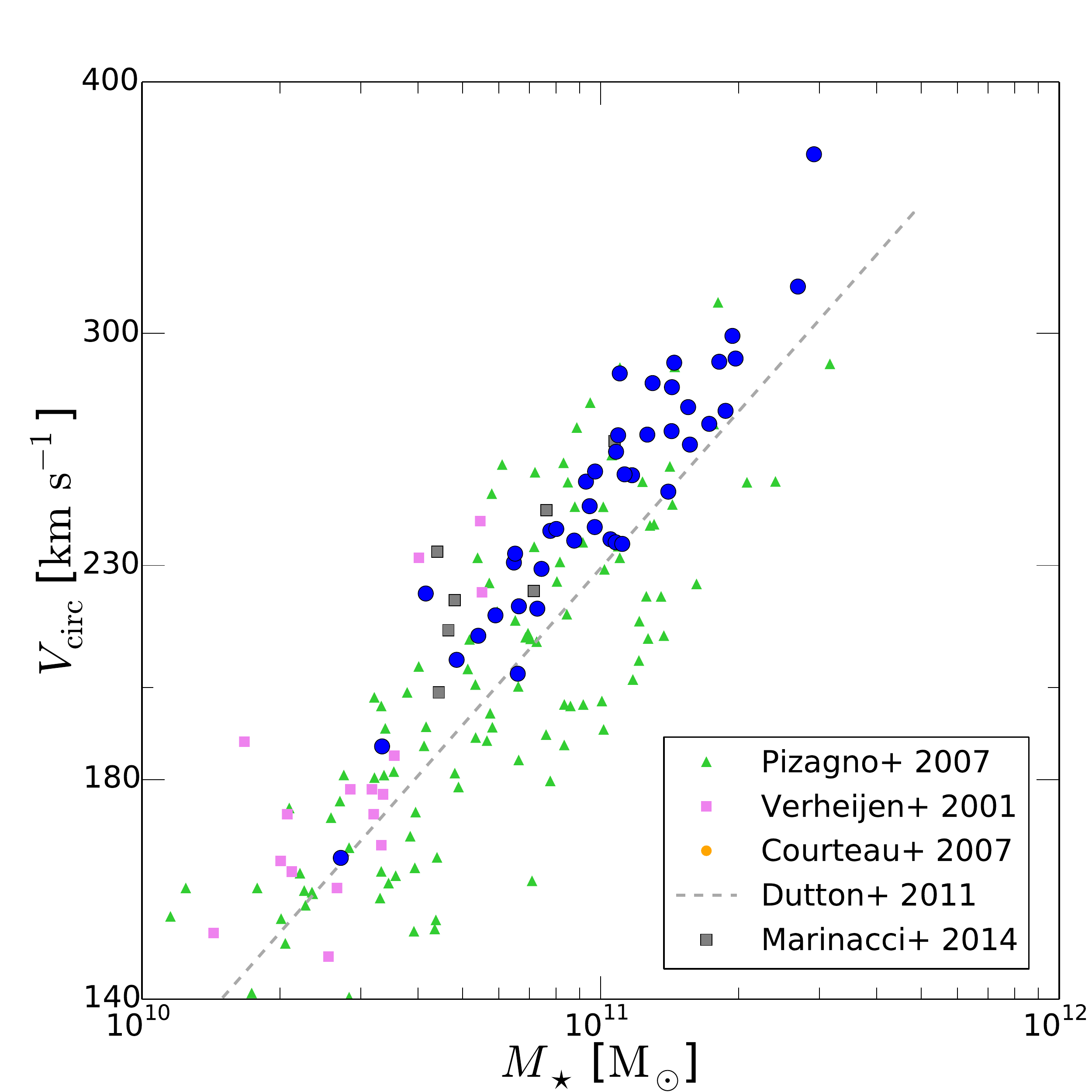}
\caption{Tully-Fisher relations. Left panel: Baryonic Tully-Fisher relation for the $42$ disk galaxies
  shown in Figure~\ref{fig:blue_galaxies}. We compare to
  observational mean trends (solid lines) with $1\sigma$ scatter
  bands~\protect\citep[][]{Avila-Reese2008,Hall2012,McGaugh2012}.  We note that this
  is only a small sub-sample of well-resolved simulated disk
  galaxies. Our simulation contains, for example, $4,177$ Milky Way
  halo analogs with virial masses in the range $10^{11.5-12.5}\msun$
  which are resolved similarly well as the galaxies presented here. We
  also include simulation points from a recent sample of
  high-resolution zoom-in simulations from~\protect\cite{Aumer2013}. Right panel: Stellar Tully-Fisher relation for the
  same sample compared to observations~\citep[][]{Verheijen2001, Pizagno2007, Courteau2007, Dutton2011}. We also
  show the theoretical predictions of \protect\cite{Marinacci2014a} for the Aquila haloes.}
\label{fig:TF_baryon}
\end{figure*}

We present a sample of ``by eye'' selected ``blue'' and ``red'' central
galaxies in Figure~\ref{fig:blue_galaxies} (blue) and
Figure~\ref{fig:red_galaxies} (red).  These images show composite g,r,i
SDSS-band light distributions. Most of the blue galaxies are star-forming disk
galaxies, whereas most of the red galaxies in the sample are ellipticals with
only little or essentially no ongoing SF.  We refrain from showing a
larger sample here, but illustrate the basic features of our galaxy population
based on these two samples. The composite images of the blue galaxies (B-1 to
B-42) in Figure~\ref{fig:blue_galaxies} reveal several interesting features.
Firstly, the sample includes barred and non-barred spirals. Star-forming
regions in spiral arms clearly stand out as blue connected knots. The opening
angle and winding of the spiral arms varies similarly to what is found
observationally along the Hubble sequence. Interesting features can also be
found in the light distribution of the redder galaxy sample (R-1 to R-42) in
Figure~\ref{fig:red_galaxies}. Since we did not make any cuts on morphology
some galaxies appear to have disk like features. A few of the galaxies like
R-2, R-7, and R-11 show shell like structures. Most likely these are remnants
of recent dry mergers as discussed above. 

Some characteristics of these galaxies are summarised in
Table~\ref{table:galaxies}, where we list virial masses, stellar masses, SMBH
masses, star formation rates, specific star formation rates,  {(g-r)} colors, r
band magnitudes, and local galaxy overdensities.  One can also see that the
selected red objects are on average more massive and brighter (r band
magnitude).  Furthermore, the {(g-r)} color values are, by selection, also
consistently higher for the red sample compared to the blue galaxies. The
reddening of the red galaxies is due to quenched SF which leads to an older
stellar population and therefore redder stellar light. The red galaxies have
massive SMBHs which cause AGN feedback, a key process in quenching massive systems.  In
fact, the SMBH population for the blue galaxies is significantly less massive
than the population of the red galaxies. We note that some of the red massive
systems are quenched so strongly that they have essentially no on-going star
formation anymore (e.g., R-2, R-3, etc.).  Galaxies are ordered by halo mass in
Table~\ref{table:galaxies}. Blue galaxies listed below B-33 are Milky Way-like
galaxies in terms of halo mass. Most of these galaxies host SMBH which are
significantly more massive than the SMBH at the Galactic center. Only B-40 has
a relatively low SMBH mass of about $9\times 10^6\msun$. All other blue galaxies in
the Milky Way's halo mass regime have more massive SMBHs.

The galaxies presented in Table~\ref{table:galaxies} all have stellar masses
above $10^{10}\msun$, and most of them have $M_\star \gtrsim 10^{11}\msun$.
This means that the stellar population in these systems is resolved with $\sim
10^5$ resolution elements. As argued above, this resolution is sufficient to
explore the internal structure of the galaxies.  We demonstrate this by
inspecting the internal mass structure of some of the blue galaxies by
constructing circular velocity profiles, $V_{\rm circ}(r)=\sqrt{G M(<r)/r}$. The results
are presented in Figure~\ref{fig:rotcurve}.  There we present circular velocity curves
for stars, gas, DM, and the total circular velocity as a function of radius. 

Late-type spiral galaxies are characterised by nearly flat circular velocity
curves without signs of a very significant bulge. The galaxies presented in
Figure~\ref{fig:rotcurve} all show a rapid rising and then nearly flat circular
velocity curve with a total amplitude depending on the total enclosed mass of
the galaxy. In all cases the contribution of the gas component is sub-dominant
compared to the stellar and DM contribution. Similar results were found
by~\cite{Marinacci2014a} using zoom-in simulations of eight Milky Way-sized
haloes from the Aquarius project~\citep[][]{Springel2008} using the Illustris
galaxy formation model. The largest gas contribution over the sample is $\sim
70\kms$, however in most cases the maximum contribution from the gas component
is around $\sim 50\kms$. The vertical lines in Figure~\ref{fig:rotcurve} show
our fiducial galaxy radius ($r_\star$) corresponding to twice the stellar
half-mass radius. The various panels of Figure~\ref{fig:rotcurve} also
demonstrate how the stellar circular velocity curve and the DM circular
velocity curve add up to a very flat total circular velocity curve. The inner
parts of the galaxy are clearly dominated by the stellar mass. Beyond $r_\star$
the DM contribution to the circular velocity curve dominates for all galaxies
over the stellar contribution. The circular velocity curves stay flat out to
$50\kpc$ for all galaxies shown in Figure~\ref{fig:rotcurve}. We also include
in each panel the same eight circular velocity curves for the Aquila haloes
taken from \cite{Marinacci2014a} (gray thin lines). We note that our mass and
spatial resolution is inferior to these simulations. However, \cite{Marinacci2014a}
demonstrated that the circular velocity curves are reasonably well converged
even if the mass resolution is changed by a factor of $64$. Nevertheless, the
initial rise and the exact location of the maximum of the circular velocity
curve can still vary to some degree \citep[see][for details]{Marinacci2014a}. A
study of the full galaxy population is provided in \cite{Genel2014},
where average circular velocity curves for the whole population, for a wide
range of masses, and several redshifts are presented. There we also study the
impact of baryons on the maximum circular velocity.

Disk galaxies are observationally found to follow the Tully-Fisher relation (TFR)~\citep[][]{Tully1977}
relating stellar luminosities (masses) to internal kinematics via the circular
velocities. It has been shown that this relation becomes even tighter when the
circular velocities are correlated with the total baryonic mass instead of the
luminosities or stellar masses, leading to a very tight baryonic TFR
(BTFR)~\citep[][]{McGaugh2012}. We note that the BTFR is a strong prediction of
Modified Newtonian Dynamics (MoND)~\citep[][]{Milgrom1983} as an alternative of
CDM. We present the BTFR of the $42$ disk like galaxies of
Figure~\ref{fig:blue_galaxies} in the left panel of Figure~\ref{fig:TF_baryon}
along with recent observations including $1\sigma$ uncertainty
bands~\citep[][]{Avila-Reese2008,Hall2012, McGaugh2012}.  We measure the total
baryonic mass within $r_\star$, and for the circular velocity we take the total
mass within that radius and calculate the associated circular
velocity~\citep[see also][]{Scannapieco2012}.  The $r_\star$ radii are shown as
dashed brown lines in the circular velocity curves of
Figure~\ref{fig:rotcurve}, demonstrating that this radius lies already within
the flat regime of the circular velocity curve. Our results for the TFR are
therefore not very sensitive to this choice.  

\begin{figure}
\centering
\includegraphics[width=0.49\textwidth]{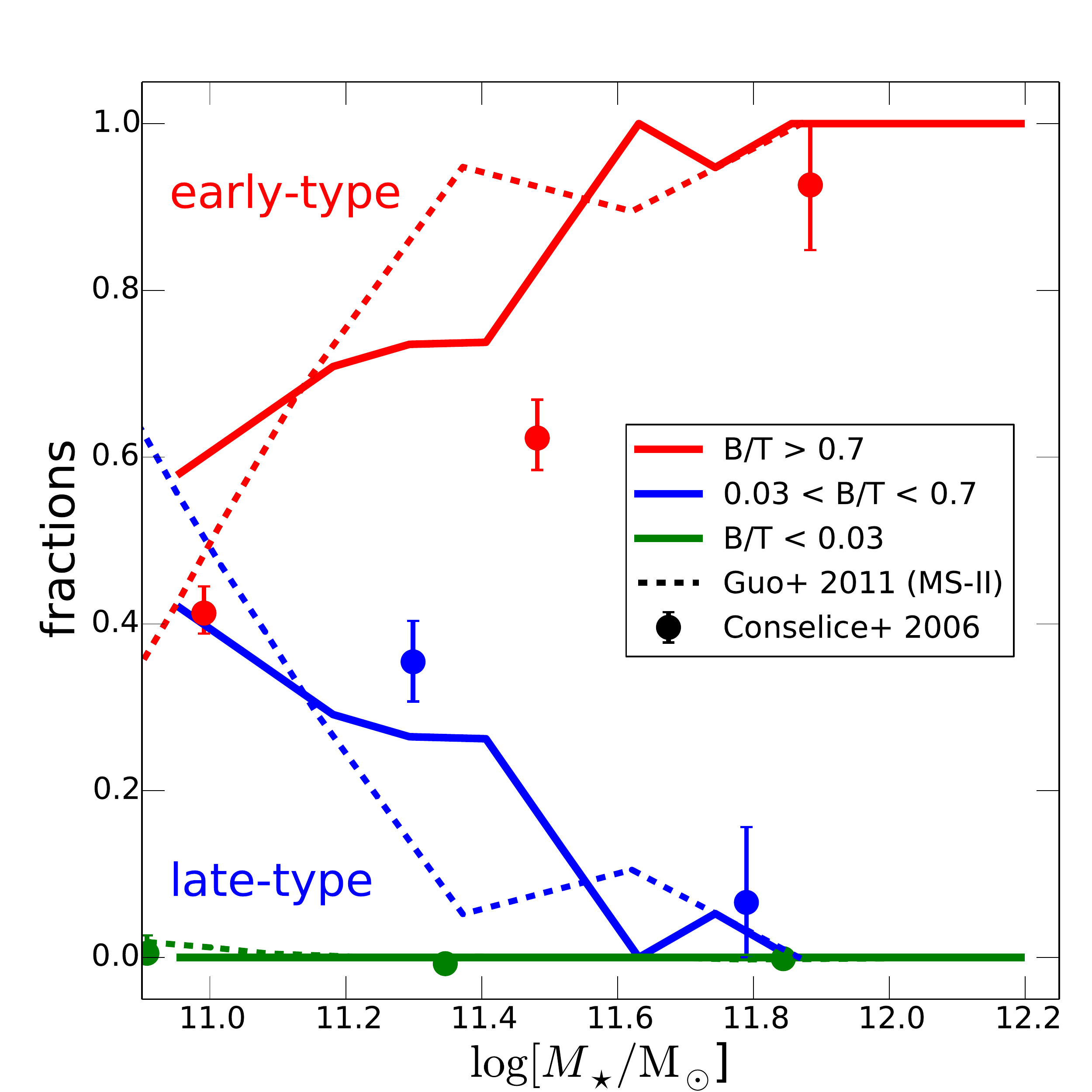}
\caption{Distribution of galaxy types based on a kinematic classification using
thresholds in bulge-to-disk (B/T) ratios. The red line represents
bulge-dominated systems like early type galaxies. The blue line represents
spiral galaxies with some bulge contribution, whereas the green line shows the
fraction of pure-disk galaxies. We compare our results to observations from
\protect\cite{Conselice2006}. We also show the predictions of the semi-analytic
model of \protect\cite{Guo2011} based on the Millennium-II (MS-II) simulation.
We can probe only massive systems due to resolution limitations. Objects with
$M_\star \lesssim 10^{11}\msun$ contain fewer than $10^5$ stellar particles
which is insufficient for a reliable galaxy type classification. Despite the
small mass range covered, we find that the simulation correctly describes the
transition from the bulge-dominated regime for more massive systems to the
disk-dominated regime for lower mass galaxies. We employ the same B/T cut as
\protect\cite{Guo2011} and find that the `crossing point' of the galaxy types
is predicted to be at nearly the same mass within the two galaxy formation
models.}
\label{fig:dtratio}
\end{figure}

The BTFR of the disk galaxies in our simulation (represented here by this small
sample) agrees well with the overall observational constraints, demonstrating
that the internal structure of the stellar disks are characterised reasonably well,
and that CDM models can reproduce the observed BTFR.  Interestingly, our model
predicts a BTFR which is closer to the tightest observational
constraints~\citep[][]{McGaugh2012} than recent state-of-the-art zoom-in
simulations~\citep[][]{Aumer2013} although our numerical resolution for these
galaxies is significantly below the resolution of the zoom-in results. The very
tight and steep BTFR is observed for a gas-rich galaxy sample, which is
expected to give a more accurate measure of the slope and the normalisation of
the BTFR than obtained from star-dominated spiral galaxies. Although our
simulation results agree with the predicted scatter of
\cite{Avila-Reese2008,Hall2012} the spread of our BTFR is still much larger
than the results of~\cite{McGaugh2012}. Also the slope is not as steep as
predicted by that study.  We note that a more detailed exploration of the BTFR
has to take into account the full sample of well-resolved spiral galaxies, and
also take into account the actual rotation velocities.

The right panel of Figure~\ref{fig:TF_baryon} shows the stellar TFR for the
same selected disk galaxy sample compared to different
observations~\citep[][]{Verheijen2001, Pizagno2007, Courteau2007,
Dutton2011};~i.e. we plot the same velocity as in the left panel now as a
function of stellar mass instead of total baryonic mass. We also show the
theoretical predictions of \cite{Marinacci2014a} for the Aquila haloes. Our
galaxies follow a similar trend as the high-resolution Aquila haloes: we
recover the correct slope and amount of scatter in the relation. However, it
seems that our results indicate slightly too high circular velocities. This is
also true for the Aquila galaxies of \cite{Marinacci2014a}. 

The blue and red galaxy samples discussed so far were selected ``by eye'' to be
representative for distinct classes of galaxies. In the following we would like
to characterise in more detail the morphological galaxy mix as a function of
stellar mass. However, we are severely limited by our mass resolution to
properly model and characterise galaxy types of systems that are only resolved
by a few tens of thousands of stellar resolution elements. In fact, reliably
identifying the type of a galaxy requires substantially more particles, and we
will not attempt to quantify the morphological type of galaxies for systems
that are resolved with less than $\sim 10^5$ stellar particles resulting in a
lowest stellar mass of about $\sim 10^{11}\msun$.  Automatically classifying
galaxy types for less well-resolved objects is rather difficult and the
obtained results are highly uncertain. 

For the well-resolved objects with $M_\star > 10^{11}\msun$ we will apply a
kinematic bulge-to-disk decomposition. Specifically, we follow \cite{Abadi2003}
and define for every star particle with specific angular momentum $j_z$ around
a selected $z$-axis a circularity parameter $j_z/j(E)$, where $j(E)$ is the
maximum specific angular momentum possible at the specific binding energy $E$
of the star. We define a $z$-axis based on the star-forming gas, or the stars,
if there is no star-forming gas in the system, which can occur in more massive
and heavily quenched systems. Having the circularities of all stellar particles
of the system we can then determine the fraction, $f_{> 0.7}$, of particles
with circularities above $0.7$. These stars are typically classified as disk
stars~\citep[see also][]{Marinacci2014a}. We then calculate the bulge-to-total
ratio (B/T) as $1-f_{> 0.7}$ for each galaxy. This gives for every galaxy a
bulge-to-total ratio for the stellar component based on the circularities of
the stars. Based on those values we can then split the distribution of galaxies
into different types. Here we follow the cuts applied in \cite{Guo2011}, where
the morphological types of galaxies based on the Millennium-I and -II (MS-I,
MS-II) simulations combined with a semi-analytic model were compared to
observations from~\cite{Conselice2006}. As described above we can probe the
morphological types only for the most massive galaxies in our sample due to the
resolution requirements,~i.e. we can probe the types only over a much smaller
range than \cite{Guo2011}, for example. 

We present our results in Figure~\ref{fig:dtratio}, where we compare our
findings to the observations of~\cite{Conselice2006} and to the predictions of
the semi-analytic model of~\cite{Guo2011} (based on the MS-II). Although
Figure~\ref{fig:dtratio} only probes the most massive systems in our
simulations, we find that we recover the observed trends for the mass
dependence of the different galaxy types. Most importantly, we find the correct
transitional behaviour from more disk-like systems at lower masses to
bulge-dominated systems towards larger stellar masses. Over this mass range we
find a similar level of agreement with the observations as the
predictions of \cite{Guo2011}. However, this should not be over-interpreted
since the comparison to observations is not unique due to the different
techniques that were applied to extract the morphological types in the
observations and the simulations. Our cuts are identical to those
applied in \cite{Guo2011}, and it is therefore interesting that we find a similar
``crossing-mass'' of about $M_\star \sim 10^{11}\msun$ as the
semi-analytic prediction.

\begin{figure}
\centering
\includegraphics[width=0.49\textwidth]{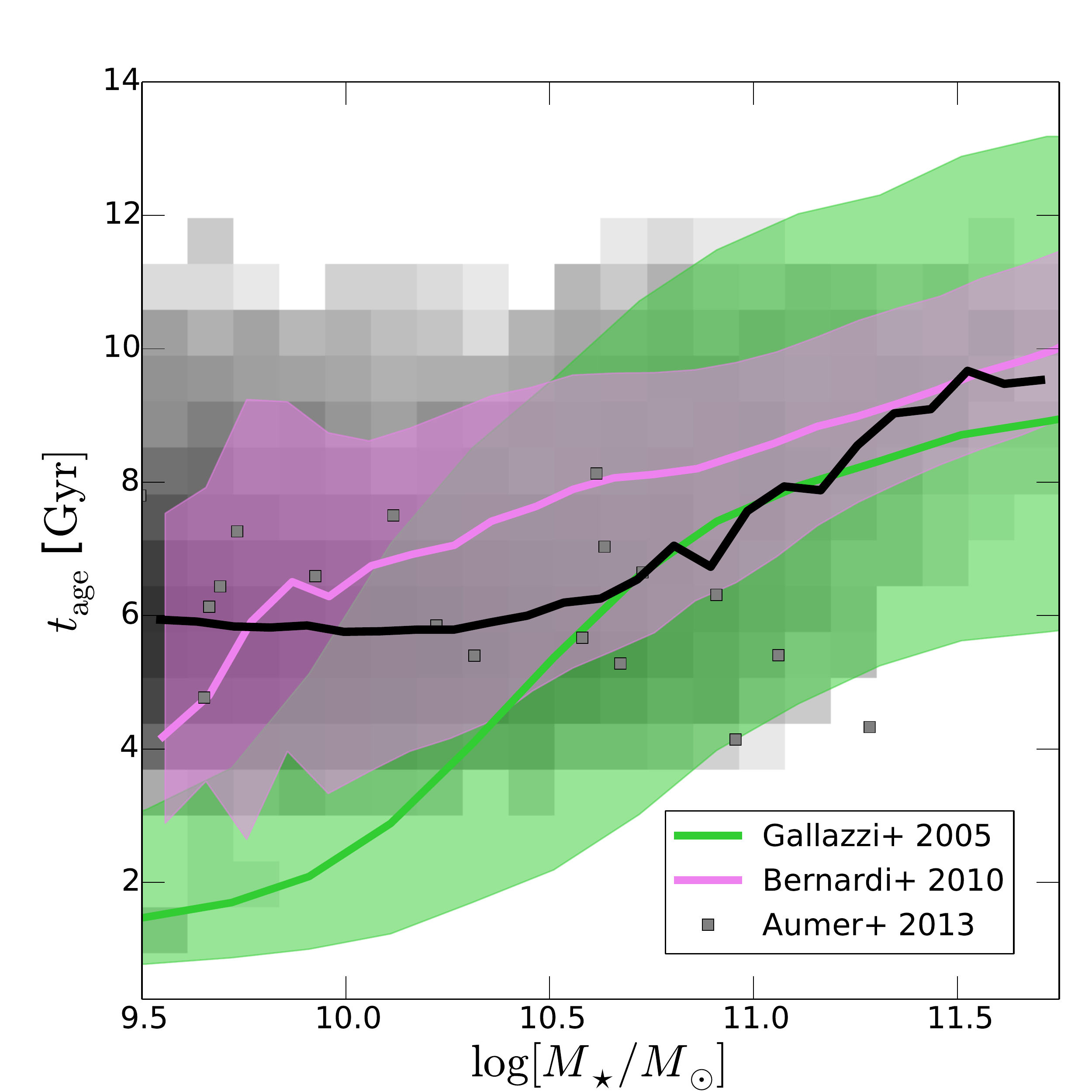}
\caption{Stellar age as a function of stellar mass. The black line
  shows the simulation median, and the background histogram all
  simulated galaxies. Colored lines show two observational samples:
  one covering a large fraction of SDSS galaxies~\protect\citep[][]{Gallazzi2005},
  and one showing a subsample of early-type
  galaxies~\protect\citep[][]{Bernardi2010}. We also include the simulation results from
  \protect\cite{Aumer2013}. Our model predicts the mean ages of galaxies correctly for systems
  more massive than $\sim 10^{10.5}\msun$. However, low mass systems tend to be too old compared to
  the full observational sample. They seem to agree better with the early-type subset. This problem of low mass galaxies being too old is shared with the model of \protect\cite{Aumer2013}, which predicts 	even older galaxies around $M_\star \sim 10^{10}\msun$.}
\label{fig:stellar_AGE}
\end{figure}

As a final point of comparison between our model and an observable of the local
Universe, we also investigate the age of the stellar populations in galaxies.
This quantity can be inferred observationally and can therefore be used to test
our model.  Simply speaking, SF in galaxies is either biased towards earlier or
later periods of structure formation, depending on the galaxy type: early-type
galaxies form most of their stellar content early, whereas late-type galaxies
show more late SF activity.  This ``downsizing'' effect, which is ultimately
caused by baryonic processes like feedback, works therefore contrary to the
hierarchical bottom-up build-up of CDM, where larger systems like clusters
assemble later.  We can inspect downsizing in our model by plotting the
mass-weighted stellar age of simulated galaxies against their stellar mass, as
shown in Figure~\ref{fig:stellar_AGE}. We compare here to two observational
datasets: one derived from a large SDSS sample~\citep[][]{Gallazzi2005}, and a
second one based on a SDSS subset biased towards early-type
galaxies~\citep[][]{Bernardi2010}. We find reasonable agreement within the
scatter and the total sample. Most importantly, we recover the correct trend of
age downsizing. However, low-mass systems ($M_\star \lesssim 10^{10.5}\msun$)
tend to form their stars too early in our simulation; i.e.~they are too old
compared to observations. We stress that this tension is shared by
semi-analytic models and other state-of-the-art hydrodynamical simulations
alike, pointing towards an open problem in low-mass galaxy
formation~\citep[][]{Weinmann2012,Henriques2013}. It remains to be seen whether new more
explicit feedback models can resolve this problem~\citep[][]{Hopkins2013}.

\section{Summary}

We have introduced the Illustris Project, a series of large-scale
hydrodynamical and DM-only simulations of galaxy formation. The highest
resolution simulation, Illustris-1, covers a volume of $(106.5\mpc)^3$, has a
dark mass resolution of ${6.26 \times 10^{6}\msun}$, and an initial baryonic
matter mass resolution of ${1.26 \times 10^{6}\msun}$.  At $z=0$ gravitational forces are
softened on scales of $710\pc$, and the smallest hydrodynamical gas cells have
a fiducial radius of $48\pc$.  The simulation follows the evolution of $2\times 1820^3$
resolution elements and additionally evolves $1820^3$ passive Monte
Carlo tracer particles reaching a total particle count of more than $18$
billion. The free parameters of our model are set to physically plausible values and
have been constrained based on the star formation efficiency using smaller
scale simulations~\citep[][]{Vogelsberger2013} and with minor modifications in
high-resolution zoom-in simulations of individual Milky Way-like
haloes~\citep[][]{Marinacci2014a, Marinacci2014b}.  They were also used in
recent magneto-hydrodynamical simulations of Milky Way-like
haloes~\citep[][]{Pakmor2014}.
 
In this first paper and a companion paper \cite{Nature2014}, we have
presented initial results focusing on the galaxy population at $z=0$. High
redshift results are presented in \cite{Genel2014}. We will summarise
in the following our main results.

\vspace{0.25cm}\noindent
{\bf Impact of baryons on the halo mass function:} 
We find that galaxy formation physics affects the halo mass function at low and
high halo masses, leading to a reduction of the overall abundance by up to
$\sim 30\%$ at both ends.  In these mass ranges baryonic feedback processes are
strongest. For intermediate mass haloes we find deviations below $\lesssim
10\%$. Our radio-mode implementation leads to a significant reduction in halo
mass also towards higher masses, which is different from previous
findings~\citep[e.g.][]{Sawala2013}, which do not include AGN feedback. We
note, however, that other implementations of radio-mode feedback may yield a
different outcome at higher halo masses.

\vspace{0.25cm}\noindent
{\bf Impact of baryons on halo masses:} 
Related to the change in the mass function we also find that individual halo
masses are significantly affected by baryonic physics: haloes less or more
massive than $\sim 10^{11}\msun$  have up to $\sim 20-30\%$ less mass in
Illustris-1 compared to matched haloes in the corresponding
DM-only simulation. These trends differ from most previous results, where the
strongest impact of baryonic physics was found towards lower halo masses
\citep[e.g.,][]{Sawala2013}. However, our energetic AGN feedback also causes
halo masses to differ substantially in more massive systems. Intermediate mass
haloes around $\sim 10^{11}\msun$ tend to be more massive in the hydrodynamical
simulation by about $\sim 10\%$.

\vspace{0.25cm}\noindent
{\bf Cosmic star formation rate density:} 
By construction, the build-up of stellar mass predicted by our galaxy formation
model agrees well with observations. We have tested our results against recent
measurements of the cosmic star formation rate density.  We find good agreement
with observations up to $z \sim 10$, although our model predicts slightly too
much present-day star formation which is dominated by galaxies with stellar
masses about $M_\star \sim 10^{10-10.5}\msun$.  At high redshifts we find good
agreement of our model with recent lower limits of the cosmic star formation
rate densities.  Beyond $z\sim 8$ our model agrees best with the shallowest
slope ($-10.9+2.5$) of recent best-fit slopes for combined $z\geq8$
measurements ($\propto (1+z)^{-10.9 \pm 2.5}$) by \cite{Oesch2013}.

\vspace{0.25cm}\noindent
{\bf Galaxy stellar mass and luminosity functions:} 
The present-day abundance of galaxies both as a function of stellar mass and
stellar luminosity agrees well with observations of the local Universe as
indicated by a comparison with SDSS based observations. We stress that for both
quantities there are significant observational measurement uncertainties in the
stellar light assignment at the bright and massive
end~\citep[see][]{Bernardi2013} that we take into account by providing two
stellar light and mass estimates. We find reasonable agreement of our
simulation with observations over a substantial mass ($M_\star \sim 10^9 -
10^{12.5}\msun$) and luminosity range ($M_{r} \sim -15.0 - -24.5$). We also
find good agreement with the stellar mass function of star-forming galaxies,
and reproduce the measured fraction of ``red and dead'' galaxies at the massive
end: for $M_\star>10^9\msun$ we find that about $52\%$ of all stellar mass at
$z=0$ is in galaxies with a low SF rate, which agrees reasonably well with
observations of~\cite{Moustakas2013},

\vspace{0.25cm}\noindent
{\bf Baryon conversion efficiency:} 
We have compared the baryon conversion efficiency of our model to recent
abundance matching results, where we recover the inferred observation results
within the $1\sigma$ uncertainties.  Baryon conversion is most efficient around
halo masses $M_{\rm 200, crit} \sim 10^{12}\msun$, and drops quickly for lower
and higher mass systems due to SN and AGN feedback. Our model predicts a
maximum efficiency of about $\sim 20\%$ at halo masses slightly above
$10^{12}\msun$ in reasonable agreement with the findings of abundance matching
techniques, although the results of \cite{Behroozi2013} peak at higher masses
than those of \cite{Moster2013}. The exact amplitude and location of the peak
predicted by our model also depends on the exact definition of stellar mass:
considering the total stellar mass of the halo, excluding satellites, peaks at
a slightly higher mass than the efficiency based on the galactic stellar mass
only. However, this uncertainty is smaller than the differences between
different abundance matching results.  We stress that the interplay of SN and
AGN feedback is crucial to reproduce this result and especially the shape of
the stellar mass to halo mass relation. 

\vspace{0.25cm}\noindent
{\bf Colours of galaxies:}
In this first paper we have studied only intrinsic {g-r} and {u-i} colours of
galaxies neglecting the important effect of dust. Given this caveat, we find a bimodal colour
distribution which has a significantly less pronounced maximum at red colours
compared to observations. The most massive galaxies with $M_\star \sim
10^{12}\msun$  peak around ${\rm u-i}\sim 2.8$. The $M_\star=10^{10.5}\msun$ to
$M_\star=10^{11.0}\msun$ mass range shows a clear bimodal distribution with
broad maxima at ${\rm u-i}\sim 1.5$ and ${\rm u-i}\sim 2.75$. Lower mass
systems around $M_\star \sim 10^{10}\msun$ show a broad maximum at ${\rm
u-i}\sim 1.4$. Our low and high mass results agree reasonably well with
observations and predictions from the recent semi-analytic models
of~\cite{Guo2011}. However, we do not match the colours of intermediate mass
systems. A more detailed study taking into account the effects of dust is
required to understand this discrepancy better. We divide the galaxy population
also into central and satellite galaxies finding that for all galaxies above
$M_\star \sim 10^9\msun$ the red part of the {g-r} and {u-i} distribution is
dominated by satellites. For galaxies more massive than $M_\star \sim
10^{11}\msun$ we find however that centrals dominate the red maximum of the
distribution in both colours. We have also inspected the red fraction of
galaxies as a function of stellar mass, specific star formation rate, and BH
mass. The most rapid transition from blue galaxies to red galaxies can be found
as a function of specific star formation rate, where our model predicts a sharp
transition from blue to red around $10^{-2}\Gyr^{-1}$. The transition as a
function of BH mass and stellar mass is significantly broader. 

\vspace{0.25cm}\noindent
{\bf The impact of environment on the galaxy population and mass and environment quenching:}
We have studied galaxy colours {(g-r)} and red fractions as a function of
stellar mass ($M_\star$) and galaxy overdensity ($1+\delta$), where $\delta$
was constructed based on the distance to the fifth nearest neighbouring galaxy
with an r band magnitude brighter than $-19.5$. We recover qualitatively the
observational trends of mass and environment quenching.  Specifically,  we
recover the observational trends of~\cite{Peng2010} when plotting the red
fraction as a function of mass and galaxy overdensity. Interestingly, we also
reproduce the drop in the red fraction around $M_\star \sim 10^{10}\msun$ for
high galaxy overdensities $1+\delta \gtrsim 10^2$, where a slightly shifted
$0.9$ red fraction contour line agrees remarkably well with our simulation
predictions. We demonstrate that this high density drop in the red fraction is
less pronounced once the environment is characterised with an estimator that is
less contaminated by a remaining halo mass dependence ($D_{1,1}$ taken
from \cite{Haas2012} indicating that $\delta$ is not ideal in probing
environmental effects alone~\cite[see also the discussion in ][]{Haas2012}.
Our model also reproduces the observed environment independence of star
formation rates for star-forming galaxies  in agreement with the findings
of~\cite{Peng2010}. This demonstrates that high density environments increase
the probability of galaxies being quenched, but does not directly affect the
star formation rates of non-quenched, star-forming, galaxies. 

\vspace{0.25cm}\noindent
{\bf Morphology of galaxies:}
The numerical resolution of Illustris is sufficient to study the internal
structure of relatively more massive galaxies ($M_\star \gtrsim 10^{11}$),
which are resolved with more than $\sim 10^5$ stellar resolution elements. We
have constructed small samples of red and blue galaxies (R-1 to R-42 and B-1 to
B-42) which are well-resolved and representative of our galaxy population. A
visual classification of these galaxies reveals that our model clearly yields a
significant population of star-forming blue disk-like galaxies co-existing with
quenched red elliptical galaxies. We have extended this classification to all
galaxies with $M_\star \gtrsim 10^{11}\msun$ using a kinematic bulge-disk
decomposition. We find that our simulation correctly describes the transition
from late type galaxies to early types as a function of stellar mass. We have
compared the relative fractions in both populations with the observational
sample from \cite{Conselice2006} and the semi-analytic prediction
of~\cite{Guo2011} finding good agreement among the three. Interestingly, our
model predicts the same transition mass $M_\star \sim 10^{11}\msun$, the mass
where the fraction of early- and late-types both equals $50\%$, as the recent
semi-analytic model of~\cite{Guo2011} using very similar cuts in the
bulge-to-disk ratios for the galaxy type classification.

\vspace{0.25cm}\noindent
{\bf Circular velocity curves and Tully-Fisher relation:}
We have constructed circular velocity curves for a few selected late-type
spiral galaxies with halo masses between $\sim 10^{12}\msun$ and $\sim
10^{13}\msun$. All selected galaxies show steeply rising circular velocity
curves which then become flat towards larger radii. The gas contribution to the
circular velocity profiles is not significant and typically is at most $\sim
50\kms$ and in rare cases $\sim 70\kms$.  Our model reproduces the observed
stellar and baryonic Tully-Fisher relation reasonably well. We have
demonstrated this based on a small subset of spiral galaxies. As a function of
stellar mass these galaxies tend to lie a few ${\rm km}\,{\rm s}^{-1}$ above
the observed trend.  The slope and spread of the baryonic Tully-Fisher
relations is reproduced when compared to the observations
of~\cite{Avila-Reese2008} and~\cite{Hall2012}.  However, we do not reproduce
the steep slope and small scatter of the relation found by~\cite{McGaugh2012}.
Nevertheless, the baryonic Tully-Fisher of our selected sample agrees slightly
better with observations than the recent high resolution simulations
of~\cite{Aumer2013}.

\vspace{0.25cm}\noindent
{\bf Stellar ages:}
Our galaxy formation model works well for most of the observational probes
considered here.  However, the stellar ages of low mass galaxies are not
predicted correctly.  Specifically, we recover the correct trend of downsizing
with respect to stellar ages, but low mass galaxies ($M_\star \lesssim
10^{10.5}\msun$) are a factor two to three too old compared to observational
constraints, which point towards very young stellar populations for these
masses.  In our simulation, these systems tend to form their stars too early.
The agreement at higher masses, where our galaxies are resolved better, is
good. Although the failure at low stellar masses could be related to resolution
effects, we doubt that this can fully explain the discrepancy. 

\vspace{0.25cm}\noindent
{\bf Summary:}
In summary we find that the Illustris simulation is able to reproduce many of
the observed trends in the local Universe reasonably well. There are still some
discrepancies related to the stellar ages of low mass galaxies and the
quenching of massive galaxies, which require further investigation.  We also
stress that it is not advisable to tune galaxy formation models such that they
reproduce any observation exactly.  For example, the systematic errors at the
bright end of the galaxy stellar mass function~\citep[][]{Bernardi2013} have to
be considered when constructing a galaxy formation model. The same is true for
reproducing the stellar to halo mass relationship, which is also uncertain
towards higher masses~\citep[][]{Kravtsov2014}. Different abundance matching
results differ also towards lower halo
masses~\citep[see][]{Moster2013,Behroozi2013}, and they might even break down
towards dwarf-scale masses~\citep[][]{Sawala2014}. Such systematic
uncertainties have to be considered when trying to match observational data
with semi-analytic models or hydrodynamical simulations. Special care needs to
be taken when invoking new physical mechanisms for feedback or star formation
to accommodate certain aspects of observational data if systematic errors have
to be considered. Reproducing observational data perfectly does not imply a
correct and physically meaningful model.

\section{Conclusion and Outlook}

We have demonstrated in this study that state-of-the-art hydrodynamical
simulations like Illustris are capable of reproducing a significant amount of
observational data both on large and on small scales. Those simulations have
now reached the mass and spatial resolution of recent large-scale DM only
simulations. Although not every observed relation is reproduced precisely by
our model, it seems that the most important aspects of galaxy formation may have
been identified. A model universe like Illustris can now be used to refine our
understanding of galaxy formation. The way to move forward is to identify
remaining problems of the model, where the predictions severely disagree with
observational data. One such prediction is the age distribution of low mass
galaxies, which conflicts with observations. Furthermore, at the massive end we
find that our AGN feedback is not efficient enough to lead to sufficient
quenching of massive galaxies. Understanding these problems will lead to better
insights into the relevant physics. 

\begin{figure}
\centering
\includegraphics[width=0.49\textwidth]{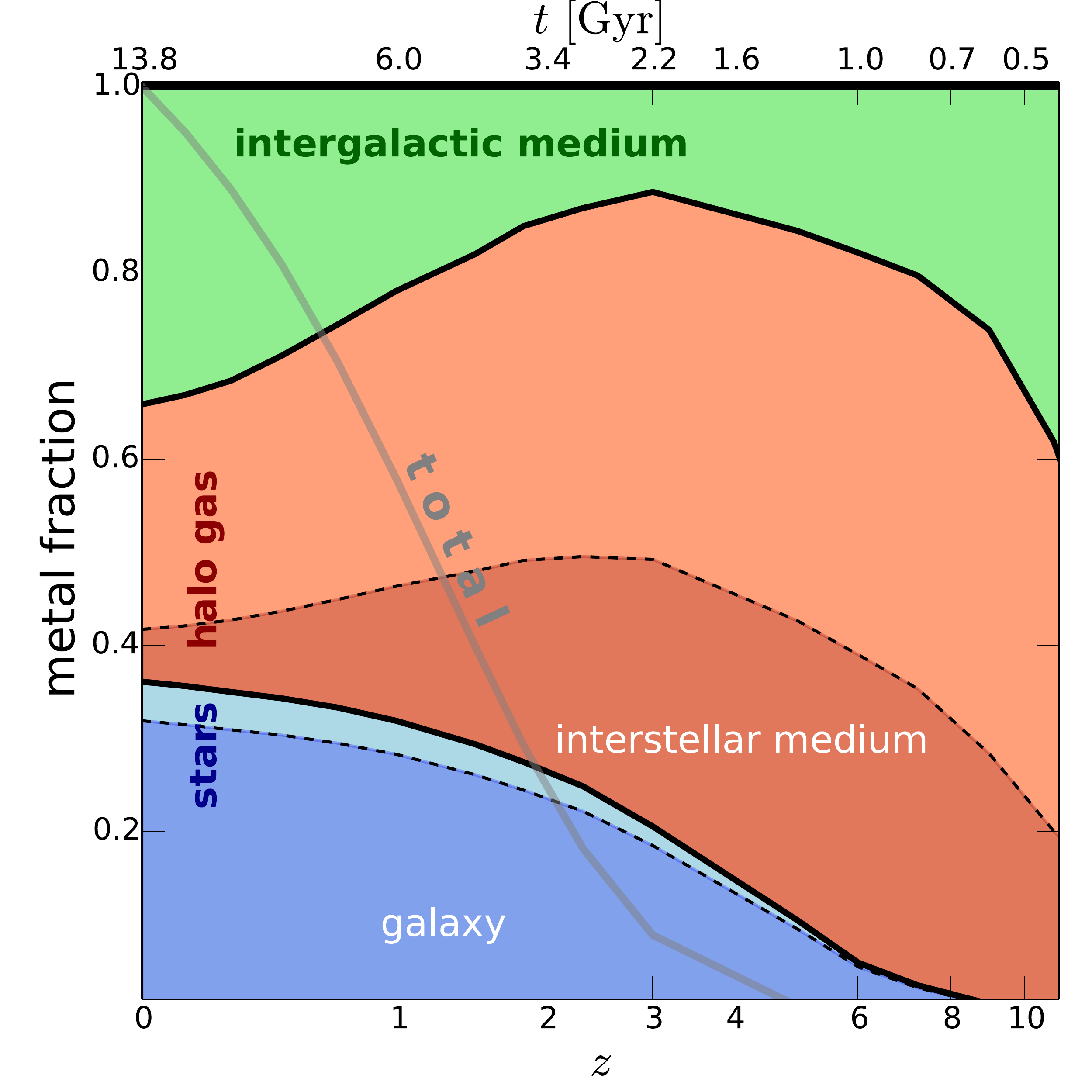}
\caption{Fractional metal content in different baryonic phases as a function of
time. We divide the gas in three main components: stars (stars bound to
haloes), halo gas (gas bound to haloes), and unbound gas (intergalactic
medium). Stars are further subdivided into stars belonging to the galaxy
(within $r_\star$) and stars outside the galaxy. A similar division is made for
the gas, where we divide between gas belonging to the interstellar medium, and
other halo gas, which is gravitationally bound.  The thick gray line shows the
increase of the total metal content on the
same linear scale normalised to unity at $z=0$. Our
simulation predicts ``metal equipartition'' at $z \sim 0$ such that
each baryon phase holds about one third of the total metal content in the
Universe. At all other redshifts the situation is rather different, for example,
at high redshifts halo gas largely dominates the metal budget. }
\label{fig:metals}
\end{figure}

Besides comparing to observational data it is crucial that galaxy formation
models also make distinct predictions that can be tested in the future to
falsify certain aspects of the model.  Our galaxy formation model, for example,
also follows the build-up and distribution of metals as described above. Metals
can be found in many different ``baryonic phases'' in the Universe. They are
produced in stars and distributed into the CGM and IGM via SN driven winds. We
can use our simulation to predict in which baryonic phase we expect metals to
be found at various redshifts.  A first coarse attempt of this is presented in
Figure~\ref{fig:metals}, where we distinguish three main baryonic phases:
stars, gravitationally bound gas, and gas which is not bound to any halo.  We
further subdivide the stellar contribution into stars which belong to the
galaxy ($r < r_\star$), and those which are part of the outer halo.  We make an
additional distinction for bound gas where we separate star-forming
interstellar gas from non-star forming gas. Cold and dense gas forms stars,
which subsequently synthesise metals. This explains why the contribution of
metals in stars is increasing and the relative contribution in the gas phase is
decreasing. SN driven winds enrich the intergalactic medium, leading to more
metals in this diffuse phase at later times. Interestingly, our simulation
predicts a kind of ``metal equipartition'' at $z \sim 0$ such that each
baryonic phase holds about one third of the total metal content in the
Universe. At all other redshifts the situation is rather different, for
example, at high redshifts halo gas largely dominates the metal budget. Our
simulation provides therefore detailed predictions of the evolving metal
abundances in all baryonic phases that can potentially be probed through
upcoming observations allowing us to understand better the complicated ``baryon
cycle'' of galaxies. It remains to be seen how well these predictions can be
tested in the future and what physical insight can be extracted through even
more detailed comparisons between simulations and observations.

\section{Data release and access}

The Illustris simulation data in the form of various group catalogues, merger
trees (subhalo-, halo-, and galaxy-based), light-cone mock observations,
pre-processed snapshot data in HDF5 format allowing access to every stored
quantity at all redshifts, galaxy images, galaxy spectra and broad-band
magnitudes, etc. will be made available in the near future. Data distribution
will be handled through \url{http://www.illustris-project.org}. There we also
provide additional visual material in the form of movies and images. The
website also hosts a simulation explorer interface with query functionality to
browse the simulation data interactively. This explorer provides $17.2$
gigapixel images of the gas distribution (density, temperature, metallicity,
etc.), stellar density, dark matter density, and SZ- and X-ray maps. It can be
used interactively to query the group catalogues, merger trees, galaxy images,
SMBH data etc. Galaxy images from an  image post-processing pipeline (Torrey et
al., in prep) are also available, resulting in  galaxy image and spectral
databases with a total data volume of about $1$ petabyte in its final
installation.

\section*{Acknowledgements}
MV thanks Qi Guo and Federico Marinacci for providing data.  VS acknowledges
support by the DFG Research Centre SFB-881 ``The Milky Way System'' through
project A1, and by the European Research Council under ERC-StG EXAGAL-308037.
GS acknowledges support from the HST grants program, number HST-AR-12856.01-A.
Support for program \#12856 (PI J. Lotz) was provided by NASA through a grant
from the Space Telescope Science Institute, which is operated by the
Association of Universities for Research in Astronomy, Inc., under NASA
contract NAS 5-26555.  LH acknowledges support from NASA grant NNX12AC67G and
NSF grant AST-1312095.  DX acknowledges support from the Alexander von Humboldt
Foundation. Simulations were run on the Harvard Odyssey and CfA/ITC clusters,
the Ranger and Stampede supercomputers at the Texas Advanced Computing Center
as part of XSEDE, the Kraken supercomputer at Oak Ridge National Laboratory as
part of XSEDE, the CURIE supercomputer at CEA/France as part of PRACE project
RA0844, and the SuperMUC computer at the Leibniz Computing Centre, Germany, as
part of project pr85je.

\label{lastpage}

\end{document}